\documentclass[10pt,epsf,preprint]{aastex}
\usepackage{epsf}
\input psfig 

\begin{document}
\title{Early-type galaxies in the SDSS. I. The sample}

\author{
Mariangela Bernardi\altaffilmark{\ref{Chicago},\ref{CMU}},
Ravi K. Sheth\altaffilmark{\ref{Fermilab},\ref{Pitt}},
James Annis\altaffilmark{\ref{Fermilab}},
Scott Burles\altaffilmark{\ref{Fermilab}},
Daniel J. Eisenstein\altaffilmark{\ref{Arizona}},
Douglas P. Finkbeiner\altaffilmark{\ref{Berkeley},\ref{Princeton},\ref{HF}},
David W. Hogg\altaffilmark{\ref{NYU}},
Robert H. Lupton\altaffilmark{\ref{Princeton}},
David J. Schlegel\altaffilmark{\ref{Princeton}}, 
Mark Subbarao\altaffilmark{\ref{Chicago}},
Neta A. Bahcall\altaffilmark{\ref{Princeton}},
John P. Blakeslee\altaffilmark{\ref{JHU}},
J. Brinkmann\altaffilmark{\ref{APO}},
Francisco J. Castander\altaffilmark{\ref{yale},\ref{chile}},
Andrew J. Connolly\altaffilmark{\ref{Pitt}}, 
Istvan Csabai\altaffilmark{\ref{Eotvos},\ref{JHU}},
Mamoru Doi\altaffilmark{\ref{Tokyo1},\ref{Tokyo2}},
Masataka Fukugita\altaffilmark{\ref{ICRR},\ref{IAS}},
Joshua Frieman\altaffilmark{\ref{Chicago},\ref{Fermilab}},
Timothy Heckman\altaffilmark{\ref{JHU}},
Gregory S. Hennessy\altaffilmark{\ref{USNO}},
\v{Z}eljko Ivezi\'{c}\altaffilmark{\ref{Princeton}},
G. R. Knapp\altaffilmark{\ref{Princeton}},
Don Q. Lamb\altaffilmark{\ref{Chicago}},
Timothy McKay\altaffilmark{\ref{UMich}},
Jeffrey A. Munn\altaffilmark{\ref{USNO}},
Robert Nichol\altaffilmark{\ref{CMU}},
Sadanori Okamura\altaffilmark{\ref{Tokyo3},\ref{Tokyo2}}, 
Donald P. Schneider\altaffilmark{\ref{PSU}},
Aniruddha R. Thakar\altaffilmark{\ref{JHU}},
and Donald G.\ York\altaffilmark{\ref{Chicago}}
}

\newcounter{address}
\setcounter{address}{1}
\altaffiltext{\theaddress}{\stepcounter{address}
University of Chicago, Astronomy \& Astrophysics
Center, 5640 S. Ellis Ave., Chicago, IL 60637\label{Chicago}}
\altaffiltext{\theaddress}{\stepcounter{address}
Department of Physics, Carnegie Mellon University, Pittsburgh, PA 15213
\label{CMU}}
\altaffiltext{\theaddress}{\stepcounter{address}
Fermi National Accelerator Laboratory, P.O. Box 500,
Batavia, IL 60510\label{Fermilab}}
\altaffiltext{\theaddress}{\stepcounter{address}
Department of Physics and Astronomy, University of Pittsburgh, Pittsburgh, PA 15620\label{Pitt}}
\altaffiltext{\theaddress}{\stepcounter{address}
Stewart Observatory, University of Arizona, 933 N. Clarry Ave., Tucson, AZ 85121\label{Arizona}}
\altaffiltext{\theaddress}{\stepcounter{address}
Department of Astronomy, University of California at Berkeley, 601 Campbell Hall, Berkeley, CA 94720\label{Berkeley}}
\altaffiltext{\theaddress}{\stepcounter{address}
Princeton University Observatory, Princeton, NJ 08544\label{Princeton}}
\altaffiltext{\theaddress}{\stepcounter{address}Hubble Fellow\label{HF}}
\altaffiltext{\theaddress}{\stepcounter{address}
Department of Physics, New York University, 4 Washington Place, New York, NY 10003\label{NYU}}
\altaffiltext{\theaddress}{\stepcounter{address}
Department of Physics \& Astronomy, The Johns Hopkins University, 3400 North Charles Street, Baltimore, MD 21218-2686\label{JHU}}
\altaffiltext{\theaddress}{\stepcounter{address}
Apache Point Observatory, 2001 Apache Point Road, P.O. Box 59, Sunspot, NM
88349-0059\label{APO}}
\altaffiltext{\theaddress}{\stepcounter{address} Yale University, P. O. Box
208101, New Haven, CT 06520\label{yale}}
\altaffiltext{\theaddress}{\stepcounter{address} Universidad de Chile, Casilla
36-D, Santiago, Chile\label{chile}}
\altaffiltext{\theaddress}{\stepcounter{address}
Department of Physics of Complex Systems, E\"otv\"os University, Budapest, H-1117 Hungary\label{Eotvos}}
\altaffiltext{\theaddress}{\stepcounter{address}
Institute of Astronomy, School of Science, University of Tokyo, Mitaka, Tokyo 181-0015, Japan\label{Tokyo1}}
\altaffiltext{\theaddress}{\stepcounter{address}
Research Center for the Early Universe, School of Science,
    University of Tokyo, Tokyo 113-0033, Japan\label{Tokyo2}}
\altaffiltext{\theaddress}{\stepcounter{address}
Institute for Cosmic Ray Research, University of Tokyo, Kashiwa 277-8582, Japan\label{ICRR}}
\altaffiltext{\theaddress}{\stepcounter{address}
Institute for Advanced Study, Olden Lane, Princeton, NJ 08540\label{IAS}}
\altaffiltext{\theaddress}{\stepcounter{address}
U.S. Naval Observatory, 3450 Massachusetts Ave., NW, Washington, DC 20392-5420\label{USNO}}
\altaffiltext{\theaddress}{\stepcounter{address}
Department of Physics, University of Michigan, 500 East University, Ann Arbor, MI 48109\label{UMich}}
\altaffiltext{\theaddress}{\stepcounter{address}
Department of Astronomy, University of Tokyo,
   Tokyo 113-0033, Japan\label{Tokyo3}}
\altaffiltext{\theaddress}{\stepcounter{address}
Department of Astronomy and Astrophysics, The Pennsylvania State University, University Park, PA 16802\label{PSU}}



\begin{abstract}
A sample of nearly 9000 early-type galaxies, in the redshift range 
$0.01 \le z \le 0.3$, was selected from the Sloan Digital Sky Survey  
using morphological and spectral criteria. This paper describes how the 
sample was selected, presents examples of images and seeing corrected 
fits to the observed surface brightness profiles, describes our method 
for estimating K-corrections, and shows that the SDSS spectra are of 
sufficiently high quality to measure velocity dispersions accurately.
It also provides catalogs of the measured photometric and spectroscopic  
parameters. In related papers, these data are used to study how 
early-type galaxy observables, including luminosity, effective radius, 
surface brightness, color, and velocity dispersion, are correlated with 
one another.  
\end{abstract}  

\section{Introduction}
Galaxies have a wide range of luminosities, colors, masses, sizes,
surface brightnesses, morphologies, star formation histories and
environments.  This heterogeneity is not surprising, given the variety
of physical processes which likely influence their formation and
evolution, including gravitational collapse, hydrodynamics,
turbulence, magnetic fields, black-hole formation and accretion,
nuclear activity, tidal and merger interactions, and evolving and
inhomogeneous cosmic radiation fields.

What \emph{is} surprising is that populations of galaxies show several
very precise relationships among their measured properties.  The properties 
we use to describe galaxies span a large ``configuration space'', but 
galaxies do not fill it.  Galaxy spectral energy distributions, when 
scaled to a fixed broad-band luminosity, appear to occupy a thin, 
one-dimensional locus in color space or spectrum space 
(e.g., Connolly \& Szalay 1999).  Spiral galaxies 
show a good correlation between rotation velocity and luminosity 
(e.g., Tully \& Fisher 1977; Giovanelli et al. 1997).  
Galaxy morphology is strongly correlated with broad-band colors, 
strengths of spectral features, and inferred star-formation histories 
(e.g., Roberts \& Haynes 1994).  

Among all galaxy families, early-type (elliptical and S0) galaxies 
show the most precise regularities (Djorgovski \& Davis 1987; 
Burstein, Bender, Faber \& Nolthenius 1997).  Early-type galaxy 
surface-brightness distributions follow a very simple, universal 
``de~Vaucouleurs'' profile (de~Vaucouleurs 1948).  Their spectral energy 
distributions appear to be virtually universal, showing very little 
variation with mass, environment, or cosmic time (e.g., van~Dokkum \&
Franx 1996; Pahre 1998).  What variations they do show are measurable 
and precise.  Early-type galaxy colors, luminosities, half-light radii, 
velocity dispersions, and surface brightnesses are all correlated 
(Baum 1959; Fish 1964; Faber \& Jackson 1976; Kormendy 1977; 
Bingelli, Sandage \& Tarenghi 1984); they can be combined into a 
two-dimensional ``Fundamental Plane'' with very little scatter 
(e.g., Dressler et al. 1987; Djorgovski \& Davis 1987; Faber et al. 1987).  

The homogeneity of the early-type galaxy population is difficult to understand 
if early-type galaxies are assembled at late times by stochastic mergers 
of less-massive galaxies of, presumably, different ages, star formation 
histories, and gas contents, as many models postulate 
(e.g., Larson 1975; White \& Rees 1978; van Albada 1982; Kauffmann 1996; 
Kauffmann \& Charlot 1998).  It is possible that the homogeneity of 
early-type galaxies points to early formation (e.g., Worthey 1994; 
Bressan et al. 1994; Vazdekis et al. 1996; Tantalo, Chiosi \& Bressan 1998); 
certainly their stellar populations appear old (e.g., Bernardi et al. 1998; 
Colless et al. 1999; Trager et al. 2000a,b; Kuntschner et al. 2001).  
Alternatively, the observable properties of the stellar content 
of early-type galaxies are fixed entirely by the properties of the 
collisionless, self-gravitating, dark-matter haloes in which we
believe such galaxies lie (e.g., Hernquist 1990).  
These halos, almost by definition, are not subject to the vagaries of 
gas dynamics, star formation, and magnetic fields; they are influenced 
only by gravity.  

It is essentially a stated goal of the Sloan Digital Sky Survey (SDSS; 
York et al. 2000; Stoughton et al. 2002) to revolutionize the study of
galaxies.  The SDSS is imaging $\pi$ steradians of the sky (Northern 
Galactic Cap) in five bands and taking spectra of $\sim 10^6$ galaxies 
and $\sim 10^5$ QSOs.  Among the $10^6$ SDSS spectra there will be roughly 
$2\times 10^5$ spectra taken of early-type galaxies; in fact $10^5$ of the 
spectroscopic fibers are being used to assemble a sample of luminous 
early-type galaxies with a larger mean redshift than the main SDSS sample 
(Eisenstein et al. 2001).  The high quality of the SDSS 5-band CCD imaging 
(Gunn et al. 1998; Lupton et al. 2001) allows secure identification of 
early-type galaxies and precise measurements of their photometric 
properties; most spectroscopic targets in the SDSS are detected in the 
imaging at signal-to-noise ratios $(S/N)>100$.

Early-type galaxy studies in the past, for technical reasons, have
concentrated on galaxies in clusters at low (e.g., 
J{\o}rgensen, Franx \& Kj{\ae}rgaard 1996; Ellis et al. 1997; 
Pahre, Djorgovski \& de Carvalho 1998a,b; Scodeggio et al. 1998; 
Colless et al. 2001; Saglia et al. 2001; Kuntschner et al. 2001; 
Bernardi et al. 2002a,b) and intermediate redshifts 
(e.g., van Dokkum et al. 1998, 2001; Kelson et al. 2000; 
Ziegler et al. 2001).  Only the large area `Seven Samurai' 
(e.g., Faber et al. 1989) and ENEAR surveys (e.g., da Costa et al. 2000) 
of nearby early-types, recent work with galaxies in the SBF survey 
(Blakeslee et al. 2001), and some studies at intermediate redshifts 
by Schade et al. (1999), Treu et al. (1999, 2001a,b) and 
van Dokkum et al. (2001), are not restricted to cluster environments.  
In constrast, the SDSS is surveying a huge volume of the local Universe, 
so the sample includes early-type galaxies in every environment from 
voids to groups to rich clusters.  
As of writing, when only a small fraction of the planned SDSS imaging 
and spectroscopy has been taken, the number ($\sim 9000$) of early-type 
galaxies with well-measured velocity dispersions and surface-brightness 
profiles in the SDSS greatly exceeds the total number in the entire 
astronomical literature to date.  

This is the first of four papers in which we use the SDSS sample to 
measure the Fundamental Plane and other early-type galaxy correlations 
in multiple bands. This first paper describes how the sample was 
selected and presents the data.  
Section~\ref{sample} describes the main properties of the SDSS database.
The criteria used to select the early-type galaxy sample are described
in Section~\ref{finalsample}. This section also presents a selection 
of images from a range of redshifts, shows de~Vaucouleur profile fits 
to the observed surface brightness profiles, and argues that the SDSS 
photometric pipeline treatment of seeing and sky-subtraction have not 
strongly compromised the estimates of the best-fitting model parameters.  
It also displays examples of the spectra from which we estimate velocity 
dispersions.  
Section~\ref{catalog} presents the final catalog of photometric and 
spectroscopic parameters which we use for our subsequent analyses of 
early-type galaxy properties, and discusses some properties of the sample.  
Many details are relegated to Appendices.  
Appendix~\ref{kcorrs} contains a discussion of the various K-corrections 
we have tried.  The way we estimate velocity dispersions is presented in 
Appendix~\ref{vmethods}.  A novel method for estimating aperture 
corrections to the velocity dispersions is discussed in Appendix~\ref{vr}.  
Finally, the covariance matrix of the errors is discussed in 
Appendix~\ref{errors}.  

Paper~II (Bernardi et al. 2003a) studies correlations between various 
pairs of observables, 
such as the Faber-Jackson and Kormendy relations. It also presents the 
luminosity function and its evolution.  
The Fundamental Plane and its dependence on waveband, color, redshift, 
and environment is studied in Paper~III (Bernardi et al. 2003b).
The co-added spectra of these galaxies are studied in Paper~IV 
(Bernardi et al. 2003c).
One of the results of that paper is a library of co-added spectra which 
contains spectra that represent a wide range of early-type galaxies.  
This library is available electronically.
These spectra indicate that the chemical composition of the early-type galaxy 
population evolves with redshift.  The chemical abundances and evolution 
are then combined with stellar population models to estimate the ages 
and metallicities of the galaxies in our sample. This paper also analyzes
correlations with color (e.g., color--magnitude and color--$\sigma$ 
relations) and discusses the effects of color gradients on measurements 
of the strength of the correlation between color and magnitude.

Except where stated otherwise, we write the Hubble constant as 
$H_0=100\,h~\mathrm{km\,s^{-1}\,Mpc^{-1}}$, and we perform our 
analysis in a cosmological world model with 
$(\Omega_{\rm M},\Omega_{\Lambda},h)=(0.3,0.7,0.7)$, where 
$\Omega_{\rm M}$ and $\Omega_{\Lambda}$ are the present-day scaled 
densities of matter and cosmological constant.  
In such a model, the age of the Universe at the present time is 
$t_0=9.43h^{-1}$ Gyr.  For comparison, an Einstein-de Sitter model has 
$(\Omega_{\rm M},\Omega_{\Lambda})=(1,0)$ and $t_0=6.52h^{-1}$~Gyr.  
We frequently use the notation $h_{70}$ as a reminder that we have 
set $h=0.7$.  Also, we will frequently be interested in the 
logarithms of physical quantities.  Our convention is to set 
$R\equiv\log_{10}R_o$ and $V\equiv \log_{10}\sigma$, where $R_o$ 
and $\sigma$ are effective radii in $~h_{70}^{-1}$~kpc and velocity 
dispersions in km~s$^{-1}$, respectively.  

\section{The SDSS database}\label{sample}
The SDSS project is described in Stoughton et al. (2002).  
The data we analyze in this paper were selected from the SDSS database 
in summer 2001.  At that time the SDSS had imaged $\sim 1,500$ square 
degrees; $\sim 65,000$ galaxies and $\sim 8000$ QSOs had both photometric 
and spectroscopic information.  
The photometric and spectroscopic data were taken with the 2.5-m 
SDSS Telescope at the Apache Point Observatory (New Mexico) between
1999 March and 2000 October. Details of the photometric and 
spectroscopic observations and data reduction procedure will be 
presented elsewhere.  Here we briefly summarize.  

\subsection{SDSS imaging data}
Images are obtained by drift scanning with a mosaic CCD camera 
(Gunn et al. 1998) which gives a field of view of 
$3\times 3~\mathrm{deg^2}$, with a spatial scale of 
$0.4~\mathrm{arcsec\,pix^{-1}}$ in five bandpasses 
($u$, $g$, $r$, $i$, $z$) with central wavelengths
(3560, 4680, 6180, 7500, 8870\AA) (Fukugita et al. 1996).
The errors in $u$ band measurements are larger than the others, 
so we will only present results in the other four bands.  
In addition, the photometric solutions we use in this paper are 
preliminary (for details, see discussion of the Early Data Release 
in Stoughton et al. 2002); we use $r^*$ rather than $r$, and 
similarly for the other bands, to denote this.

The effective integration time is 54~sec.  The raw CCD images are
bias-subtracted, flat-fielded and background-subtracted.  Pixels
contaminated by the light of cosmic rays and bad columns are masked.
Astronomical sources are detected and overlapping sources are
de-blended. The data are flux-calibrated by comparison with a set of 
overlapping standard-star fields calibrated with a 0.5-m ``Photometric 
Telescope'' (Hogg et al. 2001; Smith et al. 2002).  
The Photometric Telescope 
is also used for measuring the atmospheric extinction coefficients 
in the five bands.  The median effective seeing (the median FWHM of 
the stellar profiles) for the observations used here is 
$1.5~\mathrm{arcsec}$.  All of this image processing is performed 
with software specially designed for reducing SDSS data 
(Lupton et al. 2001).  The uncertainty in the $r^*$-band 
zero-point calibration is $<0.01$~mag; the uncertainty in the sky 
background subtraction is less than about 1\%.  

Surface photometry measurements are obtained by fitting a set of 
two-dimensional models to the images. The model fits account for the 
effects of seeing, atmospheric extinction, and Galactic extinction 
(this last uses the results of Schlegel, Finkbeiner, \& Davis 1998). 
The SDSS model for seeing is described in Stoughton et al. (2002). 
Briefly, the PSF is expected to have a Fourier Transform 
$\propto \exp[-(kr_{\rm PSF}/2)^{5/3}]$ characteristic of Kolmogorov 
turbulence (e.g., Saglia et al. 1993).  This PSF is approximated as 
the sum of Gaussians (this parametrization allows for a substantial 
reduction in the processing time required to deconvolve the effects 
of seeing).

The SDSS photometric pipeline fits two models to the two-dimensional 
image of each object in each band: 
a pure de~Vaucouleurs profile and a pure exponential profile 
(e.g. Stoughton et al. 2002). Briefly, the photon counts are binned 
into a number of radial and twelve angular bins. Then, two-dimensional 
de~Vaucouleurs and exponential models, convolved with the seeing, are 
fitted to the cumulative binned counts. The pipeline does not include 
more complicated models (e.g., bulge plus disk) because the 
computational expense to compute them is not justified for the majority 
of the detected objects.  Algorithms which fit bulge and disk components 
(which are desirable to fit on large galaxies) to the surface brightness 
profiles are currently being developed within the collaboration, but are 
not yet available. We present examples of $r^*$-band images and mean 
surface brightness profiles in Section~\ref{checkphoto}.

The SDSS image processing software provides several global photometric
parameters, for each object, which are obtained independently in each 
of the five bands.  Because we are interested in early-type galaxies, 
we use primarily the following: 
\begin{itemize}
\item
The ratio $b/a$ of the lengths of the minor and major axes 
   of the observed surface brightness profile.
\item
The effective radius (or half-light radius) $r_{\rm dev}$ along the 
   major axis and 
\item
the total magnitude $m_{\rm dev}$; these are computed by fitting a 
   two-dimensional version of de~Vaucouleurs (1948) $r^{1/4}$ model 
   to the observed surface brightness profile.  
   (The fitting procedure accounts for the effects of seeing---we test 
   the accuracy of this procedure below.)  
\item
Likelihood parameters ${\tt deV\_L}$ and ${\tt exp\_L}$ that indicate 
   how well the de Vaucouleurs or exponential models, when convolved with 
   a model for the seeing, fit the observed light profile. 
\item
The {\it model} magnitude $m_m$; this is the total magnitude calculated 
   by using the (de Vaucouleurs or exponential) model which fits the galaxy 
   profile best in the $r^*$-band.  The {\it model} magnitudes in the other 
   four bands are computed using that $r^*$ fit as filter; in effect, this 
   measures the colors of a galaxy through the same aperture. 
\item
The Petrosian magnitude $m_p$ is also computed; this is the flux within 
   $2r_p$, where $r_p$ is defined as the angular radius at which the ratio 
   of the local surface brightness at $r$ to the mean surface brightness 
   within a radius $r$ is 0.2 (Petrosian 1976).  
\item
The Petrosian radii $r_{50}$ and $r_{90}$; these are the angular 
   radii containing 50\% and 90\% of the Petrosian light, respectively. 
\end{itemize}

Although our analyses on early-type galaxy properties presented here
and in the companion papers have been performed with both the de~Vaucouleurs 
fit parameters and the Petrosian quantities, in most of our analyses only 
the results of the de~Vaucouleurs fits are presented.  This is because the
de~Vaucouleurs model appears to be a very good fit to the early-type
galaxy surface-brightness profiles in the SDSS sample and because it
is conventional, in the literature on early-type galaxies, to use these
quantities.  
On the other hand, for reasons given in Stoughton et al. (2002), 
unless stated otherwise, galaxy colors are always computed using 
{\it model} magnitudes.

\subsection{SDSS spectroscopic data}\label{spectro}
The SDSS takes spectra only for a target subsample of objects.  
Target selection criteria are described in Stoughton et al. (2002) 
and Strauss et al. (2002).  Spectra are obtained using a multi-object 
spectrograph which observes 640 objects at once. 
Each spectroscopic plug plate, 1.5 degrees in radius, has 640 fibers, 
each $3~\mathrm{arcsec}$ in diameter. Two fibers cannot be closer than 
$55~\mathrm{arcsec}$ due to the physical size of the fiber plug.  
Typically $\sim 500$ fibers per plate are used for galaxies, 
$\sim 90$ for QSOs, and the remaining for sky spectra and 
spectrophotometric standard stars.  

Each plate typically has three to five spectroscopic exposures of 
fifteen minutes, depending on the observing conditions (weather, moon); 
a minimum of three exposures is taken to ensure adequate cosmic ray 
rejection. For galaxies at $z \le 0.3$ the median spectrum $S/N$ 
per pixel is 16 (see Figure~\ref{fig:vmeth} in Appendix~\ref{vmethods}).
The wavelength range of each spectrum is $3900-9000$~\AA. 
The instrumental dispersion is $\log_{10}\lambda=10^{-4}$dex/pixel 
which corresponds to 69~km~s$^{-1}$ per pixel.  (There is actually some 
variation in this instrumental dispersion with wavelength, which we 
account for; see Figure~\ref{fig:resol} and associated discussion 
in Appendix~\ref{vmethods}.)  
The instrumental resolution of galaxy spectra, measured from the 
autocorrelation of stellar template spectra, ranges from 85 to 
105 km~s$^{-1}$, with a median value of 92 km~s$^{-1}$.  

A highly automated software package has been designed for reducing 
SDSS spectral data. 
The raw data are bias-subtracted, flat-fielded, wavelength calibrated, 
sky-lines removed, co-added, cleaned from residual glitches 
(cosmic rays, bad pixels), and flux calibrated.  The spectro-software
classifies objects by spectral type and determines emission and absorption
redshifts. (Redshifts are corrected to the heliocentric reference frame.) 
The redshift success rate for objects targeted as galaxies is $>99$\% 
and errors in the measured redshift are less than about $10^{-4}$.  
Once the redshift as been determined the following quantites are 
computed:  Absorption-line strengths (Brodie \& Hance 1986; 
Diaz, Terlevich \& Terlevich 1989; Trager et al. 1998),
equivalent widths of the emission lines, and 
eigen-coefficients and classification numbers of a PCA analysis 
(Connolly \& Szalay 1999).  
Some information about the reliability of the redshift and the quality of 
the spectrum is also provided.  We present examples of the spectra 
in Section~\ref{exspec}.  

The SDSS pipeline does not provide an estimate of the line-of-sight 
velocity dispersion, $\sigma$, within a galaxy, so we compute 
it separately for the early-type galaxy sample (see Section~\ref{vdisp}). 

\section{The sample}\label{finalsample}

\subsection{Selection criteria}\label{selection}
The main goal of this series of papers is to study the properties of 
early-type galaxies using the main galaxy sample of the SDSS database.  
Therefore, one of the crucial steps in our study is the separation of 
galaxies into early and late types.  We want to select objects whose 
spectra are good enough to compute the central velocity dispersion.  
In addition, because we wish to study the colors of the galaxies in 
our sample, we must not use color information to select the sample.  
To reach our goal, we have selected galaxies which satisfy the 
following criteria:
\begin{itemize}
 \item concentration index $r_{90}/r_{50} > 2.5$ in $i^*$;
 \item the likelihood of the de Vaucouleurs model is at least 1.03 times 
       the likelihood of the exponential model;
 \item spectra with PCA classification numbers $a < -0.1$, typical of 
       early-type galaxy spectra (Connolly \& Szalay 1999);
 \item spectra without masked regions (the SDSS spectroscopic pipeline 
  outputs a warning flag for spectra of low quality; 
  we only chose spectra for which this flag was set to zero); 
 \item $S/N > 10$;
 \item redshift $< 0.3$.
\end{itemize}

Section~\ref{vdisp} describes how we estimate the velocity dispersion 
$\sigma$ for galaxies which satisfy the above criteria.  
We consider velocity dispersion estimates smaller than about 
70~km~s$^{-1}$ to be unreliable (see Appendix~\ref{vmethods}).  
Therefore, in the final sample presented in Section~\ref{catalog}, 
galaxies with $\sigma\le 70$~km~s$^{-1}$ have been excluded.  
The results of this and the companion papers are not significantly 
different if we change the cut-off on velocity dispersions to 
100~km~s$^{-1}$.  

About 9,000 objects satisfied all the above criteria.  

As stated earlier, the SDSS pipeline does not output disk-to-bulge 
ratios from fits to the light profiles.  The first two requirements above 
attempt to select profile shapes which are likely to be those of 
spheroidal systems.  The $i^*$ band measurements tend to be less noisy 
than $g^*$ or $z^*$, so we chose to use the $i^*$ band estimate of how 
centrally concentrated the light is.  
The second requirement reduces approximately to requiring that 
${\tt deV\_L/(deV\_L+exp\_L)}>0.5$; that is, the surface brightness 
profile should be better fit by the de~Vaucouleurs model than by an 
exponential.  

The spectra of late-type galaxies show emission lines, so examining 
the spectra (e.g., using the PCA classification) is a simple way of 
removing such objects from the sample.  
Because the aperture of an SDSS spectroscopic fiber ($3~\mathrm{arcsec}$) 
samples only the inner parts of nearby galaxies, and because the spectrum 
of the bulge of a nearby late-type galaxy can resemble that of an 
early-type galaxy, it is possible that some nearby late-type galaxies 
could be mistakenly included in the sample 
(e.g., Kochanek, Pahre \& Falco 2000).  Most of these will have been 
excluded by the first two cuts on the shape of the light profile.  To 
check this, we visually inspected all galaxies with 
$r_{\rm dev} > 8~\mathrm{arcsec}$.  About $\sim 50$ of 225 
(i.e., about 20\%) looked like late-types, and so we removed them.  
Note that weak emission lines, such as H$_{\alpha}$ and/or O~II, are 
still present in the early-type galaxy spectra in our sample.  For 
example, the median H$_{\alpha}$ equivalent width for galaxies in our 
sample is $-1.4$~\AA, but $\sim 5$\% of the galaxies have equivalent 
widths larger than $1$~\AA\ and $\sim 10$\% of these (i.e. 0.5\% of 
the sample) have equivalent widths in the range $4-10$~\AA.
Similarly, although the median equivalent width of O~II is $1.5$~\AA, 
$\sim 15\%$ of the galaxies show O~II equivalent widths larger than 
$4$~\AA\ and $\sim 10\%$ of these (i.e., 1.5\% of the sample) have 
equivalent widths ranging from $8$ to $15$~\AA.

The reason for our redshift cut is as follows.  
The SDSS main galaxy sample from which we select our early-type 
galaxies is apparent magnitude limited in $r^*$, where the limits are 
defined using Petrosian magnitudes $m_p$.  This means that the main 
galaxy sample is not magnitude limited in the other bands.  
Blanton et al. (2001) describe the cuts in $m_p$ one must apply in the 
other bands to obtain complete magnitude limited samples.  They also 
note that $m_{\rm dev}\approx 2.5\log_{10}0.8 + m_p$.  
Since we will almost always be working with $m_{\rm dev}$ rather than 
$m_p$, our cuts are slightly different from theirs; the cuts we use 
to define magnitude limited samples in the different bands are 
summarized in Table~\ref{tab:photerr} (also see Figure~\ref{fig:appmz}).  

At $z\ge 0.3$, however, most early-type galaxies in the SDSS database 
were targeted using different selection criteria than were used for the 
main SDSS galaxy sample (see Strauss et al. 2002; they make-up the 
Luminous Red Galaxy sample described by Eisenstein et al. 2001).  
In the interest of keeping our sample as close to being magnitude 
limited as possible, we restricted our sample to $z\le 0.3$.
In addition, one might expect an increasing fraction of the early-type 
population at higher redshifts to have emission lines:  if so, then 
our removal of emission line objects amounts to a small but redshift 
dependent selection effect.  Since our sample is restricted to $z\le 0.3$, 
this bias should be small.

\begin{figure}
\centering
\epsfxsize=\hsize\epsffile{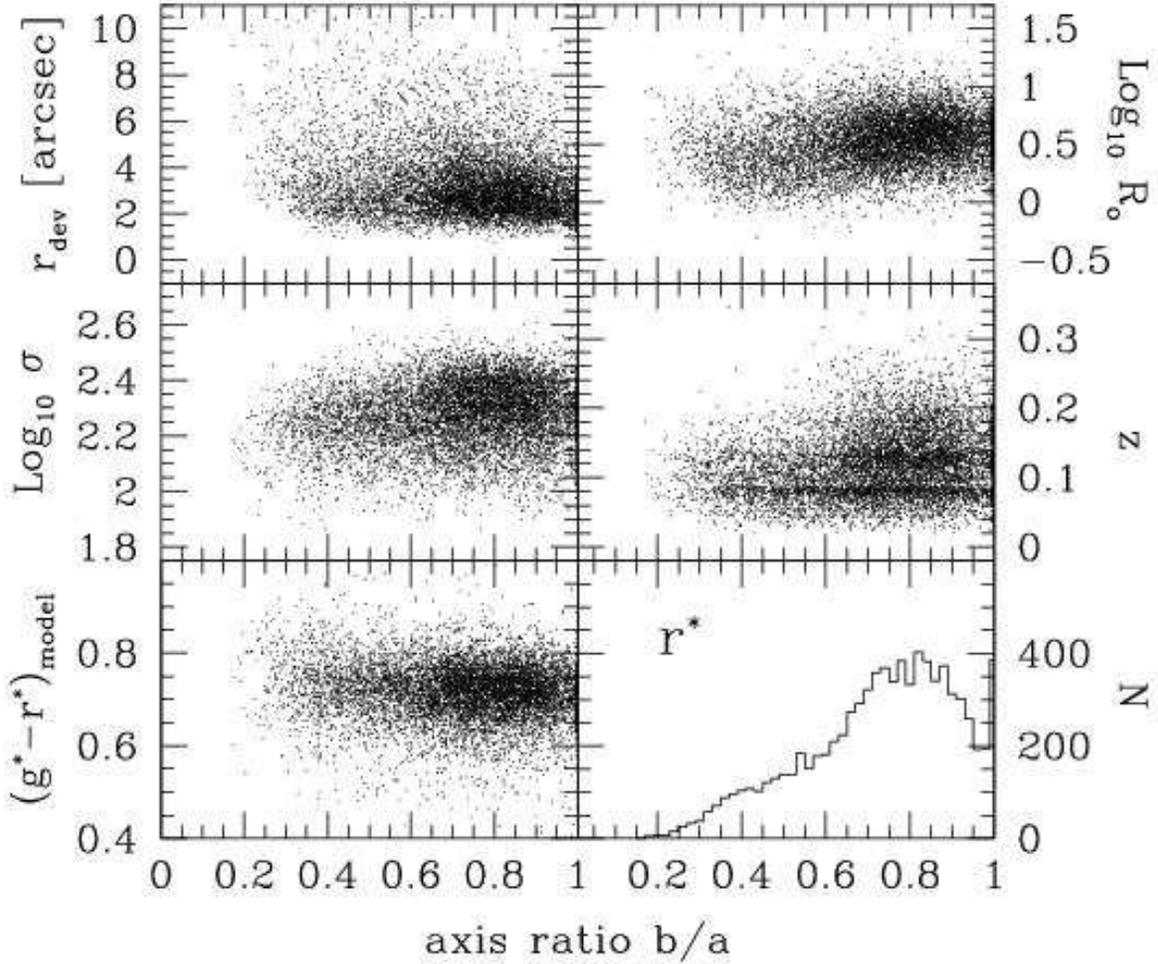}
\caption{Effective angular sizes $r_{\rm dev}$, effective 
circular physical sizes $R_o$, velocity dispersions $\sigma$, 
redshifts $z$, and $(g^*-r^*)$ colors as a function of axis ratio 
$b/a$ for the galaxies in our $r^*$ sample.  Bottom right panel 
shows that the typical axis ratio is $b/a\approx 0.8$.  
There is only a weak tendency for galaxies with small $r_{\rm dev}$ 
to be rounder, suggesting that the estimate of the shape is not 
compromised by seeing (typical seeing is about 1.5 arcsec).  
Results in the other bands are similar.}
\label{abscatter}
\end{figure}

Before moving on, it is worth pointing out that there is a 
morphologically based selection cut which we could have made but 
didn't.  Elliptical galaxies are expected to have axis ratios greater 
than about $0.6$ (e.g., Binney \& Tremaine 1987).  Since we have axis 
ratio measurements of all the objects in our sample, we could have 
included a cut on $b/a$.  The bottom right panel of 
Figure~\ref{abscatter} shows the distribution axis ratios $b/a$ in 
our $r^*$ sample:  about 20\% of the objects in it have $b/a\le 0.6$.  
(The spike at $b/a=1$ is artificial---the pipeline reports large 
uncertainties in the fitted value of $b/a$ for these objects.)

Our combination of cuts on the shapes of the light profiles and spectral 
features mean that these objects are unlikely to be late-type galaxies.  
Indeed, a visual inspection of a random sample of the objects with axis 
ratios smaller than 0.6 shows that they look like S0s.  
The bottom left panel of Figure~\ref{abscatter} shows that $b/a$ does 
not correlate with color:  in particular, the colors of the most 
flattened objects are not bluer than in the rest of the sample.  
Also, recall that all objects with $r_{\rm dev}\ge 8$ arcsec were 
visually inspected and these, despite having $b/a\le 0.6$ (top left panel), 
did not appear peculiar.  
In addition, $b/a$ does not correlate with surface brightness or 
apparent magnitude.  Galaxies with small angular sizes $r_{\rm dev}$ 
are assigned large values of $b/a$ only slightly more often than 
average (top left panel; the median $b/a$ is 0.79, 0.78, 0.76, and 0.7 
for $r_{\rm dev}$ in the range 1--2, 2--3, 3--4 and greater than 4 arcsec).  
However, there is a weak trend for the objects at higher $z$ to be 
rounder (middle right panel).  This trend with $z$ may be related to 
the magnitude limit of our sample rather than reflecting problems 
associated with the fits to the observed light profiles.  (The two 
bands at $z\sim 0.08$ and $z\sim 0.13$ show that the effects of 
large scale clustering in our sample are significant.)  
Section~\ref{LandR} describes how we convert from the observed 
half-light radius $r_{\rm dev}$ to an estimate of the physical 
half-light radius $R_o$, and Section~\ref{vdisp} discusses how we 
estimate velocity dispersions.   
Objects with smaller values of $R_o$ tend to be more flattened (top 
right panel), and to have slightly smaller velocity dispersions (middle 
left); because of the magnitude limit, these objects drop out of our 
sample at higher redshifts.  Requiring that $b/a\ge 0.6$ would remove 
such objects from our sample completely.  

In Paper~III we study the Fundamental Plane populated by the galaxies 
in this sample.  Excluding all objects with $b/a < 0.6$ has no effect 
on the shape of this Plane.  So, in the interests of keeping our sample 
as close to being magnitude limited as possible, we chose not to make 
an additional selection cut on $b/a$.  

\subsection{The photometric data of the sample}\label{phot}

\subsubsection{Conversion to restframe luminosities and sizes}\label{LandR}

\begin{figure}
\centering
\epsfxsize=\hsize\epsffile{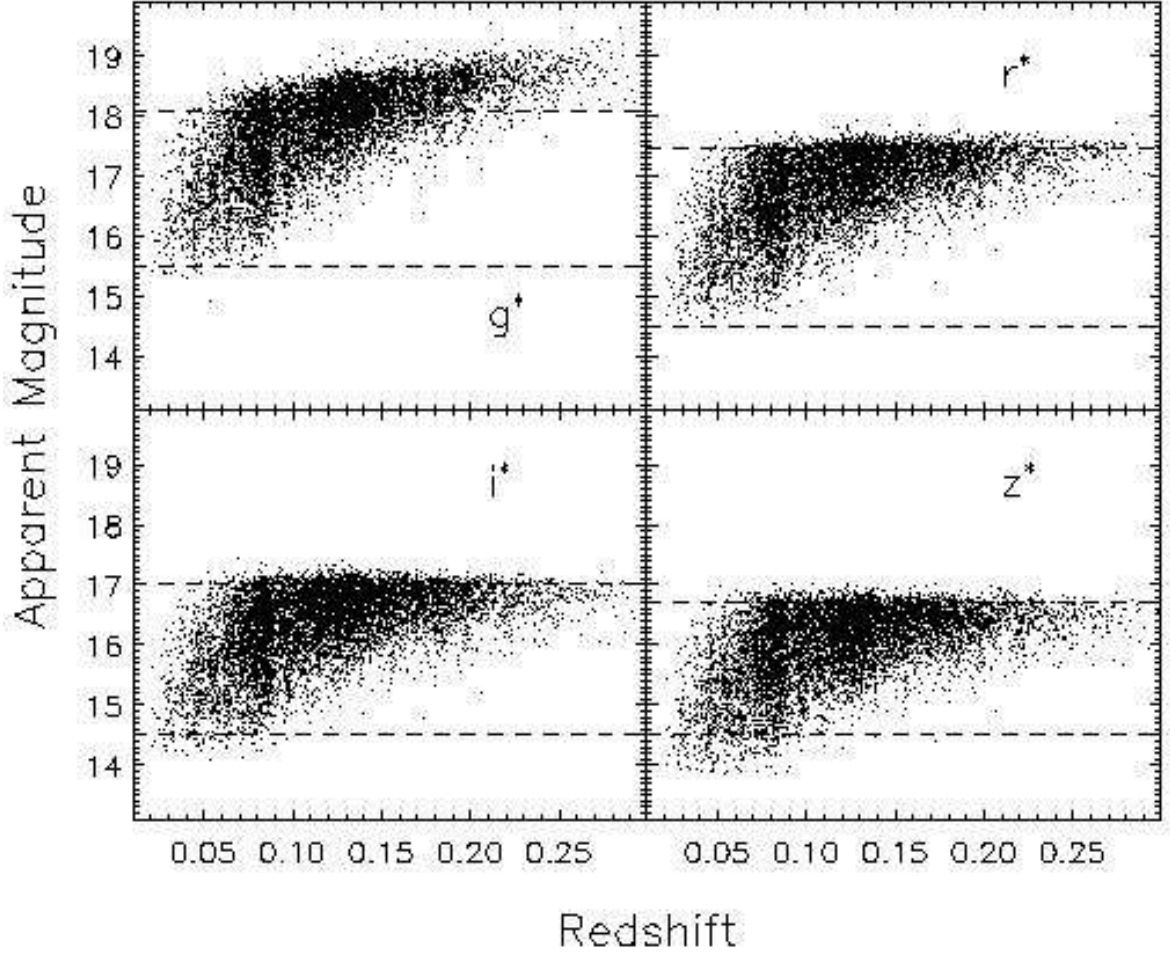}
\caption{Apparent magnitude ($m_{\rm dev}$) in the $g^*$, $r^*$, $i^*$, 
and $z^*$ bands, versus redshift for galaxies which satisfied our 
selection criteria.  In each band, magnitude limited samples are 
defined by including only those galaxies which fall between the dashed 
lines shown;  these magnitude limits are summarized in 
Table~\ref{tab:photerr}.}
\label{fig:appmz}
\end{figure}

\begin{figure}
\centering
\epsfxsize=\hsize\epsffile{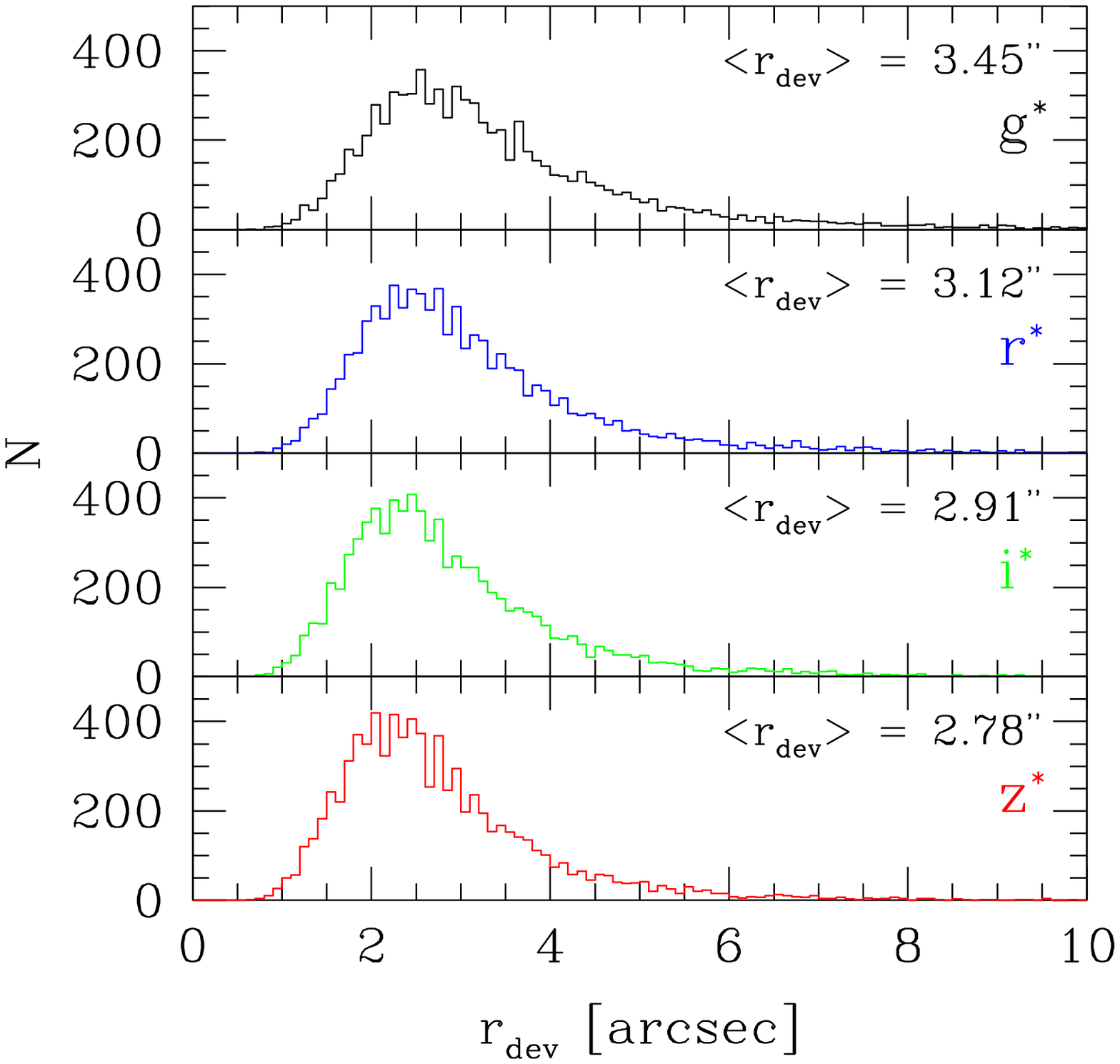}
\caption{Distribution of (seeing-corrected) effective angular sizes 
of galaxies in our sample.  Typical seeing is about 1.5 arcsec.  
The distribution of effective radii in all the bands are very 
similar, although the radii are slightly larger in the bluer bands.}
\label{angre}
\end{figure}

Figure~\ref{fig:appmz} shows the distribution of the $m_{dev}$ apparent
magnitudes in our sample as a function of redshift.  
The dashed lines in each panel show the magnitude limits summarized 
in Table~\ref{tab:photerr}; the magnitude limited samples span the 
redshift range $0.01\le z\le 0.3$.  

To convert the apparent magnitude $m$ to an absolute magnitude $M$ we 
must assume a particular cosmology and account for the fact that at 
different redshifts an observed bandpass corresponds to different 
restframe bands (the K-correction).  We write the Hubble constant today 
as $100h$ km~s$^{-1}$Mpc$^{-1}$ and use 
$(\Omega_M,\Omega_{\Lambda},h)=(0.3,0.7,0.7)$.  
Most of our sample is at $cz\ge 9000$ km~s$^{-1}$;  
since line-of-sight peculiar velocities are not expected to exceed more 
than a few thousand km~s$^{-1}$, we feel that it is reasonable to assume 
that all of a galaxy's redshift is due to the Hubble recession velocity.  
This means that we can compute the absolute magnitude in a given band by 
 $M = m -5\log_{10}[D_L(z;\Omega_M;\Omega_{\Lambda})] - 25 - K(z)$, 
where $m$ is the apparent magnitude, $D_{\rm L}$ is the luminosity 
distance in Mpc (from, e.g., Weinberg 1972; Hogg 1999), and 
$K(z)$ is the K-correction for the band.  

Because we have five colors and a spectrum for each galaxy, we could, 
in principle, compute an empirical K-correction for each galaxy.  
This requires a good understanding of the accuracy of the SDSS photometry 
and spectroscopy, and should be possible when the survey is closer to 
completion.  Rather than follow the procedure adopted by the 2dFGRS 
(Madgwick et al. 2002), or a procedure based on finding the closest 
template spectrum to each galaxy and using it to compute the K-correction 
(e.g., Lin et al. 1999 for the CNOC2 survey), we use a single redshift 
dependent template spectrum to estimate the K-correction.  In effect, 
although this allows galaxies at different redshift to be different, it 
ignores the fact that not all galaxies at the same redshift are alike.  
As a result, the absolute luminosities we compute are not as accurate as 
they could be, and this can introduce scatter in the various correlations 
we study below.  Of course, using a realistic K-correction is important, 
because inaccuracies in $K(z)$ can masquerade as evolutionary trends.  

For the reasons discussed more fully in Appendix~\ref{kcorrs}, our 
K-corrections are based on a combination of Bruzual \& Charlot (2003) 
and Coleman, Wu \& Weedman (1980) prescriptions.  Specifically, the 
K-corrections we apply were obtained by taking a Bruzual \& Charlot  
model for a $10^{11}M_\odot$ object which formed its stars with an 
IMF given by Kroupa (2001) in a single solar metallicity and abundance 
ratio burst 9~Gyr ago, computing the difference between the K-correction 
when evolution is allowed and ignored, and adding this difference to the 
K-corrections associated with Coleman, Wu \& Weedman early type galaxy 
template.  The results which follow are qualitatively similar for a 
number of other K-correction schemes (see Appendix~\ref{kcorrs} for 
details).  

In addition to correcting the observed apparent magnitudes 
to absolute magnitudes, we must also apply two corrections to convert 
the (seeing corrected) effective angular radii, $r_{\rm dev}$, output 
by the SDSS pipeline to physical radii.  
First, we define the equivalent circular effective radius 
 $r_o\equiv\sqrt{b/a}\,r_{\rm dev}$.  
(Although the convention is to use $r_e$ to denote the effective radius, 
we feel that the notation $r_o$ is better, since it emphasizes that the 
radius is an effective circular, rather than elliptical aperture.)  
The reason we must make a second correction is shown in 
Figure~\ref{angre}.  The different panels show the distribution of 
(seeing corrected) effective angular sizes $r_{\rm dev}$ of the galaxies 
in the different bands.  Notice that $r_{\rm dev}$ for most of the 
objects is larger than the typical seeing scale of 1.5 arcsec:  we 
discuss this further in Section~\ref{checkphoto}.  
Comparison of the mean sizes in the different panels (the text in the 
top right corner of each panel) shows that galaxies appear slightly 
larger in the bluer bands.  

Because our sample covers a reasonably large range in redshift, this 
trend means we must correct the effective sizes to a fixed restframe 
wavelength.  Therefore, when converting from effective angular size 
$r_o$ to effective physical size $R_o$ we correct $r_o$ (and the 
Petrosian radii $r_{50}$ and $r_{90}$) in each band by linearly 
interpolating from the observed bandpasses to the central rest wavelength 
of each filter.  The typical correction is of the order of 4\%, although 
it is sometimes as large as 10\%.  In this respect, this correction is 
analogous to the K-correction we would ideally have applied to the 
magnitude and surface brightness of each galaxy.  
Note that the dependence of size on wavelength is also important for 
analyses of color gradients and the color--magnitude relation presented 
in Paper~IV.  

Our study will also require the effective surface brightness 
$\mu_o\equiv -2.5\log_{10}I_o$, where $I_o$ is the mean surface 
brightness within the effective radius $R_o$ (as opposed to the surface 
brightness at $R_o$).  In particular, we set 
$\mu_o = m_{\rm dev} + 2.5\log_{10} (2\pi r^2_o) - K(z) - 10\log_{10}(1+z)$.  
Note that this quantity is K-corrected, and also corrected for the 
cosmological $(1+z)^4$ dimming. Our earlier remarks about the K-correction 
are also relevant here.

\subsubsection{Reliability of photometric parameters}\label{checkphoto}

\begin{figure}
\centering
\epsfxsize=\hsize\epsffile{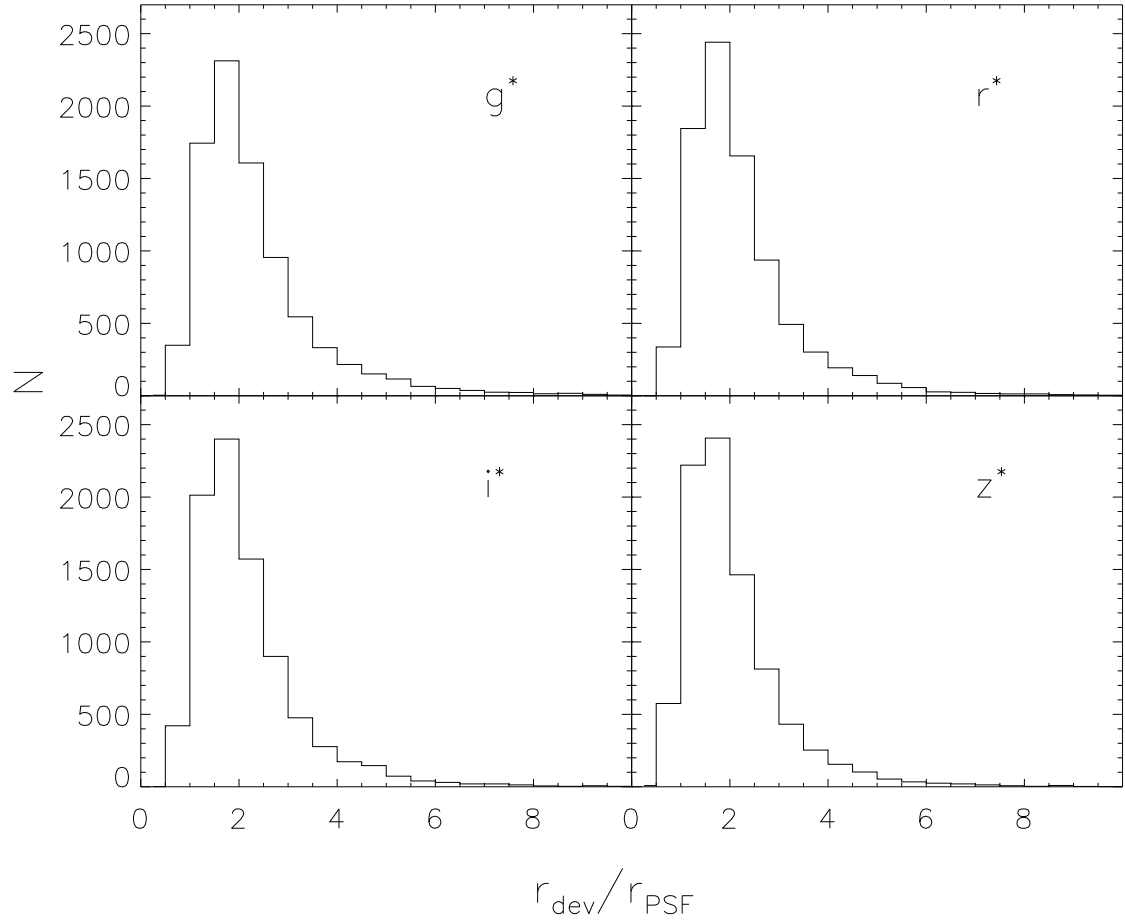}
\caption{Distribution of ratio between the effective radius 
$r_{\rm dev}$ and $r_{\rm PSF}$, the FWHM of the PSF; the 
effective radius is typically about a factor of two larger 
than the seeing.}
\label{fig:seeinghist}
\end{figure}

There is some discussion in the literature about the effects of seeing 
on estimates of the effective radius, and hence the effective surface 
brightness.  These effects are unimportant if the seeing scale, 
$r_{\rm PSF}$ (the FWHM in arcsec), is an order of magnitude smaller 
than the effective radius (e.g., Saglia et al. 1993, 1997).  
Figure~\ref{fig:seeinghist} shows histograms of 
$r_{\rm dev}/r_{\rm PSF}$ in the $g^*$, $r^*$, $i^*$ and $z^*$ bands:  
the effective radius is typically only a factor of two larger than 
the seeing.  Therefore, the accuracy of our estimates of the effective 
radii depend crucially on the correction for seeing being accurate.  

\begin{figure}
\centering
\epsfxsize=\hsize\epsffile{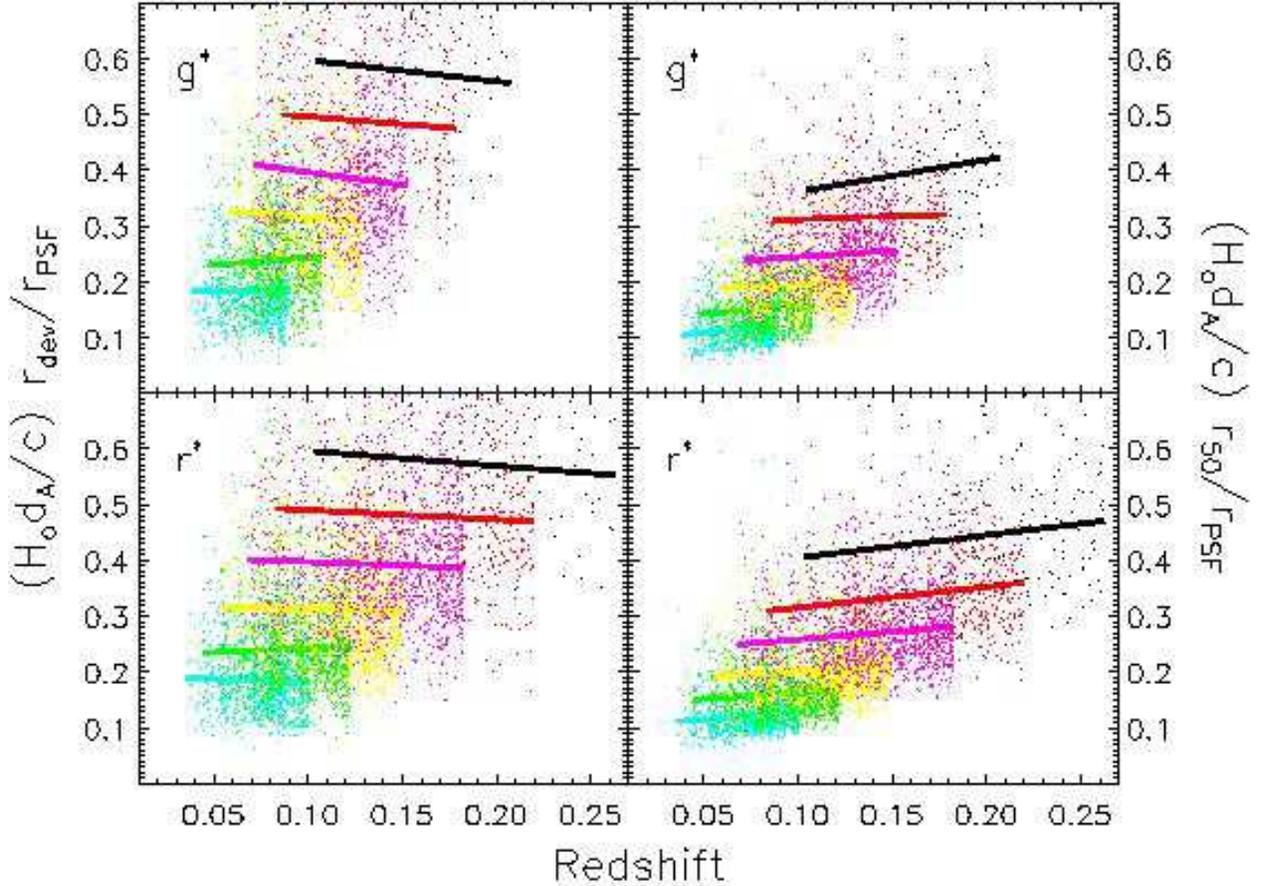}
\vspace{-1cm}
\caption{Effect of seeing on estimates of the half-light radius in 
the $g^*$ and $r^*$ bands (top and bottom panels).  
Each panel shows the ratio of $[H_0d_A(z)/c]$ times the angular 
half-light radius to $r_{\rm PSF}$ as a function of redshift, in 
volume limited catalogs chosen to be 0.5~mags wide.  The factor 
$H_0d_A(z)/c$ removes the expected scaling of angular diameter with 
redshift.  In the panels on the left, the half-light radius shown is 
$r_{\rm dev}$, which has been corrected for seeing, whereas the panels 
on the right use $r_{50}$, which is not seeing-corrected.  
In the panels on the right, there is a clear trend for the half-light 
radii to increase with redshift, suggesting that seeing is affecting 
the estimates.  This trend is absent in the panels on the left.  
(The panels on the left show a slight decrease with redshift---this 
may be a consequence of luminosity evolution.)}
\label{fig:seeing}
\end{figure}

To illustrate that the de~Vaucouleurs model fits output by the SDSS 
photometric pipeline are reasonable, Figure~\ref{fig:seeing} shows the 
ratio of $[H_0d_A(z)/c]$ times the half-light radius to $r_{\rm PSF}$ 
in the $g^*$ and $r^*$ bands (top and bottom panels) as a function of 
redshift, in a few volume limited catalogs.  The catalog limits were 
chosen to be 0.5~mags wide, so the galaxies in any given catalog 
should all be quite similar to each other.  (The factor $H_0d_A(z)/c$ 
removes the expected scaling of angular diameter with redshift.)  
It is important to use volume limited samples to make the measurement, 
because size is known to correlate with luminosity (more luminous 
galaxies are larger on average), and in a magnitude limited sample 
such as ours, only the more luminous objects are present at higher 
redshifts.  

In the panels on the left, the half-light radius is estimated by 
$r_{\rm dev}$, which has been corrected for seeing, whereas the 
panels on the right use $r_{50}$, which is not seeing-corrected.  
Seeing tends to increase the half-light radius, and indeed, in the 
panel on the right, there is a clear trend for the half-light radii 
to increase with redshift.  This trend is absent in the panels on the 
left, suggesting that the fitting procedure accounts quite well for 
the effects of seeing.  The panels on the left even show a slight 
decrease with redshift---we argue later (see discussion of 
Figure~\ref{fig:XzR} in Section~\ref{catalog}) that this may be a 
consequence of luminosity evolution.  

\begin{figure}
\centering
\epsfxsize=\hsize\epsffile{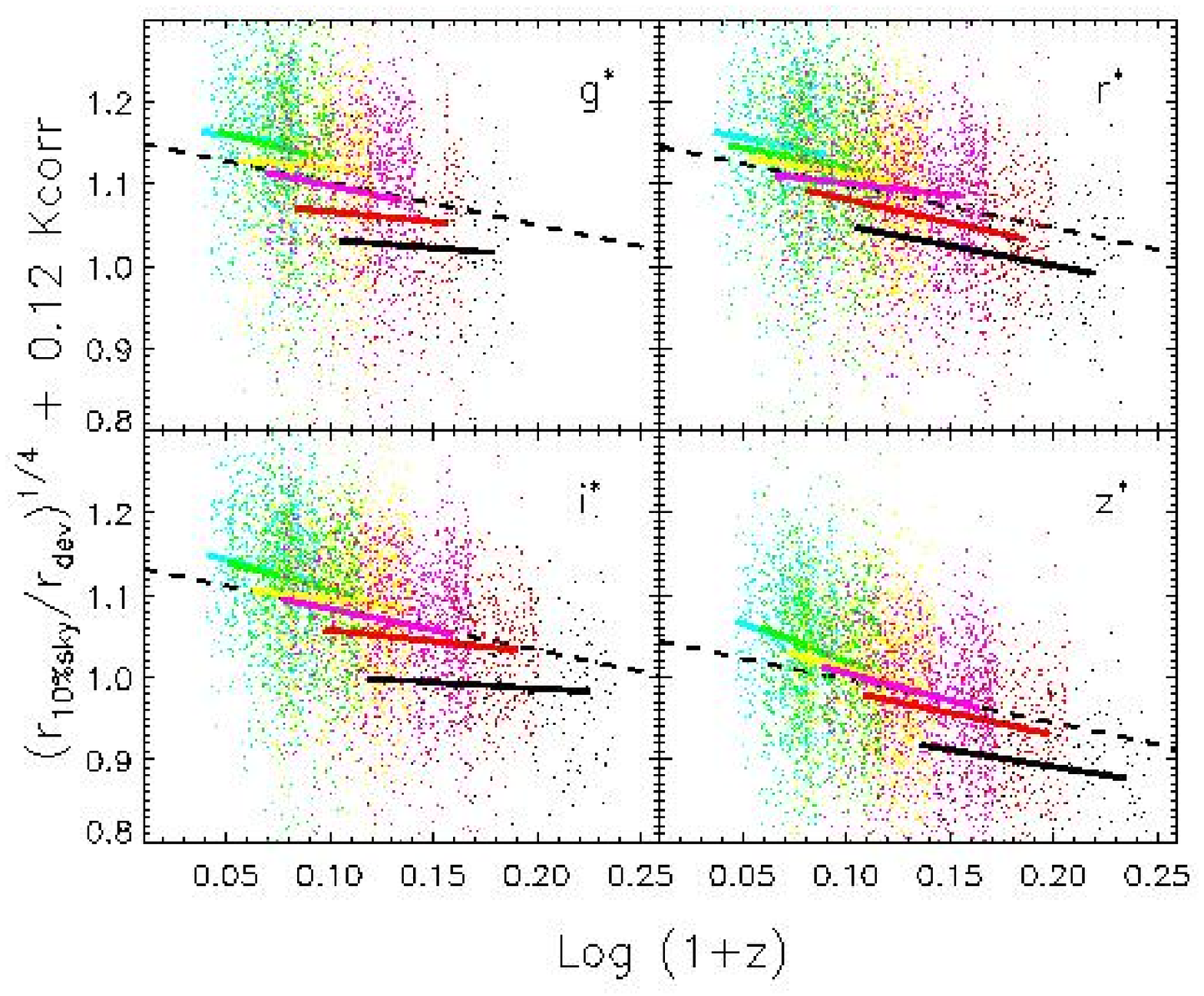}
\vspace{-1 cm}
\caption{The ratio of the radius $r_{\rm 10\% sky}$ at which the 
fitted surface brightness is 10\% of the sky level, and the 
effective radius $r_{\rm dev}$, as a function of redshift, for a 
few volume limited subsamples.  In all four bands, this ratio 
decreases with increasing redshift, although the trend is slightly 
weaker in $g^*$ than in the other bands.  Dashed line shows the 
expected scaling if there is no luminosity or size evolution.  
The trend for the combination of all the samples is substantially 
steeper than that within any one volume limited sample, suggesting 
that the typical size increases strongly with luminosity.  }
\label{fig:re10sky}
\end{figure}

We have argued that our estimates of $r_{\rm dev}$ are probably not 
strongly compromised by seeing.  In our dataset, $r_{\rm dev}$ is 
about a factor of two smaller than the radius $r_{\rm 10\%}$ at which 
the fitted surface brightness is 10\% of the sky level.  Therefore, 
errors in the sky subtraction may also compromise our estimates of 
$r_{\rm dev}$.  If the same object is moved to higher redshift but 
the sky level is kept fixed, then one expects 
$7.67\,(r_{10\%}/r_{\rm dev})^{1/4} = {\rm const}-4\ln(1+z)$ 
because cosmological surface brightness dimming scales as $(1+z)^4$.  
Therefore, a plot of $(r_{10\%}/r_{\rm dev})^{1/4}$ versus $\ln(1+z)$, 
measured from volume limited catalog, should have a slope of $-4/7.67$.  
However, because each volume limited catalog spans a large range in 
redshift, we must also account for the surface brightness dimming 
associated with the fact that redshift dependent $K$-corrections must 
be made to estimate fixed restframe luminosities.  Therefore, we expect 
the scaling with redshift to be slightly stronger:  
 $(r_{10\%}/r_{\rm dev})^{1/4} = 
  {\rm const}-0.52\ln(1+z) - 0.052\,{\rm ln}(10)\,K(z)$.

Figure~\ref{fig:re10sky} shows how 
 $(r_{\rm 10\%}/r_{\rm dev})^{1/4}+0.12\,K(z)$ 
depends on $\ln(1+z)$, in the same volume limited catalogs 
as before; solid lines show the mean trend in the faintest (top) 
to the most luminous (bottom) catalogs.  
Comparison with the dashed line, which has a slope of $-0.52$, 
shows that the scaling with redshift is close to that expected.  
(Note that it is important to use volume limited catalogs to make 
the measurement:  the trend in any given volume limited subsample is 
shallower than the trend one would infer by first combining together 
all the subsamples and then making the measurement.  To see why this 
happens, suppose that the mean size at fixed luminosity scales as 
$R_o\propto L^\alpha$.  Then the magnitude limit induces a correlation 
which should scale approximately as 
$r_{\rm dev}\propto R_o/z \propto L^\alpha/z \propto z^{2\alpha}/z$.  
If $\alpha>1/2$ then $r_{\rm dev}$ increases with $z$, 
so $r_{10\%}/r_{\rm dev}$ decreases with $z$ simply because of the 
magnitude limit.)  

Some of the small differences in slope between the dashed and solid 
lines can be attributed to the effects of luminosity evolution.  
By fixing $L$, we select objects which are slightly smaller at 
higher $z$ (c.f. Figure~\ref{fig:XzR}), hence have slightly smaller 
values of $r_{\rm dev}$ and so larger values of $r_{10\%}/r_{\rm dev}$ 
than expected; so they appear to decline less rapidly than 
const$-0.52\,z$.  
%
Luminosity evolution is expected to be stronger in the bluer bands, 
and the $g^*$ band relation does appear to be slightly shallower than 
the other bands.  
Thus, as was the case for $r_{\rm dev}/r_{\rm PSF}$, we see redshift 
dependent trends in $r_{10\%}/r_{\rm dev}$ which we believe are 
physically reasonable.  The sky-subtraction procedure does not appear 
to have introduced any obvious artifacts into our estimates of 
$r_{\rm dev}$.  

\begin{figure}
\centering
\epsfxsize=0.9\hsize\epsffile{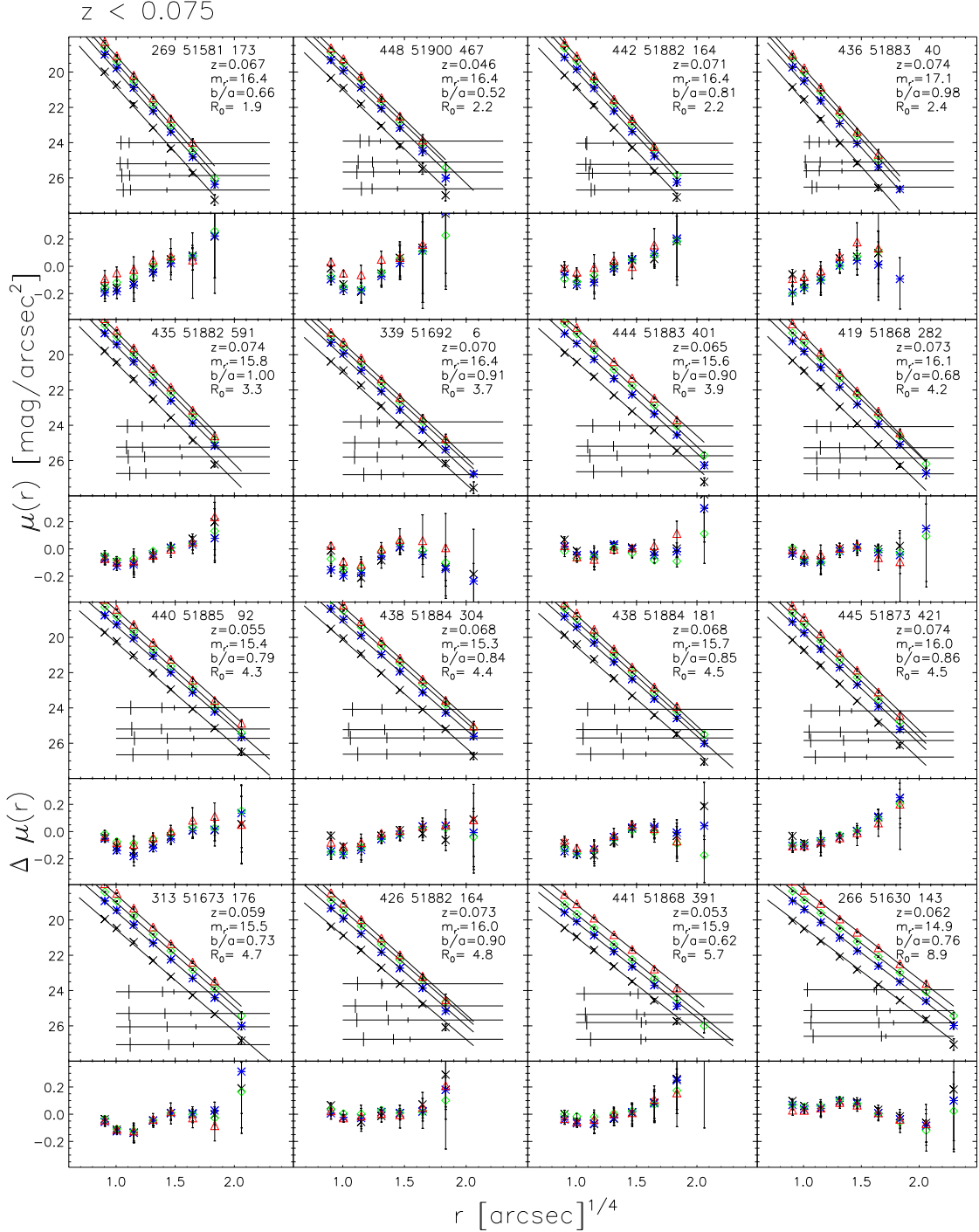}
\caption{Examples of surface brightness profiles selected randomly from 
galaxies in our sample which are in the redshift range $z < 0.075$.  
(Upper panels) Text in each panel shows plate number, MJD, and fiber ID of the 
object, the redshift $z$, the apparent $r^*$ magnitude $m_{r^*}$, 
the fitted axis ratio $b/a$, and the physical half-light radius $R_o$.  
Panels are ordered by physical radius $R_o$, which increases from 
top to bottom, and from left to right.  Symbols show the surface 
brightness profiles in the $g^*$, $r^*$, $i^*$ and $z^*$ bands (bottom to 
top in each of the upper panels),
and solid lines show the de~Vaucouleur fits which, when convolved 
with the seeing, provide the best fit to the symbols.  
Horizontal lines show 1\% of the surface 
brightness of the sky in the different bands, and 
large, medium and small tickmarks show $r_{\rm PSF}$, $r_{\rm dev}$ 
and $r_{10\%}$, the radius at which the fitted surface brightness 
equals 10\% of the sky. (Bottom panels) The residuals (observed minus fit) 
in the different bands (crosses, stars, diamonds and triangles show the 
$g^*$, $r^*$, $i^*$ and $z^*$ band results, respectively).}
\label{fig:deVfits}
\end{figure}

Finally, we turn to an inspection of how well the de~Vaucouleurs model 
actually fits the observed surface brightness profiles.  
Figure~\ref{fig:deVfits} shows angular averaged surface brightness 
profiles in the $g^*$, $r^*$, $i^*$ and $z^*$ bands (bottom to 
top in each of the upper panels), for a random selection of 
galaxies in our sample, ordered by redshift and size.  
The text in each panel shows the plate number, MJD, and fiber ID 
number of the observation, the redshift $z$, 
the fitted apparent $r^*$ magnitude $m_{r^*}$, the fitted axis ratio 
$b/a$, and the effective physical radius $R_o$, computed from the 
fitted angular radius $r_{\rm dev}$, the fitted axis ratio and the 
redshift, as described previously.  
The solid lines show de~Vaucouleurs profiles which, when convolved 
with the pipeline's model for the seeing, provide the best fit 
to the symbols (because the fits are actually not to these azimuthally 
averaged one-dimensional profiles, but to the two dimensional profile, 
they serve mainly as guides to the eye.)   The  
horizontal lines show the surface brightness at 1\% of the sky in the 
different bands.  Long, medium and short tick marks 
on these lines show $r_{\rm PSF}$, $r_{\rm dev}$ and $r_{10\%}$, the 
radius at which the fitted surface brightness equals 10\% of the sky.  

The bottom panels show the residuals (observed minus fit) in the 
different bands (crosses, stars, diamonds and triangles show the 
$g^*$, $r^*$, $i^*$ and $z^*$ band results, respectively).  The residuals 
are systematically low on small scales, but comparison with the 
large tick marks, which indicate the size of the seeing disk, 
show that this discrepancy is because we are comparing fits which 
have been corrected for seeing with the observed surface brightnesses 
which are not seeing-corrected.  On larger scales (typically larger 
than twice the half-light radii), the residuals are also large, 
indicating that there may be problems with the sky subtraction.  
In linear flux units, the residuals tend to a constant value 
which is typically smaller than 0.3\% of the sky level (recall that 
the horizontal lines in the panels show the surface brightness at 
1\% of sky).  Although this is not an entirely fair comparison, since 
the fits are not to performed on these azimuthally averaged 
one-dimensional profiles, but to the two dimensional profile, 
this test suggests that the fit is reliable, but that the zero-point 
level of the sky may be systematically biased by $\sim 0.3$\%.  
Note that the residuals are larger in the $g^*$ and $z^*$ bands; but 
this is consistent with the larger errors we quote on the photometry 
in these bands.

\begin{figure}
\centering
\epsfxsize=\hsize\epsffile{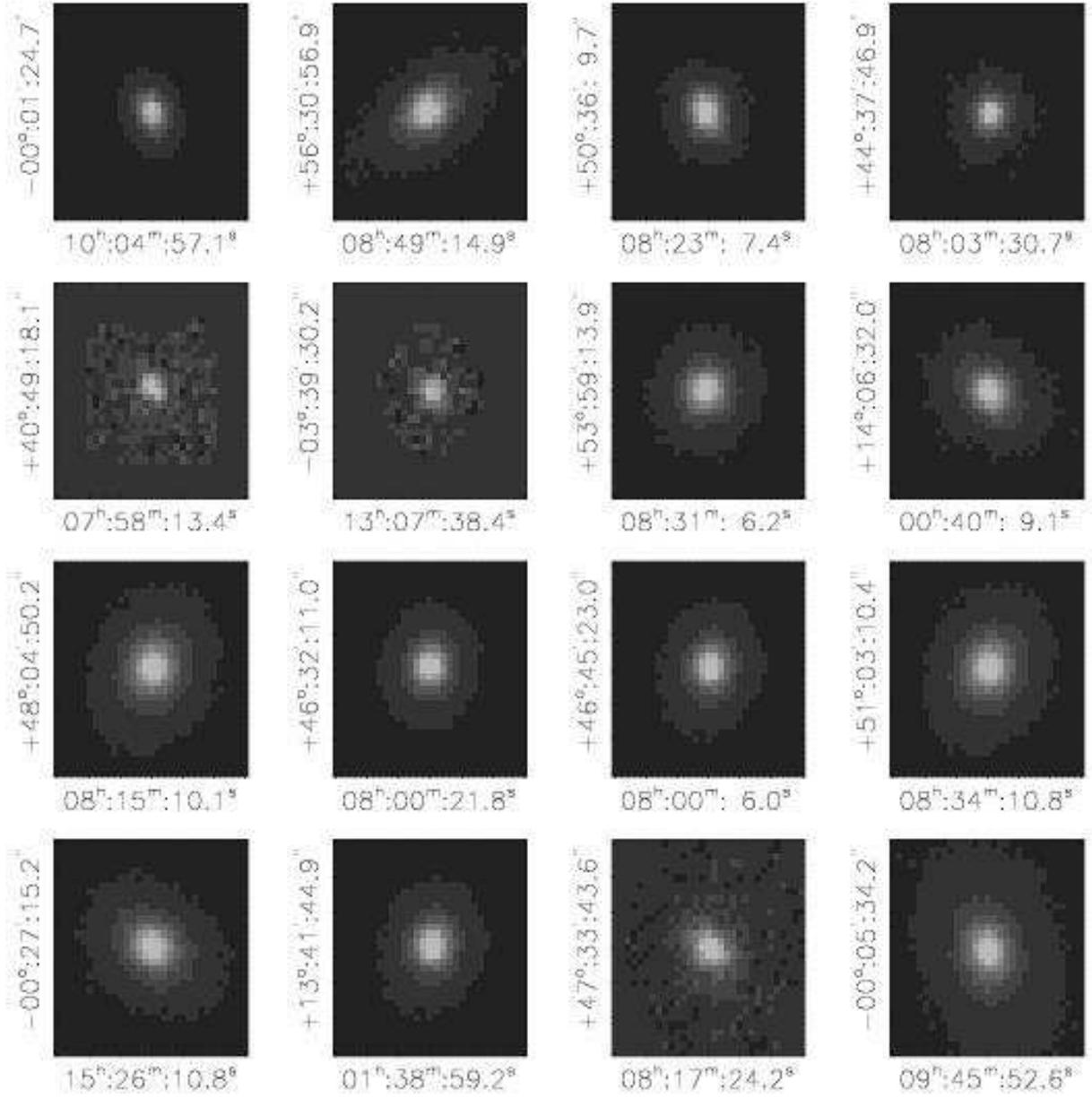}
\caption{$r^*$-band images for the $z<0.075$ objects shown in 
Figure~\ref{fig:deVfits}.  The labels on the x- and y-axes show the 
coordinates (RA and DEC in J2000) of the galaxy centers.  }
\label{fig:images}
\end{figure}

Figure~\ref{fig:images} shows the $r^*$ images of the same objects 
whose profiles are shown in Figure~\ref{fig:deVfits}.  The images 
are each $11.88\times 11.88$~arcsec$^2$ (i.e., 30 pixels on a side).  
One might expect some of the worst fits are due to the presence of a 
disk component which our de~Vaucouleur model does not account for.  
However, comparison of the images and the fits shows no obvious 
correlation between goodness-of-fit and, e.g., $b/a$.  

\begin{figure}
\epsfxsize=\hsize\epsffile{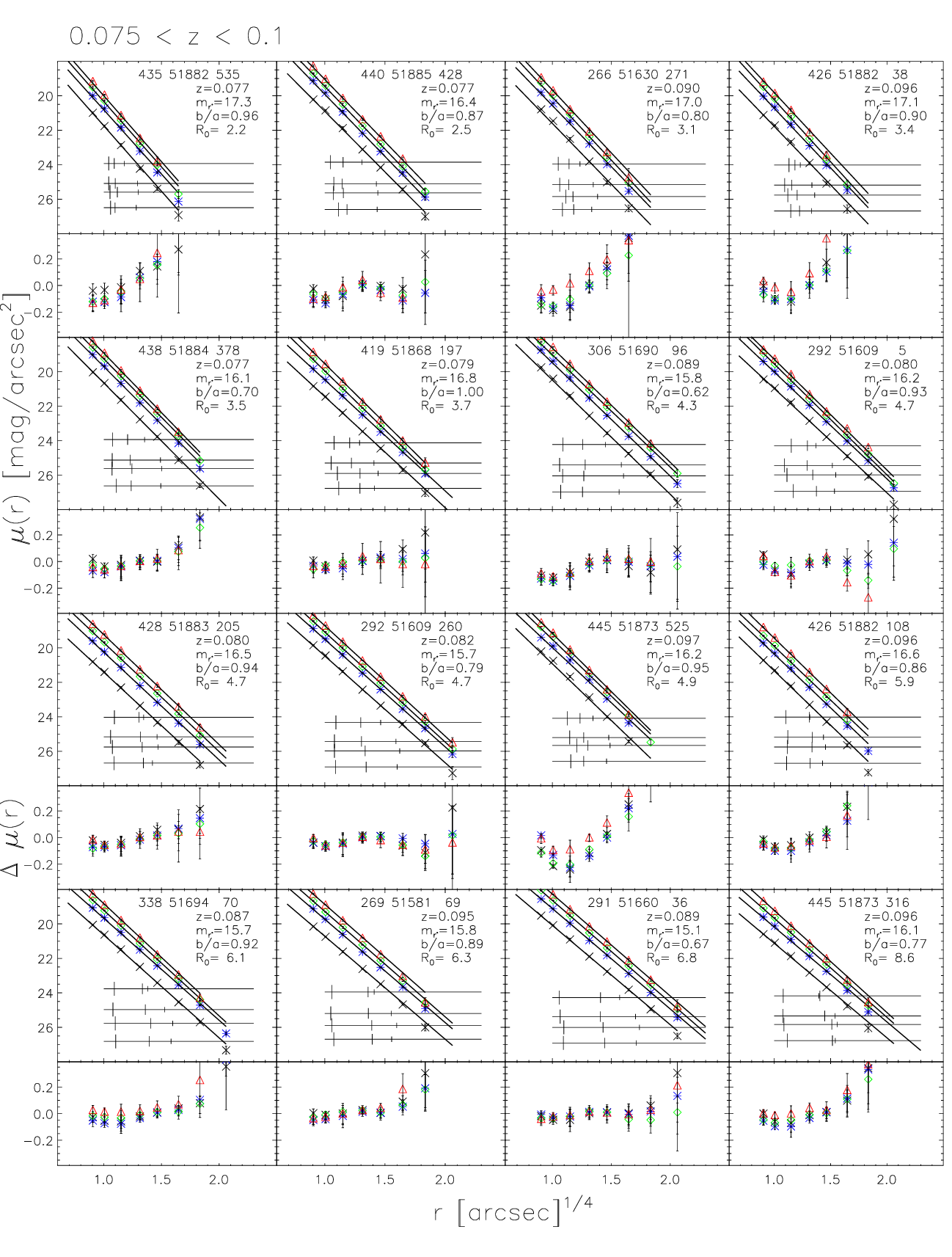}
\begin{center}
Fig. \ref{fig:deVfits}. -- Continued. Redshift range $0.075 < z < 0.1$.
\end{center}
\end{figure}

\begin{figure}
\epsfxsize=\hsize\epsffile{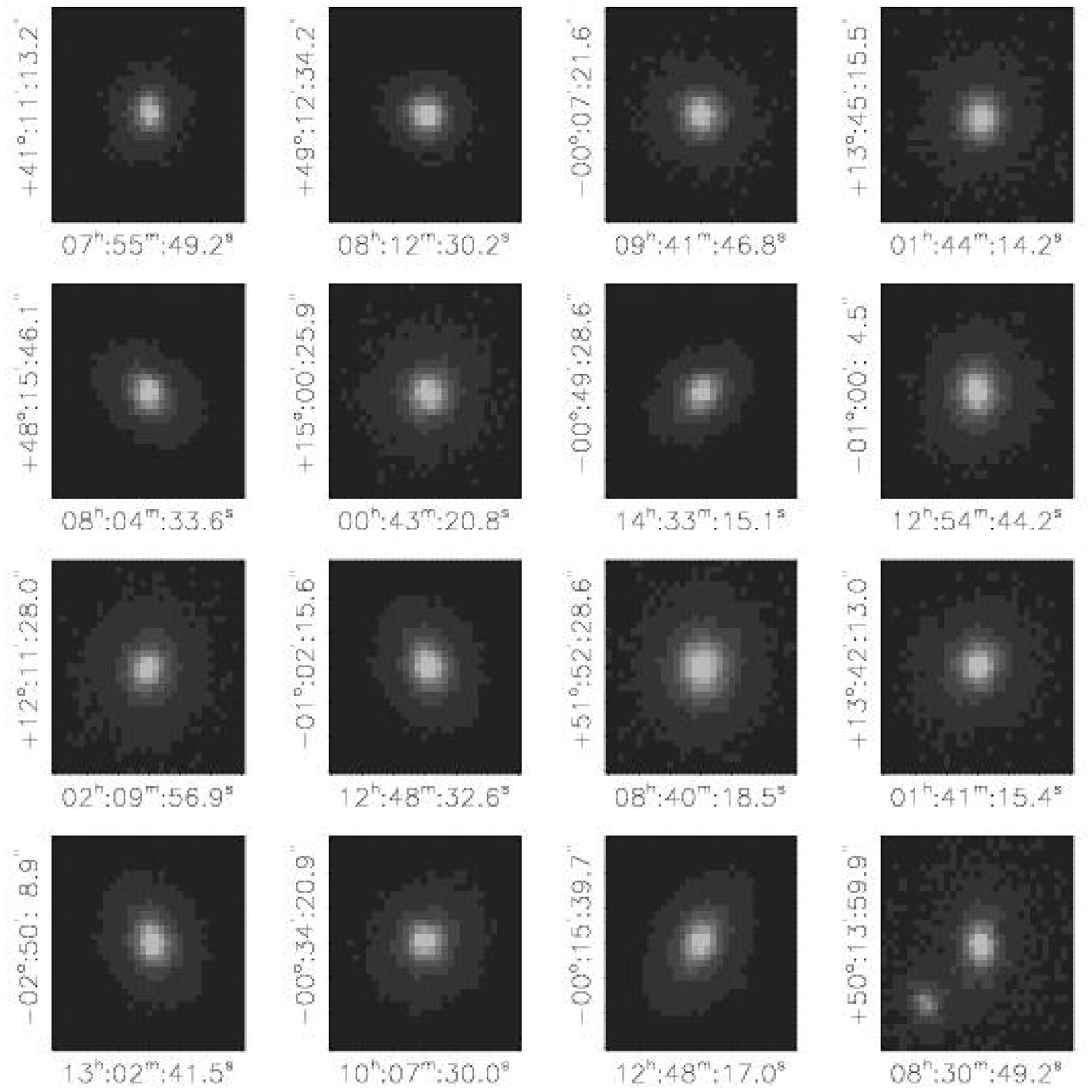}
\begin{center}
Fig. \ref{fig:images}. -- Continued. Redshift range $0.075 < z < 0.1$.
\end{center}
\end{figure}

\begin{figure}
\epsfxsize=\hsize\epsffile{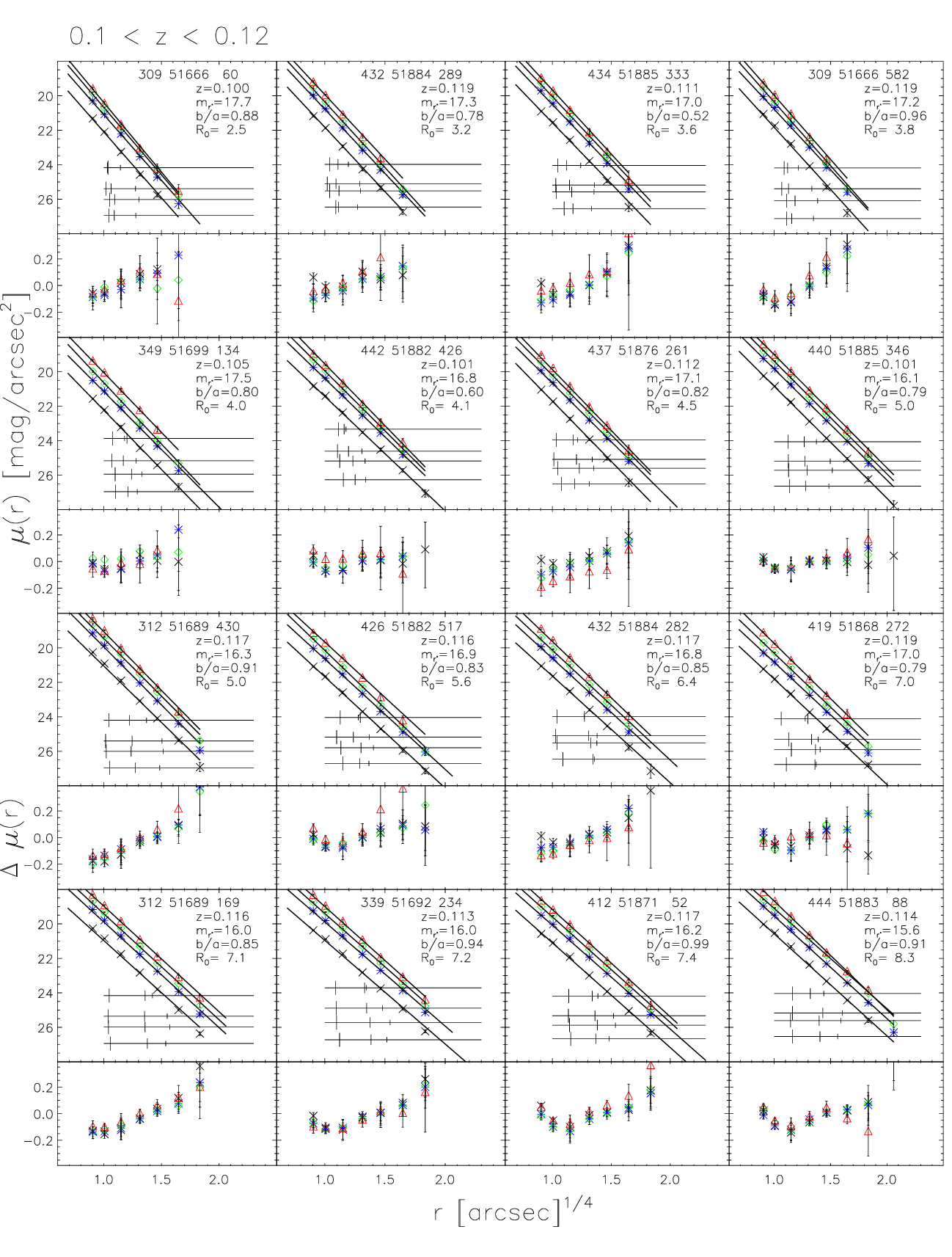}
\begin{center}
Fig. \ref{fig:deVfits}. -- Continued. Redshift range $0.1 < z < 0.12$.
\end{center}
\end{figure}

\begin{figure}
\epsfxsize=\hsize\epsffile{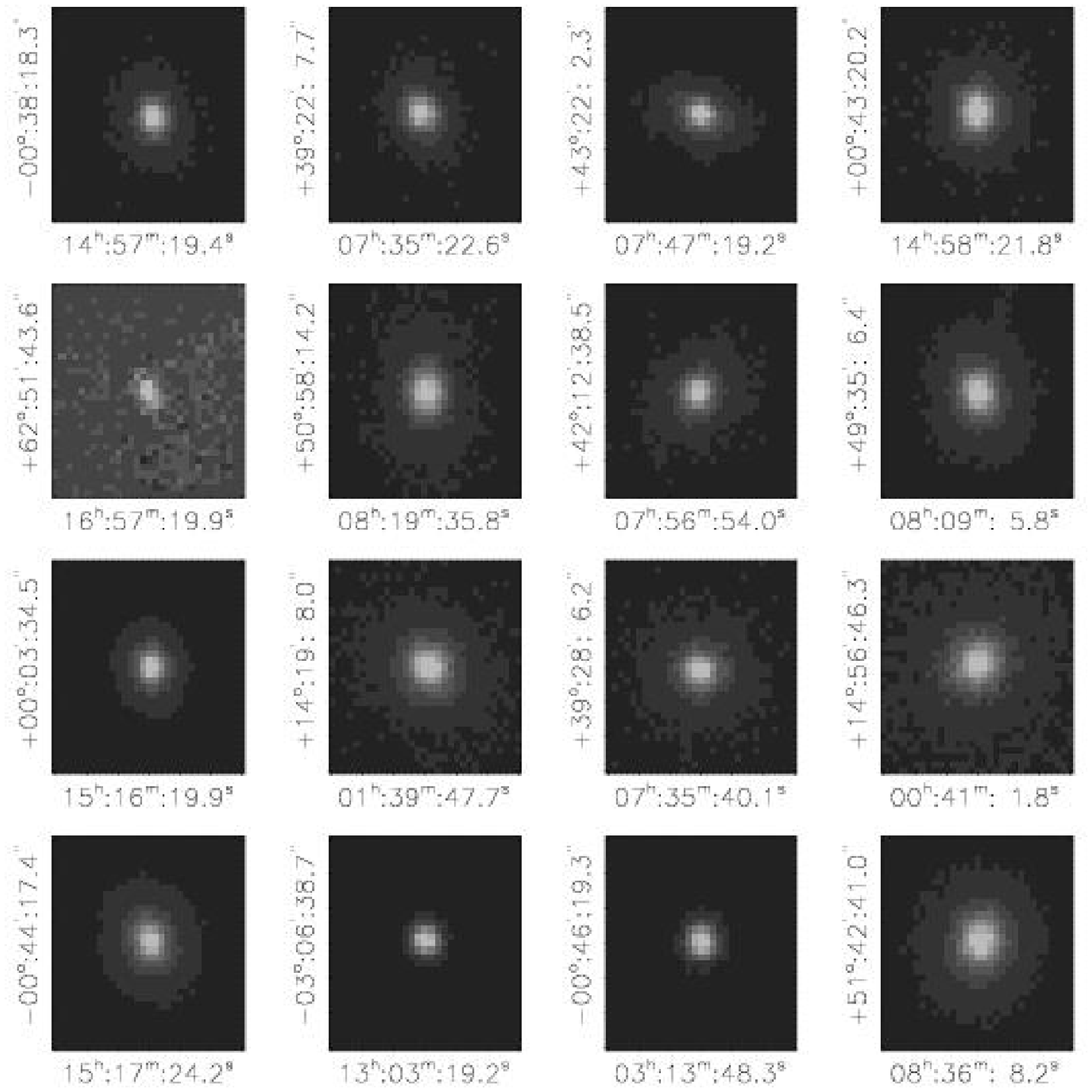}
\begin{center}
Fig. \ref{fig:images}. -- Continued. Redshift range $0.1 < z < 0.12$.
\end{center}
\end{figure}

\begin{figure}
\epsfxsize=\hsize\epsffile{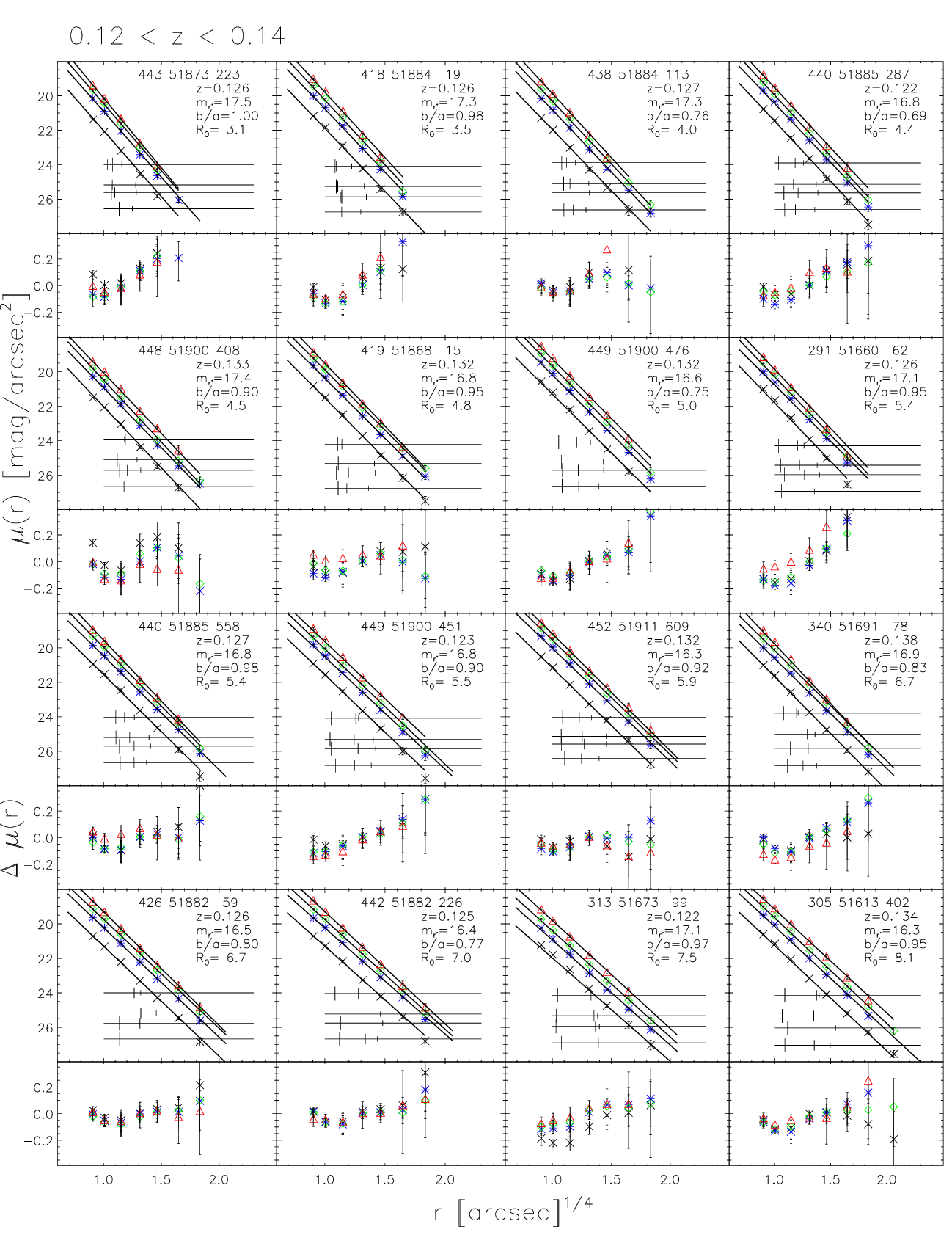}
\begin{center}
Fig. \ref{fig:deVfits}. -- Continued. Redshift range $0.12 < z < 0.14$.
\end{center}
\end{figure}

\begin{figure}
\epsfxsize=\hsize\epsffile{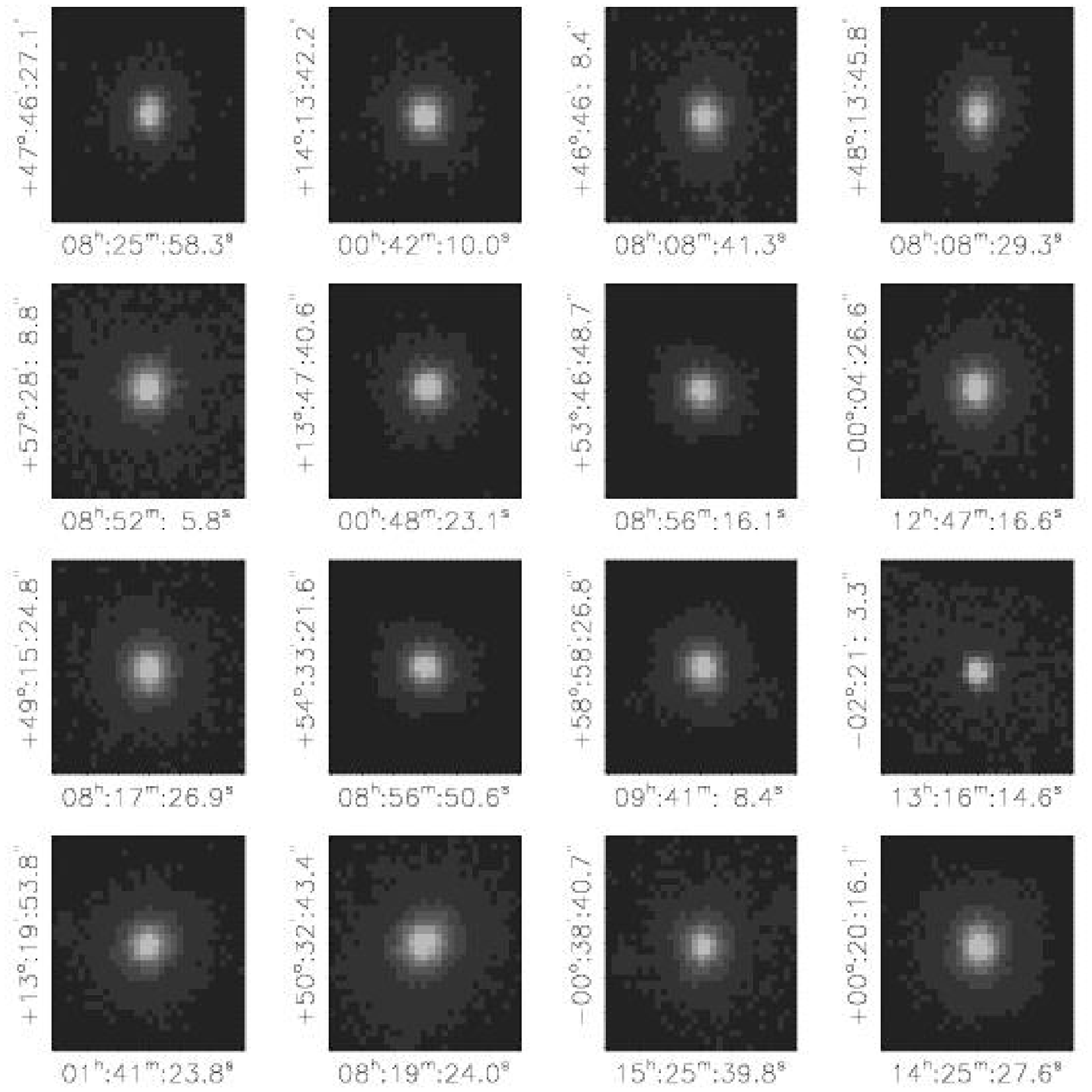}
\begin{center}
Fig. \ref{fig:images}. -- Continued. Redshift range $0.12 < z < 0.14$.
\end{center}
\end{figure}

\begin{figure}
\epsfxsize=\hsize\epsffile{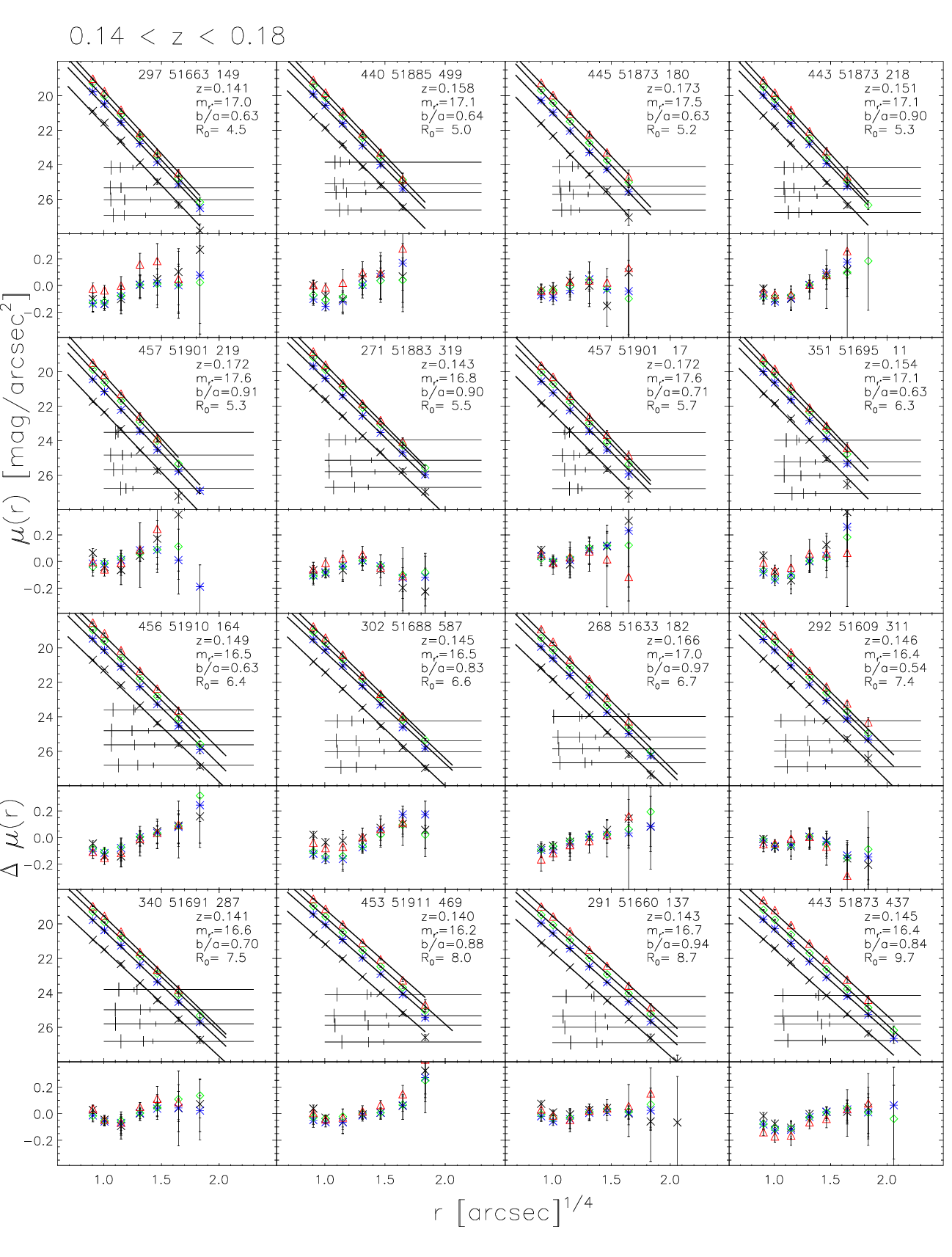}
\begin{center}
Fig. \ref{fig:deVfits}. -- Continued. Redshift range $0.14 < z < 0.18$.
\end{center}
\end{figure}

\begin{figure}
\epsfxsize=\hsize\epsffile{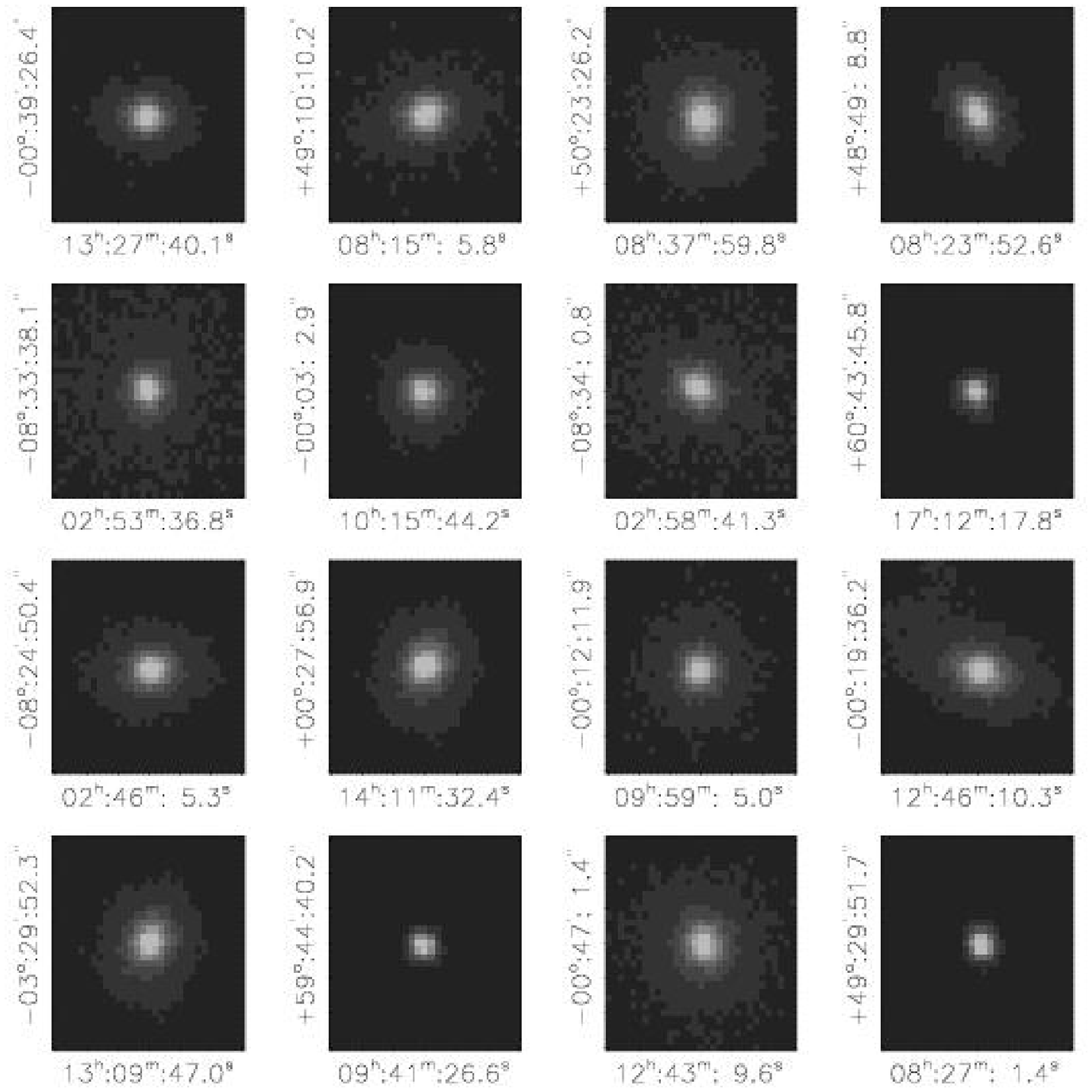}
\begin{center}
Fig. \ref{fig:images}. -- Continued. Redshift range $0.14 < z < 0.18$.
\end{center}
\end{figure}

\begin{figure}
\epsfxsize=\hsize\epsffile{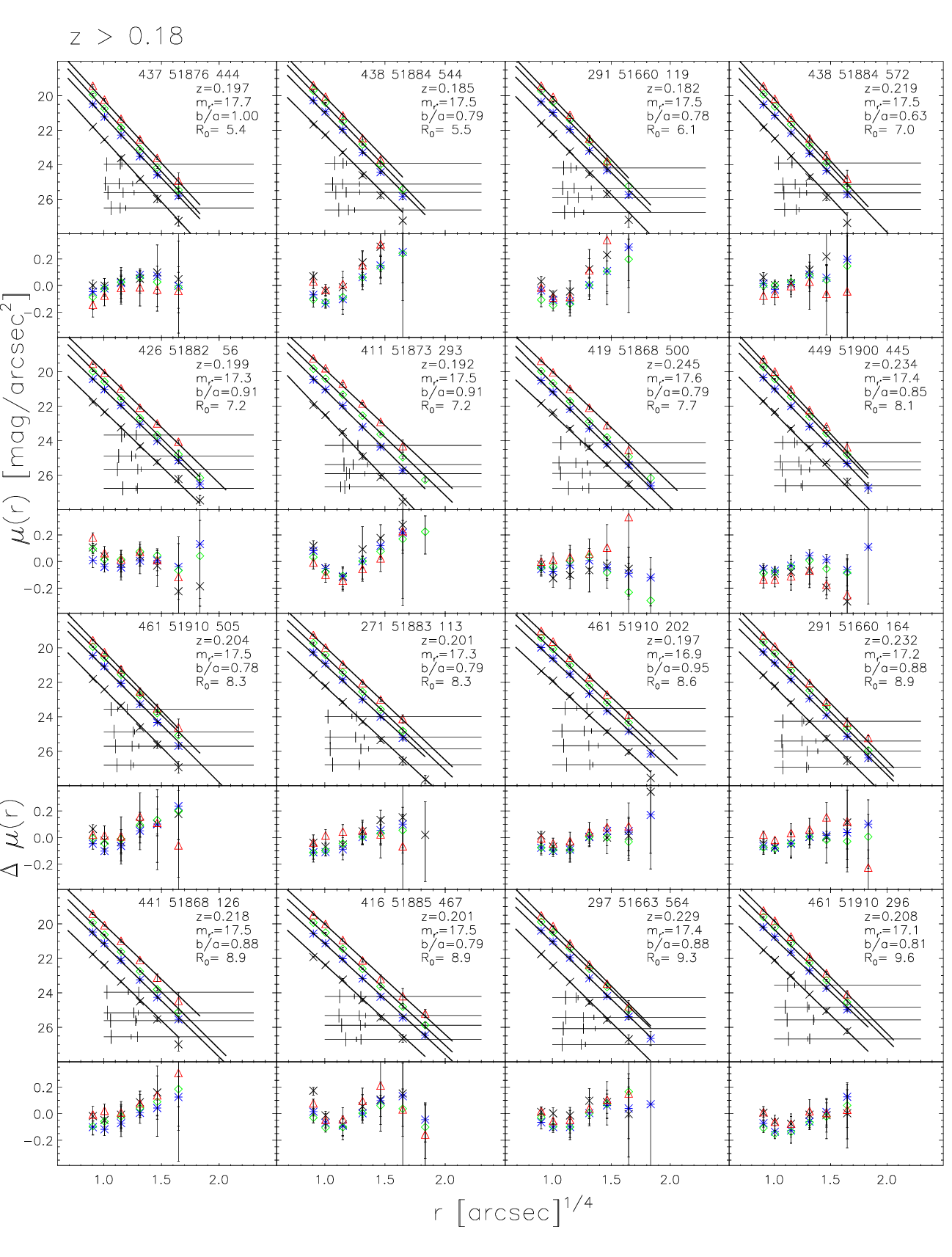}
\begin{center}
Fig. \ref{fig:deVfits}. -- Continued. Redshift range $z > 0.18$.
\end{center}
\end{figure}

\begin{figure}
\epsfxsize=\hsize\epsffile{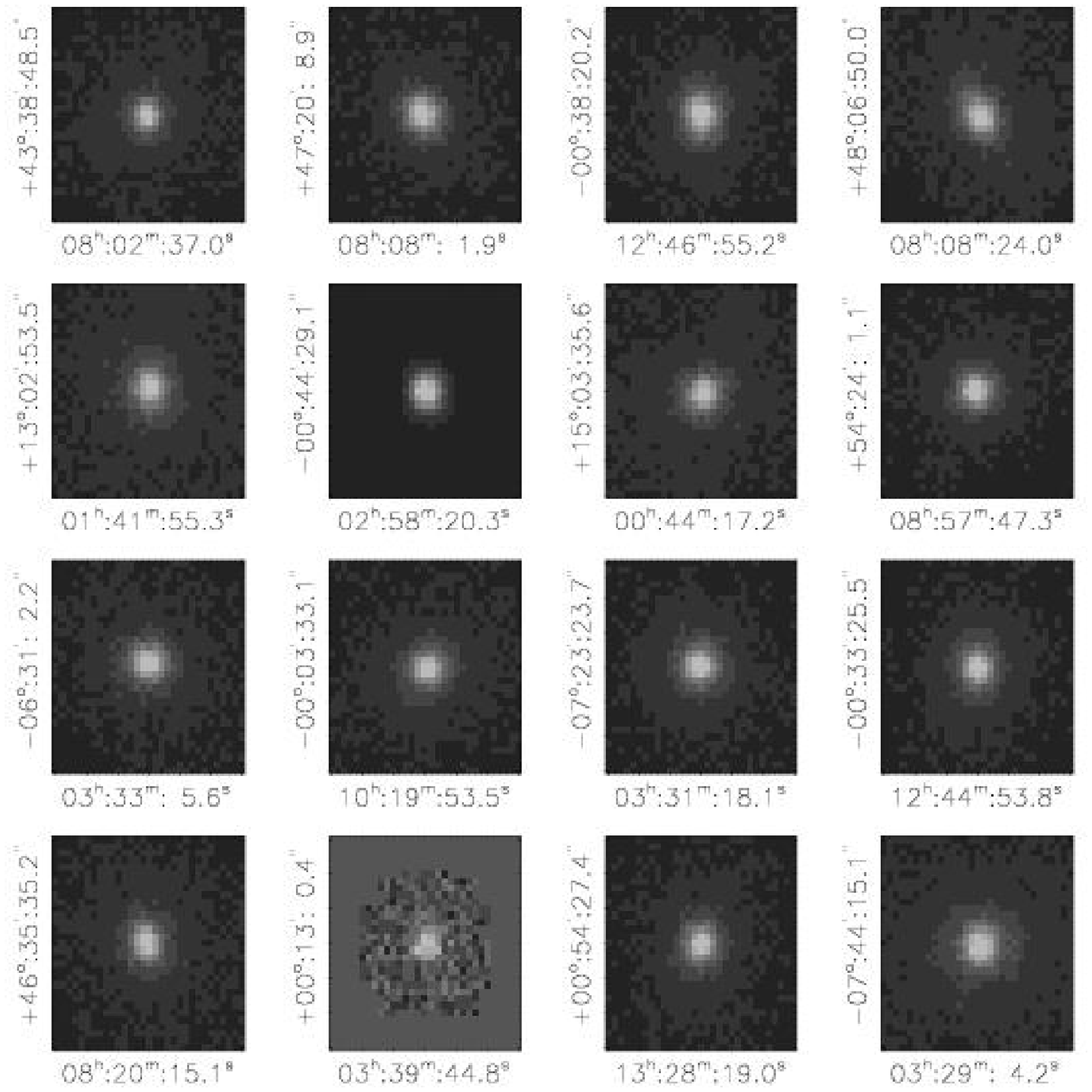}
\setcounter{figure}{8}
\begin{center}
Fig. \ref{fig:images}. -- Continued. Redshift range $z > 0.18$.
\end{center}
\end{figure}

In summary, we have performed a number of tests of the SDSS photometric 
pipeline reductions and have found little evidence that they are 
significantly biased.  Therefore, in what follows, the apparent 
magnitudes we use are those output by the SDSS photometric pipeline.  
To convert to luminosities, we K-correct (and use the inverse square 
law).  The angular sizes are output from the SDSS photometric pipeline.  
We correct these to a fixed restframe size, and we correct for $a/b$.  
Both these corrections are standard.  
Thereafter, surface brightness is defined from the restframe luminosity 
and size with no additional corrections.  
The uncertainties on the photometric parameters we use in this paper, 
and the magnitude limits in each band, are summarized in 
Table~\ref{tab:photerr}.  There is some subtlety in obtaining them 
from the format available in the SDSS database; this is discussed 
in Appendix~\ref{errors}.  

\begin{table}[t]
\centering
\caption[]{Photometric parameters and median errors of the objects in 
our sample.\\}
\begin{tabular}{cccccccccc}
\hline &&&\\
Band & $m_{\rm min}$ & $m_{\rm max}$ & $\delta m_{\rm dev}$ 
     & $\delta\log r_o$ & $\delta\log {\rm I_o}$ 
     & $\delta m_{\rm m}$ & $\delta m_{\rm pet}$ 
     & $\delta\log r_{50}$ & $\delta\log r_{90}$\\ 
     & mag & mag & mag & dex & dex & mag & mag & dex & dex \\
\hline &&&\\
$g^*$ & 15.50 & 18.10 & 0.03 & 0.02 & 0.04 & 0.03 & 0.04 &  0.02 & 0.02 \\
$r^*$ & 14.50 & 17.45 & 0.02 & 0.01 & 0.03 & 0.02 & 0.02 &  0.01 & 0.01 \\  
$i^*$ & 14.50 & 17.00 & 0.02 & 0.01 & 0.03 & 0.02 & 0.02 &  0.01 & 0.01 \\ 
$z^*$ & 14.50 & 16.70 & 0.05 & 0.03 & 0.09 & 0.05 & 0.05 &  0.03 & 0.03 \\  
\hline &&&\\
\end{tabular}
\label{tab:photerr} 
\end{table}


\subsection{The spectroscopic data of the sample}\label{spec}

\subsubsection{Estimating the velocity dispersions}\label{vdisp}
The SDSS pipeline does not provide an estimate of the line-of-sight 
velocity dispersion, $\sigma$, within a galaxy, so we must compute 
it separately for the early-type galaxy sample. 
The observed velocity dispersion $\sigma$ is the result 
of the superposition of many individual stellar spectra, each of which 
has been Doppler shifted because of the star's motion within the 
galaxy.  Therefore, it can be determined by analyzing the integrated 
spectrum of the whole galaxy.  A number of objective and accurate methods 
for making velocity dispersion measurements have been developed 
(Sargent et al. 1977; Tonry \& Davis 1979; 
Franx, Illingworth \& Heckman 1989; Bender 1990; Rix \& White 1992).  
Each of these methods has its own strengths, weaknesses, and biases.  
Appendix~\ref{vmethods} describes how we combined these different 
techniques to estimate $\sigma$ for the galaxies in our sample.  

The velocity dispersion estimates we use in what follows are obtained 
by fitting the restframe wavelength range $4000- 7000$~\AA, and then 
averaging the estimates provided by the {\it Fourier-fitting} and 
{\it direct-fitting} methods to define what we call $\sigma_{\rm est}$.  
(We do not use the cross-correlation estimate because of its behavior 
at low $S/N$ as discussed in Appendix~\ref{vmethods}.  Note 
that the $S/N$ of the SDSS spectra depend on wavelength.)  The error on 
$\sigma_{\rm est}$ is determined by adding in quadrature the errors on 
the two estimates (i.e., the Fourier-fitting and direct-fitting) which 
we averaged.  The resulting error is between $\delta\log\sigma\sim 0.02$~dex
and 0.06~dex, depending on the signal-to-noise of the spectra, with 
a median value of $\sim 0.03$~dex. 
A few galaxies in our sample have been observed more than once.  
The scatter between different measurements is $\sim 0.04$~dex, 
consistent with the amplitude of the errors on the measurements 
(see Figure~\ref{fig:vrptd}).  Based on the typical $S/N$ of the SDSS 
spectra and the instrumental resolution, we chose 70 km~s$^{-1}$ as a 
lower limit on the velocity dispersions we use in this paper.  

Following J{\o}rgensen et al. (1995) and Wegner et al. (1999), 
we correct $\sigma_{\rm est}$ to a standard relative circular 
aperture defined to be one-eighth of the effective radius:  
\begin{equation}
{\sigma_{\rm cor}\over\sigma_{\rm est}} = 
 \left(r_{\rm fiber}\over r_o/8\right)^{0.04},
 \label{appcorr}
\end{equation}
where $r_{\rm fiber}=1.5$ arcsec and $r_o$ is the effective 
radius of the galaxy measured in arcseconds.  In principle, we should 
also account for the effects of seeing on $\sigma_{\rm est}$, just 
as we do for $r_o$.  However, because the aperture correction depends 
so weakly on $r_o$ (as the 0.04 power), this is not likely to be a 
significant effect.  In any case, most galaxies in our sample have 
$r_o\ge 1.5$ arcsecs (see Figure~\ref{angre}).  

Note that this correction assumes that the velocity dispersion 
profiles of early-type galaxies having different $r_o$ are similar.  
At the present time, we do not have measurements of the profiles of 
any of the galaxies in our sample, so we cannot test this assumption.  
In Paper~II of this series we will argue that the galaxies in our sample 
evolve very little; this means that if we select galaxies of the same 
luminosity and effective radius, then a plot of velocity dispersion 
versus redshift of these objects should allow us to determine if the 
aperture correction above is accurate.  The results of this exercise 
are presented in Appendix~\ref{vr}.

\subsubsection{Examples of early-type galaxy spectra}\label{exspec}
\begin{figure}
\centering
\epsfxsize=\hsize\epsffile{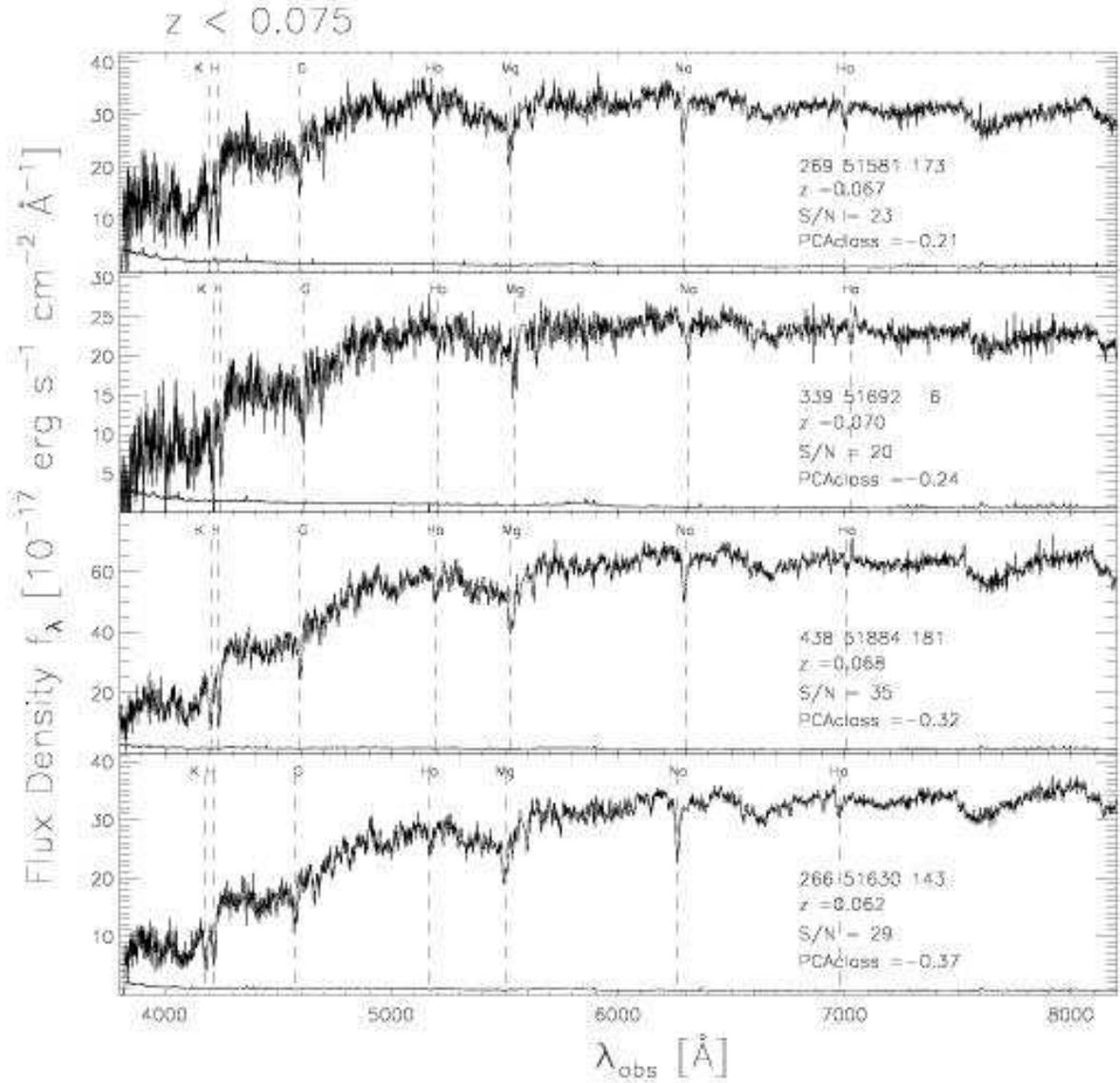}
\caption{Spectra of galaxies which lie along the diagonals of 
Figures~\ref{fig:deVfits} and~\ref{fig:images}.  Text in each panel 
gives the plate number, MJD and fiber ID, the redshift $z$, 
the signal-to-noise ratio, and the PCA classification number.  
Some common absoption features are also indicated.  The line along 
the bottom of each panel shows the error in the flux density in each 
pixel.  Here, galaxies span the redshift range $z < 0.075$.}
\label{fig:spectra}
\end{figure}

To illustrate the typical quality of the spectra in our sample, 
Figure~\ref{fig:spectra} shows spectra of the galaxies which lie 
along the diagonals in Figure~\ref{fig:deVfits} (panels from top to 
bottom here are for objects which are top left to bottom right in 
figure~\ref{fig:deVfits}).  All the spectra in our sample have PCA 
classification numbers $a < -0.1$, typical of early-type galaxy 
spectra (Connolly \& Szalay 1999), although a few of the spectra do 
show weak emission lines.  

\begin{figure}
\epsfxsize=\hsize\epsffile{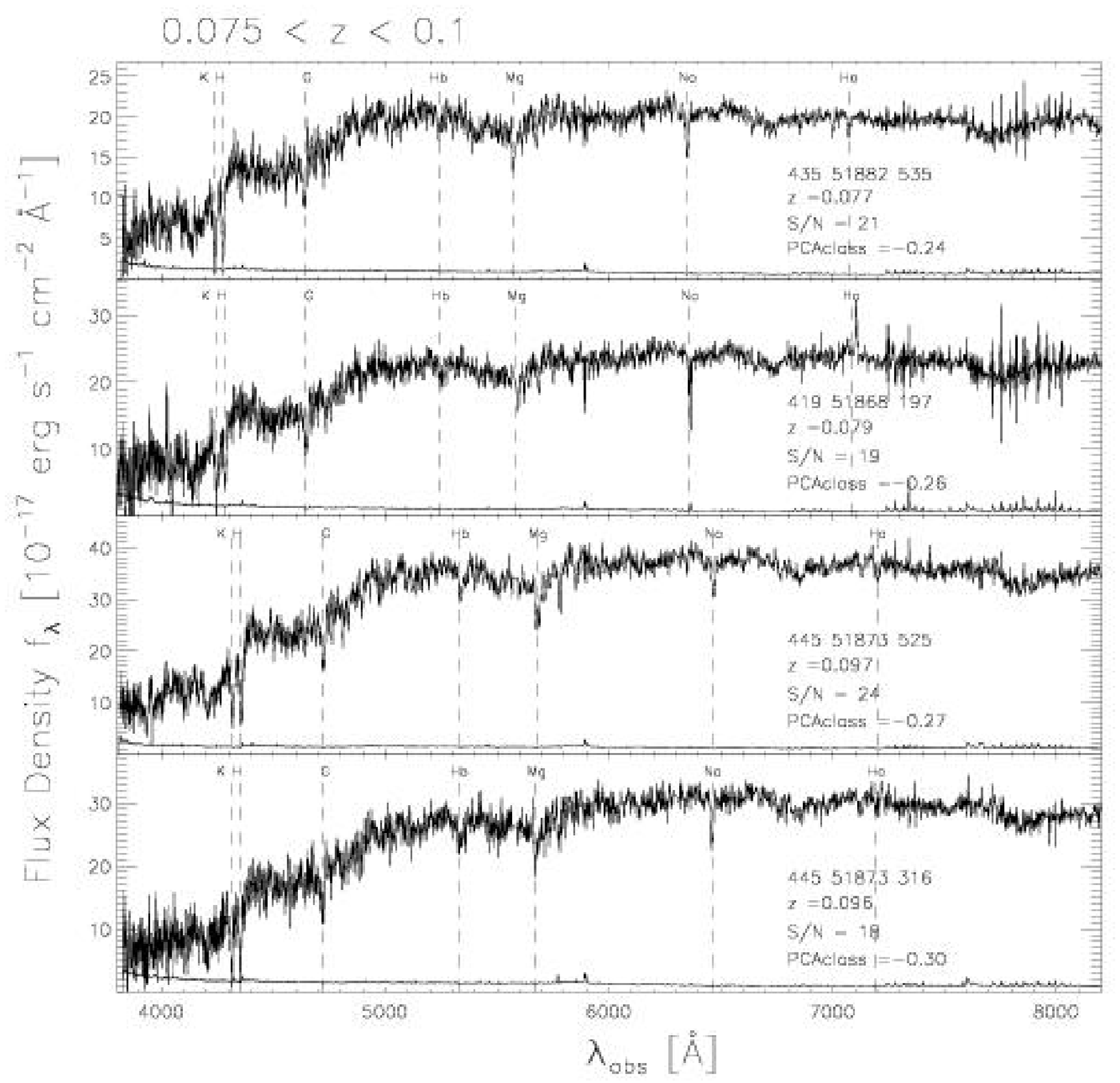}
\begin{center}
Fig. \ref{fig:spectra}. -- Continued. Redshift range $0.075 < z < 0.1$.
\end{center}
\end{figure}

\begin{figure}
\epsfxsize=\hsize\epsffile{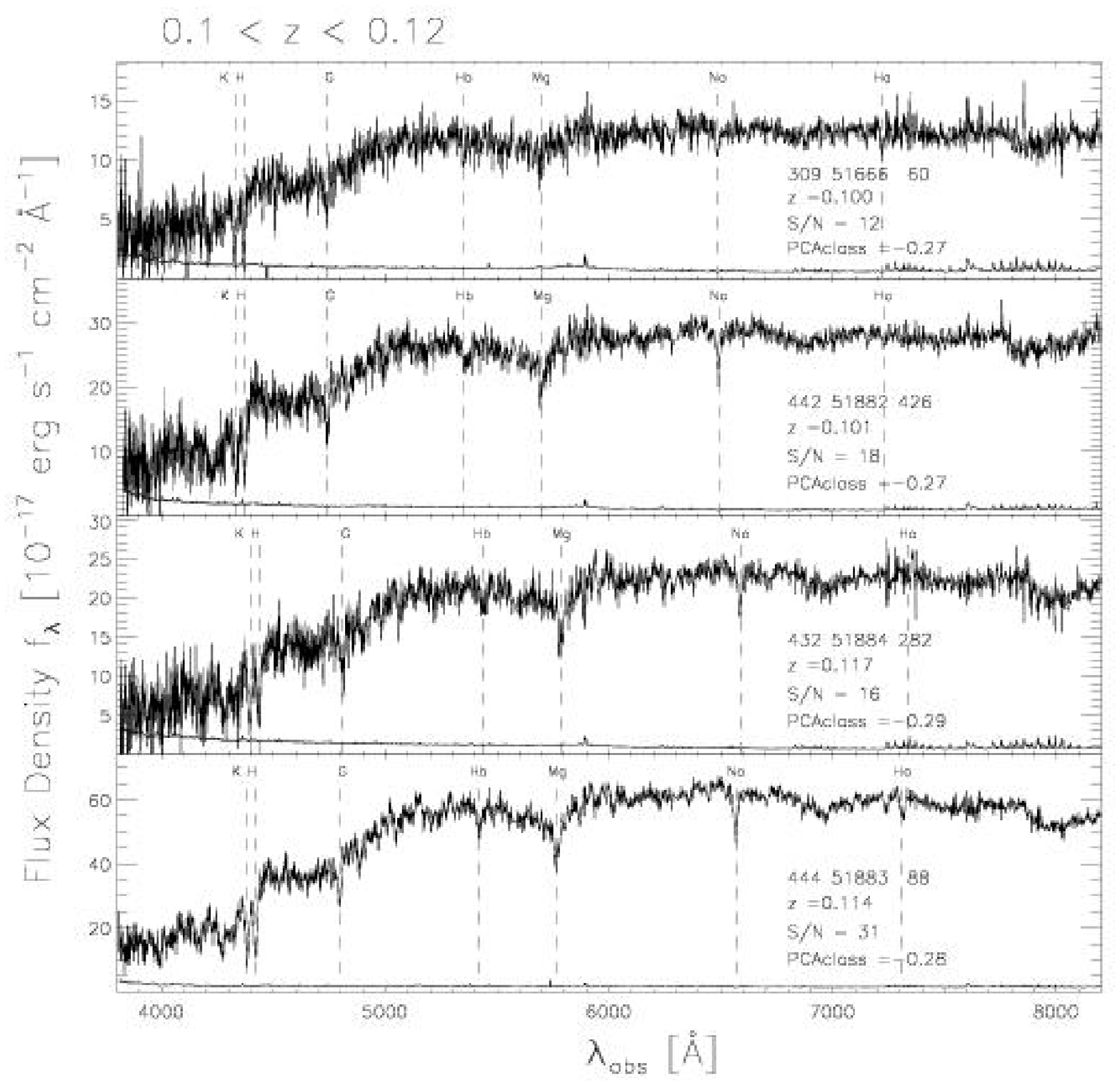}
\begin{center}
Fig. \ref{fig:spectra}. -- Continued. Redshift range $0.1 < z < 0.12$.
\end{center}
\end{figure}

\begin{figure}
\epsfxsize=\hsize\epsffile{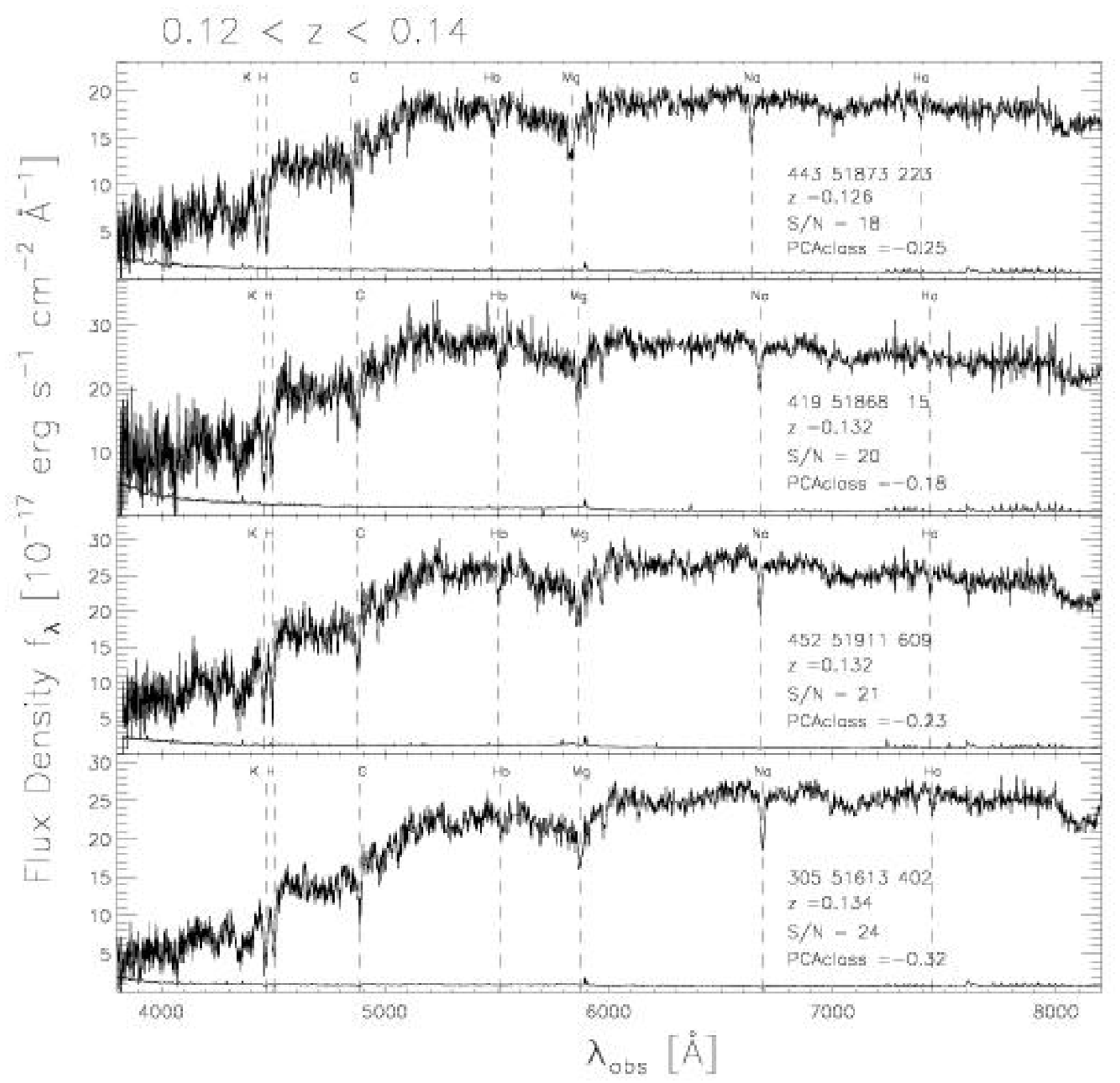}
\begin{center}
Fig. \ref{fig:spectra}. -- Continued. Redshift range $0.12 < z < 0.14$.
\end{center}
\end{figure}

\begin{figure}
\epsfxsize=\hsize\epsffile{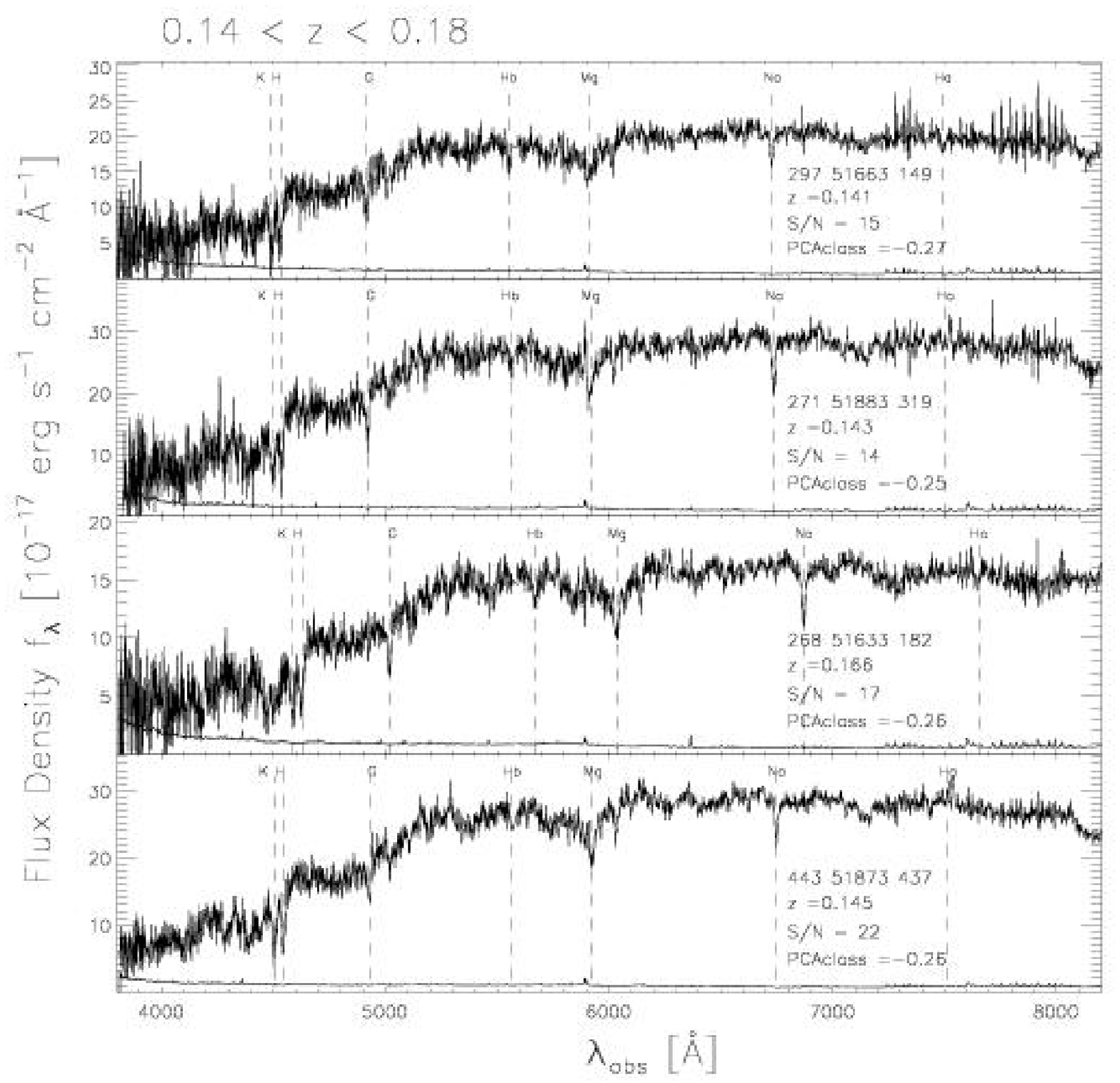}
\begin{center}
Fig. \ref{fig:spectra}. -- Continued. Redshift range $0.14 < z < 0.18$.
\end{center}
\end{figure}

\begin{figure}
\epsfxsize=\hsize\epsffile{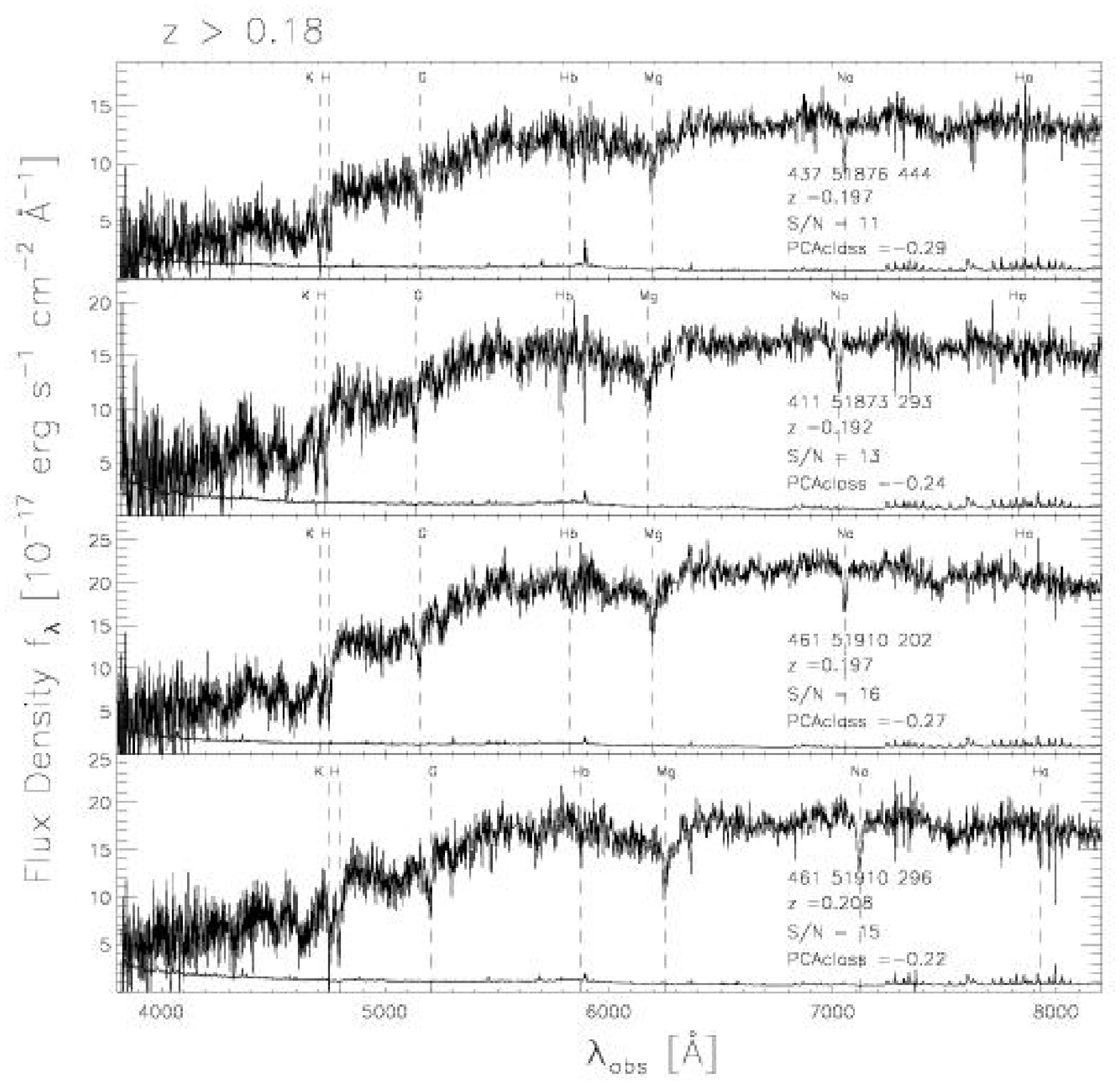}
\begin{center}
Fig. \ref{fig:spectra}. -- Continued. Redshift range $z > 0.18$.
\end{center}
\end{figure}

The typical signal-to-noise ratio of the spectra in our sample is 16, 
although $S/N$ does vary with wavelength.  
Therefore, individual spectra cannot be used to make reliable estimates 
of, e.g., the Lick indices.  However, by co-adding spectra of similar 
objects (i.e., galaxies with similar luminosities, sizes, velocity 
dispersions, and redshifts) it is possible to construct composite 
spectra which have considerably better $S/N$.  
This is the subject of Paper~IV.  

\subsection{An estimate of the local density}\label{environ}
In subsequent papers, we will be interested in how the properties of 
the galaxies in our sample depend on their local environment.  
To do so, we must come up 
with a working definition of environment.  The set of galaxies in the 
SDSS photometric database is much larger than those for which the survey 
actually measures redshifts.  Some of these galaxies may well lie close 
to galaxies in our sample, in which case they will contribute to the 
local density.  We would like to find some way of accounting for such 
objects when we estimate the local density.  

For a subset of the galaxies in our sample, the colors expected of a 
passively evolving early-type were used to select a region in 
$g^*$--$r^*$ versus $r^*$--$i^*$ color space at the redshift of the 
galaxy of interest.  
All galaxies within 0.1 magnitudes in color of this point were included 
if they were: 
a) within $1h^{-1}$ Mpc of the main galaxy, and 
b) brighter than $-20.25$ in $M_{i^*}$. 
(The box in color space is sufficiently large that the difference 
between this techinique, and using the observed colors themselves 
to define the selection box is not important.)  
These two cuts are made assuming every galaxy in the color-color 
range is at the redshift of the galaxy of interest.  
The end result of this is that each galaxy in the subsample is 
assigned a number of neighbors.  Note that, because of the selection 
on color, our estimate of the local density is actually an estimate 
of the number of neighbors which have the same colors as early-type 
galaxies.  In the papers which follow, we will often present results 
for different redshift bins.  When we do, it is important to bear in 
mind that this procedure for assigning neighbours is least secure in 
the lowest redshift bin (typically $z\le 0.08$).  

\begin{figure}
\centering
\epsfxsize=1.2\hsize\epsffile{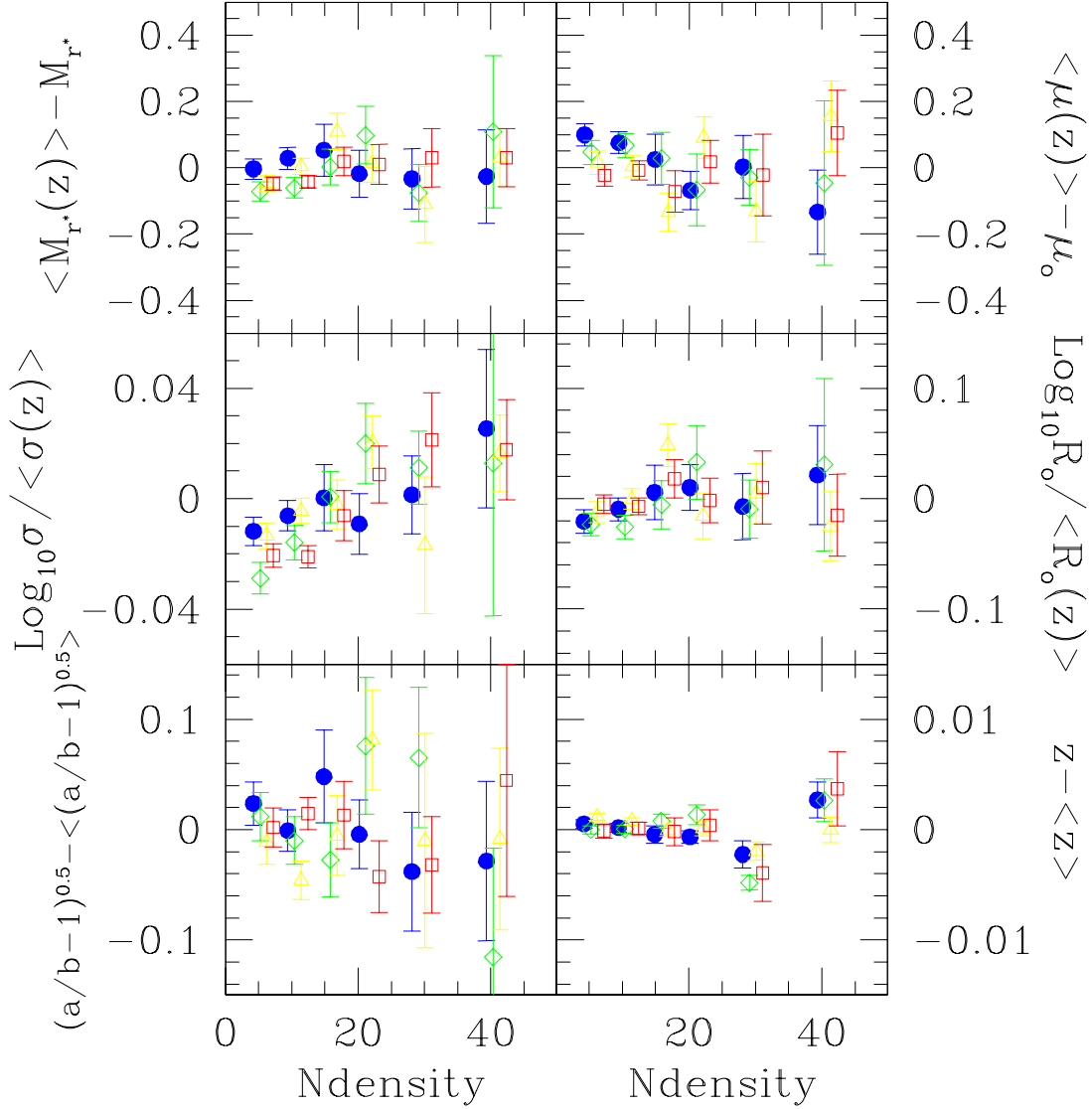}
\vspace{-1.cm}
\caption{Luminosities, surface brightnesses, velocity dispersions, 
sizes, axis ratios, and mean redshifts, as a function of nearby 
early-type neighbours.  The different symbols for each bin in density 
show averages over galaxies in different redshift bins:  circles, 
diamonds, triangles, and squares are for galaxies with redshifts in the 
range $0.075< z\le 0.1$, $0.1< z\le 0.12$, $0.12< z\le 0.14$, and 
$0.14< z\le 0.18$. Although the velocity dispersions appear to increase 
with increasing local density, the increase is small.}  
\label{lrvden}
\end{figure}

Figure~\ref{lrvden} shows how the luminosities, surface brightnesses, 
sizes, velocity dispersions and (a combination of the) axis ratios depend 
on environment.  The different symbols for each bin in density show 
averages over galaxies in different redshift bins:  circles, diamonds, 
triangles, and squares are for galaxies with redshifts in the range 
$0.075< z\le 0.1$, $0.1< z\le 0.12$, $0.12< z\le 0.14$, and 
$0.14< z\le 0.18$.  Error bars show the error in the mean value for 
each bin.  Symbols for the higher redshift bins have been offset slightly 
to the right.  

\begin{figure}
\centering
\epsfxsize=\hsize\epsffile{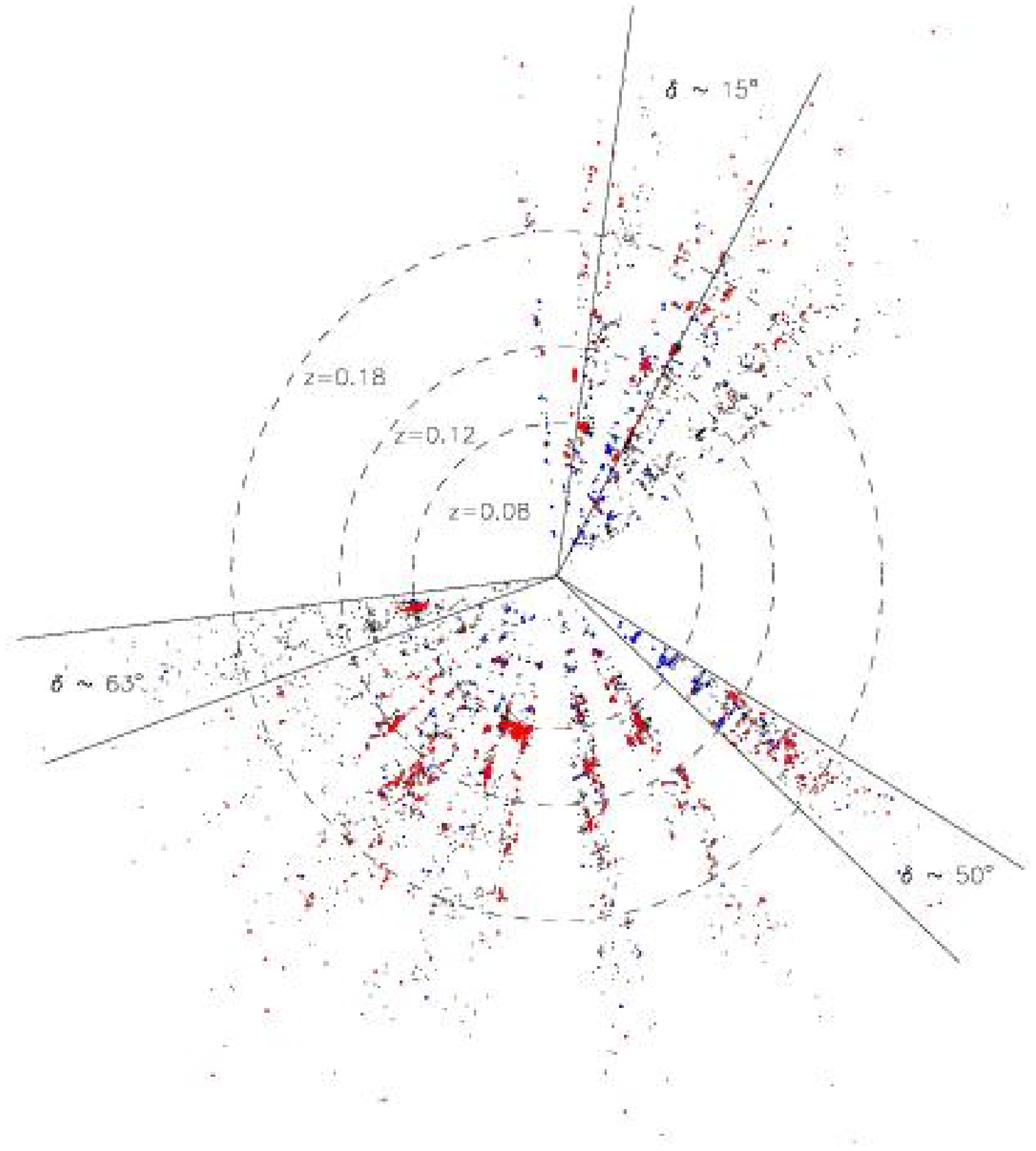}
\caption{Pie-diagram distribution of our sample.  Most of the sample 
is at low declination ($|\delta|\le 2^\circ$), but three wedges are at 
higher declinations (as indicated).  Right ascension increases 
clockwise, with the zero at twelve o'clock.  Galaxies with many ($\ge 15$) 
and a few ($\le 2$) near neighbours are shown with red and blue dots, 
whereas those in the intermediate regime, or for which the local number 
of neighbours was not determined, are shown with black dots.  }
\label{fig:pie}
\end{figure}

For any given set of symbols, the bottom right panel shows that the 
mean redshift in each bin in density is not very different from the 
mean redshift averaged over all bins.  This suggests that our procedure 
for estimating the local densities is not biased.  The other panels 
show corresponding plots for the other observed parameters.  
When the number of near neighbours is small, the luminosities, sizes 
and velocity dispersions all increase slightly as the local density 
increases, whereas the surface brightnesses decrease slightly.  
All these trends are very weak.  The bottom right panel shows 
$\sqrt{a/b - 1} - <\sqrt{a/b - 1}>$, where $b/a$ is the axis ratio.  
[Some authors (e.g. de Zeeuw \& Franx 1991) suggest that this is a good 
approximation of the ratio of rotational to random motions within the galaxy,
($v/\sigma$)$_{\rm iso}$, for oblate, rotationally flattened galaxies.  
Bright ellipticals have been shown to have a mean
($v/\sigma$)$/$($v/\sigma$)$_{\rm iso} \sim 0.4$.] 
There are no obvious trends with environment.  
It is difficult to say with certainty that the trends with environment 
in the top four panels of Figure~\ref{lrvden} are significant.  
A more efficient way of seeing if the properties of galaxies depend 
on environment is to show the residuals from the Fundamental Plane.  
This is done in Paper III.  

Figure~\ref{fig:pie} shows a redshift-space pie-diagram distribution 
of our sample.  Most of the sample is at low declination 
($|\delta|\le 2^\circ$); in addition, there are three wedges from 
three different disconnected regions on the sky.  Red and blue symbols 
denote galaxies which were classified as being in dense and underdense 
regions (as described in Section~\ref{environ} below), whereas black 
symbols show galaxies in groups of intermediate richness, or for which 
the local density was not determined.

\section{The early-type galaxy catalog}\label{catalog}
The observed parameters of each galaxy in our early-type galaxy sample 
are given in Table~\ref{tab:obs}. The different columns show the 
ra, dec, redshift $z$ and its error estimate, 
the signal-to-noise ratio $S/N$ of the spectrum, 
the apparent half light radius $r_{\rm dev}$ and error estimate, 
the fitted apparent magnitude $m_{dev}$ and error estimate, 
the model magnitude $m_m$ and error estimate, 
and the axis ratio $b/a$ and error estimate.  
The photometric parameters in this table are those measured in the 
$r^*$ band.  The complete version of this table, and similar tables 
of the catalog parameters in the $g^*$, $i^*$ and $z^*$ bands are 
available in the electronic edition of the Journal.  
Because we will be interested in the distribution of colors for a 
galaxy sample defined in, e.g., the $r^*$ band, the tables present the 
parameters of the same galaxies in all the bands.  To select the 
magnitude limited samples we use in subsequent papers refer to 
Table~\ref{tab:photerr}.  The errors $\delta m_{\rm dev-photo}$ 
and $\delta m_{\rm mod-photo}$ 
are the error estimates output by the SDSS photometric pipeline.  
As discussed in Appendix~\ref{errors}, they are {\it not} appropriate 
estimates of the errors in $m_{\rm dev}$ and $m_{\rm model}$.  
Our procedure for making more appropriate error estimates is described 
in Appendix~\ref{errors}. 

\renewcommand{\arraystretch}{.6}
\begin{deluxetable}{ccccccccccccc}
\label{tab:obs}
\rotate
\tablewidth{0pc}
\tablecaption{The SDSS early-type galaxies catalog: 
              Observed parameters}
\tablehead{
\colhead{RA (J2000)}  & \colhead{DEC (J2000)}  &
\colhead{$z$} & \colhead{$\delta z$} & \colhead{$S/N$} &
\colhead{$r_{\rm dev}$} & \colhead{$\delta r_{\rm dev}$} & 
\colhead{$m_{\rm dev}$} & \colhead{$\delta m_{\rm dev-photo}$} &
\colhead{$m_{\rm mod}$} & \colhead{$\delta m_{\rm mod-photo}$} & 
\colhead{$b/a$} & \colhead{$\delta b/a$} 
\\
\colhead{(deg)} & \colhead{(deg)} & \colhead{} & \colhead{} & \colhead{} &
\colhead{(arcsec)} & \colhead{(arcsec)} & \colhead{(mag)} & \colhead{(mag)}
  & \colhead{(mag)} & \colhead{(mag)} & \colhead{} & \colhead{}
}
\startdata
113.810738 &  36.307522 & 0.25100 & 0.00009 &12 & 3.45 & 0.15 &  17.234 &   0.009 &  17.234 &   0.009 & 0.82 & 0.03 \\
115.321098 &  37.511093 & 0.08513 & 0.00008 &19 & 2.55 & 0.04 &  16.400 &   0.009 &  16.400 &   0.009 & 0.99 & 0.01 \\
114.201271 &  36.971313 & 0.07832 & 0.00008 &15 & 2.16 & 0.05 &  17.202 &   0.009 &  17.202 &   0.009 & 0.71 & 0.02 \\
114.306282 &  37.111927 & 0.13049 & 0.00008 &20 & 2.48 & 0.05 &  16.828 &   0.008 &  16.828 &   0.008 & 0.68 & 0.01 \\
113.294510 &  36.370544 & 0.21756 & 0.00007 &13 & 2.18 & 0.08 &  17.451 &   0.008 &  17.451 &   0.008 & 0.80 & 0.03 \\
113.376007 &  36.478168 & 0.13365 & 0.00008 &14 & 2.45 & 0.08 &  17.304 &   0.008 &  17.304 &   0.008 & 0.63 & 0.02 \\
113.335358 &  36.648064 & 0.14016 & 0.00007 &21 & 2.01 & 0.04 &  16.871 &   0.006 &  16.871 &   0.006 & 0.68 & 0.01 \\
113.902443 &  37.023502 & 0.20225 & 0.00008 &12 & 4.17 & 0.17 &  16.998 &   0.011 &  16.998 &   0.011 & 0.96 & 0.03 \\
114.022346 &  37.099522 & 0.17845 & 0.00008 &14 & 2.07 & 0.08 &  17.453 &   0.011 &  17.453 &   0.011 & 0.76 & 0.03 \\
113.925407 &  37.113167 & 0.18365 & 0.00009 &15 & 2.71 & 0.08 &  17.053 &   0.010 &  17.053 &   0.010 & 0.80 & 0.02 \\
114.060677 &  37.211426 & 0.09610 & 0.00008 &13 & 2.11 & 0.07 &  17.279 &   0.010 &  17.279 &   0.010 & 0.85 & 0.03 \\
113.827904 &  37.380821 & 0.18387 & 0.00008 &12 & 4.82 & 0.18 &  16.898 &   0.009 &  16.898 &   0.009 & 0.97 & 0.03 \\
113.945900 &  37.708260 & 0.07476 & 0.00006 &19 & 2.98 & 0.05 &  16.882 &   0.009 &  16.882 &   0.009 & 0.34 & 0.01 
\enddata
\tablecomments{The complete version of this table is in the electronic
edition of the Journal.  The printed edition contains only a sample.  
This table lists the $r^*$ band photometric parameters.
Similar tables showing parameters in the $g^*$, $i^*$ and $z^*$ bands 
are also available in the electronic edition.}
\end{deluxetable}

\renewcommand{\arraystretch}{.6}
\begin{deluxetable}{cccccccccccccc}
\label{tab:phys}
\rotate
\tablewidth{0pc}
\tablecaption{The SDSS early-type galaxies catalog: Physical parameters}
\tablehead{
\colhead{RA}  & \colhead{DEC}  &
\colhead{$V$} & \colhead{$\delta V$} &
\colhead{$R$} & \colhead{$\delta R$} & 
\colhead{$\mu_{o}$} & \colhead{$\delta \mu_{o}$} & 
\colhead{$M_{\rm dev}$} & \colhead{$\delta M_{\rm dev}$} & 
\colhead{$M_{\rm mod}$} & \colhead{$\delta M_{\rm mod}$} & 
\colhead{Kcorr} & \colhead{Nden}
\\
\colhead{(deg)} & \colhead{(deg)} & \colhead{(dex)} & \colhead{(dex)} &
\colhead{(dex)} & \colhead{(dex)} & \colhead{(mag)} & 
\colhead{(mag)} & \colhead{(mag)} & \colhead{(mag)} & \colhead{(mag)} & 
\colhead{(mag)} & \colhead{(mag)} & \colhead{}
}
\startdata
113.810738 & 36.307522 & 2.501 & 0.034 &  0.979 & 0.020 & 19.873 & 0.060 & -23.587 &  0.033 & -23.587 &  0.033 &  0.309&   4 \\
115.321098 & 37.511093 & 2.133 & 0.029 &  0.600 & 0.007 & 19.918 & 0.020 & -21.648 &  0.018 & -21.648 &  0.018 &  0.104&   4 \\
114.201271 & 36.971313 & 2.055 & 0.045 &  0.414 & 0.012 & 19.993 & 0.034 & -20.643 &  0.021 & -20.643 &  0.021 &  0.093&   3 \\
114.306282 & 37.111927 & 2.363 & 0.022 &  0.664 & 0.010 & 19.620 & 0.030 & -22.268 &  0.019 & -22.268 &  0.019 &  0.159&   2 \\
113.294510 & 36.370544 & 2.307 & 0.040 &  0.871 & 0.017 & 19.944 & 0.052 & -22.978 &  0.029 & -22.978 &  0.029 &  0.268&   4 \\
113.376007 & 36.478168 & 2.319 & 0.026 &  0.619 & 0.016 & 19.810 & 0.046 & -21.851 &  0.026 & -21.851 &  0.026 &  0.163&   2 \\
113.335358 & 36.648064 & 2.266 & 0.024 &  0.541 & 0.010 & 18.866 & 0.028 & -22.407 &  0.017 & -22.407 &  0.017 &  0.173&   5 \\
113.902443 & 37.023502 & 2.352 & 0.045 &  1.110 & 0.019 & 20.886 & 0.059 & -23.229 &  0.034 & -23.229 &  0.034 &  0.244& 999 \\
114.022346 & 37.099522 & 2.388 & 0.038 &  0.749 & 0.018 & 19.872 & 0.054 & -22.440 &  0.031 & -22.440 &  0.031 &  0.212&   7 \\
113.925407 & 37.113167 & 2.433 & 0.029 &  0.856 & 0.014 & 19.934 & 0.041 & -22.914 &  0.025 & -22.914 &  0.025 &  0.217&   6 \\
114.060677 & 37.211426 & 2.173 & 0.034 &  0.542 & 0.015 & 20.219 & 0.044 & -21.058 &  0.027 & -21.058 &  0.027 &  0.114&   1 \\
113.827904 & 37.380821 & 2.347 & 0.040 &  1.094 & 0.018 & 20.965 & 0.055 & -23.071 &  0.030 & -23.072 &  0.030 &  0.217&   4 \\
113.945900 & 37.708260 & 2.090 & 0.028 &  0.394 & 0.009 & 19.687 & 0.024 & -20.851 &  0.019 & -20.851 &  0.019 &  0.088&   3
\enddata
\tablecomments{RA and DEC are in J2000.  
We have set $V=\log_{10}\sigma$ with $\sigma$ in km~s$^{-1}$, 
$R=\log_{10}R_o$ with $R_o$ in kpc~$h_{70}^{-1}$, Kcorr is the 
K-correction and Nden is the number of near neighbors which we use to 
estimate the local density.  Objects for which the number of neighbors 
was not determined have been assigned Nden=999, and we have set 
Nden=100 even if the number of neighbors is larger than 100.  
The complete version of this table is in the electronic edition of 
the Journal.  The printed edition contains only a sample.  
This table lists the $r^*$ band photometric parameters.
Similar tables showing parameters in the $g^*$, $i^*$ and $z^*$ bands 
are also available in the electronic edition.}
\end{deluxetable}

Table~\ref{tab:phys} shows restframe quantities we computed from 
the quantities in Table~\ref{tab:obs}.  
The different columns show the ra, dec, 
aperture corrected velocity dispersion $\log_{10}\sigma$ and its 
error estimate,  
the physical size $\log_{10}R_o$ and error estimate,  
the surface brightness $\mu_o$ and error estimate,  
the absolute magnitude $M_{\rm dev}$ and error estimate, 
the absolute model magnitude $M_{\rm mod}$ and error estimate, 
the K-correction, 
and the number of neighbours which we use as an estimate of the 
local density.  
As for Table~\ref{tab:obs}, a complete version of this table is 
in the electronic edition of the Journal, as are similar tables 
showing parameters in the $g^*$, $i^*$ and $z^*$ bands.  
(The errors in $M_{\rm dev}$ and $M_{\rm model}$ listed in this table 
were derived from the estimates $\delta m_{\rm photo}$ of the previous 
table.  As discussed in Appendix~\ref{errors}, they are also the 
appropriate estimates for the error on $m_{\rm dev}$ and $m_{\rm mod}$.)

\begin{figure}
\centering
\epsfxsize=\hsize\epsffile{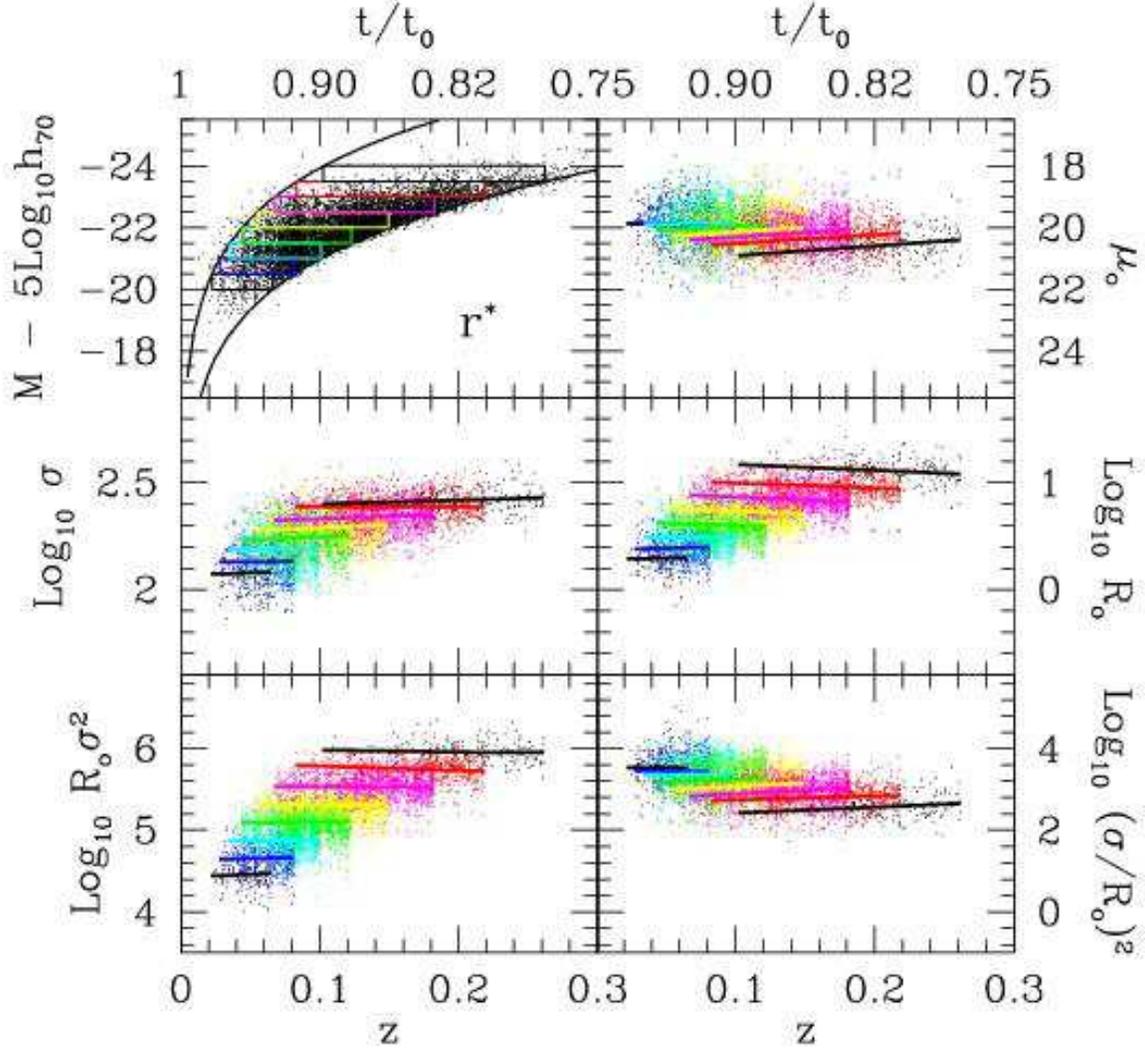}
\vspace{1cm}
\caption[]{Surface brightnesses, velocity dispersions, sizes, masses 
and densities of galaxies as a function of redshift, for a few bins 
in $r^*$ luminosity.  Top left panel shows volume limited catalogs 
which do not overlap in $r^*$ luminosity, dots in the other panels 
show the galaxies in the volume-limited subsamples defined by the 
top-left panel, and solid lines show the mean trend with redshift in 
each subsample.  Results in $g^*$, $i^*$ and $z^*$ are similar. }
\label{fig:XzR}
\end{figure}

Some properties of our $r^*$ sample as a function of redshift $z$ 
are shown in Figure~\ref{fig:XzR}.  The panels show the K-corrected 
absolute magnitude $M$, the K-corrected effective surface brightness 
$\mu_o$, also corrected for cosmological surface brightness dimming, 
the effective circular radius $R_o$ in $h_{70}^{-1}$~kpc, corrected to 
a standard restframe wavelength, 
the aperture corrected velocity dispersion $\sigma$ in km~s$^{-1}$, 
and two quantities which are related to an effective mass and density, 
all plotted as a function of redshift.  

The bold lines in the top left panel show the effect of the apparent 
magnitude cuts.  There is, in addition, a cut at small velocity 
dispersion ($\sim 70$~km~s$^{-1}$) which, for our purposes here, is 
mostly irrelevant.  The apparent magnitude cuts imply complex 
$z$-dependent cuts on the other parameters we observe.  In the papers 
which follow, we attempt to account correctly for the selection effects 
that the magnitude cuts imply.  

For small intervals in luminosity, our sample is complete over a 
reasonably large range in redshifts.  To illustrate, the thin boxes 
in the top-left panel show bins in absolute magnitude of width 0.5 mags 
over which the sample is complete.  The solid lines in the other panels 
show how the median surface-brightnesses, sizes, velocity dispersions, 
masses and densities of galaxies in each $r^*$ luminosity bin change as a 
function of redshift.  Although all these quantities depend on luminosity, 
the figure shows that, at fixed luminosity, there is some evidence for 
evolution:  at fixed luminosity the average surface brightness is 
brightening.  The size at fixed luminosity decreases at a rate which is 
about five times smaller than the rate of change of $\mu_o$.  This 
suggests that it is the luminosities which are changing and not the sizes.  
(To see why, suppose that the average size at fixed absolute magnitude is 
 $\langle \log_{10}[R_o/R_*(z)]\rangle=s[M-M_*(z)]$, where $R_*$ and $M_*$ 
are characteristic values, and $s$ is the slope of this mean relation.  
If the characteristic luminosity increases with $z$, but the 
characteristic size remains constant, $R_*(z)=R_*(0)$, and the 
slope of the relation also does not change, then the mean size at 
fixed $M$ decreases with $z$.  The surface brightness is 
$\mu_o\propto M + 5\log_{10}R_o$, so the mean $\mu_o$ at fixed $M$ 
changes five times faster than the mean $R_o$, at fixed luminosity.  
This argument is also relevant to Figures~\ref{fig:seeing} 
and~\ref{fig:re10sky}, since those plots use $r_{\rm dev}$ which differs 
from $R_o$ by two small correction factors [c.f. Section~\ref{LandR}].)
In Paper~II we argue that this trend is qualitatively consistent with 
that expected of a passively evolving population.  

\section{Summary}\label{discuss}
This is the first of four papers which study the properties of 
early-type galaxies at relatively low redshifts ($z\le 0.3$).  
This paper describes how we selected the sample from the SDSS database
using objective criteria and present the measured photometric and 
spectroscopic parameters for $\sim 9000$ early-type galaxies 
(Table~\ref{tab:obs}).  
The database contains the redshift of these galaxies as well as reliable 
measurements of a number of photometric 
properties (luminosities, sizes, surface brightnesses, colors)  
in the $g^*$, $r^*$, $i^*$ and $z^*$ bands.  
(Data in the $u^*$ band is also available, but, because it is noisier, 
we did not use it here.) 

The galaxies in our sample span the redshift range $0.01\le z\le 0.3$.  
Therefore, differences between 
observed and restframe wavelengths are not negligible.  
We discussed a number of methods for estimating the appropriate 
K-correction which must be applied to obtain luminosities which 
sample the same restframe wavelength range (Appendix~\ref{kcorrs}).  
Since the half-light radii also depend on wavelength 
(Figure~\ref{angre}), we also corrected these to sizes at fixed 
restframe wavelengths.  
Finally, we estimated the internal velocity dispersions of these 
galaxies (Appendix~\ref{vmethods}), which, following common practice, 
we corrected to an aperture of $r_e/8$ (c.f. equation~\ref{appcorr}).  
The resulting catalog listing the restframe photometric and spectroscopic 
parameters of $\sim 9000$ early-type galaxies (Table~\ref{tab:phys}) 
is useful for measuring correlations between these various observables, 
such as the Faber-Jackson $L-\sigma$ relation, the Kormendy $R_o-I_o$ 
relation, the $L-R_o$ relation, the Fundamental Plane $R_o-I_o-\sigma$ 
relation, and the color--magnitude relation.  

The sample is essentially magnitude limited (Table~\ref{tab:photerr}, 
Figure~\ref{fig:appmz}), and the galaxies in it populate a range of 
environments (Figure~\ref{fig:pie}).  
Therefore, the sample can be used for studying how the correlations 
discussed above depend on environment and redshift.  
For instance, Figure~\ref{fig:XzR} shows evidence for weak evolution 
in the luminosities, and Figure~\ref{lrvden} shows evidence for 
some weak environmental dependences.  These will be quantified 
in Papers~II and~III.  

Neither the K-corrections nor the aperture corrections we apply to 
the velocity dispersions are ideal.  By the time the Sloan Digital 
Sky Survey is complete, the uncertainties in the K-corrections, which 
prevent us at the present time from making precise quantitative 
statements about the evolution of the luminosities and colors, will be 
better understood.  In addition, the size of the sample will have 
increased by more than an order of magnitude.  This larger sample will 
allow us to use the data directly to estimate the correct aperture 
correction which should be applied to the velocity dispersions.  
This is discussed more fully in Appendix~\ref{vr}.

\vspace{1cm}

We would like to thank S. Charlot for making his stellar population 
synthesis predictions for the SDSS filters available to the 
collaboration and N. Benitez for making his package available.  
We thank M. Strauss for helpful discussions.  

Funding for the creation and distribution of the SDSS Archive has been 
provided by the Alfred P. Sloan Foundation, the Participating Institutions, 
the National Aeronautics and Space Administration, the National Science 
Foundation, the U.S. Department of Energy, the Japanese Monbukagakusho, 
and the Max Planck Society. The SDSS Web site is http://www.sdss.org/.

The SDSS is managed by the Astrophysical Research Consortium (ARC) for
the Participating Institutions. The Participating Institutions are The
University of Chicago, Fermilab, the Institute for Advanced Study, the
Japan Participation Group, The Johns Hopkins University, Los Alamos
National Laboratory, the Max-Planck-Institute for Astronomy (MPIA),
the Max-Planck-Institute for Astrophysics (MPA), New Mexico State
University, University of Pittsburgh, Princeton University, the United
States Naval Observatory, and the University of Washington.

{}

\appendix 
\section{K-corrections}\label{kcorrs}
When converting the observed apparent magnitude to the rest-frame 
absolute magnitude of an object, we must account for the fact that 
the SDSS filters measure the light from a fixed spectral range in 
the observers rest-frame; therefore, they measure different parts of 
the rest-frame spectrum of galaxies at different redshifts.  
Correcting for this is known as the K-correction.  
One way to make this correction is to assume that all the galaxies at a 
given redshift are similar, and to use an empirically determined template 
spectrum, measured from a few accurately measured spectra, to estimate 
the K-correction.  Using a mean color to estimate the K-correction 
is not ideal.  When the survey is closer to completion it should become 
possible to make this correction on an object-by-object basis.  

Empirically determined template spectra for early-type galaxies at 
low redshifts exist (e.g. Coleman, Wu \& Weedeman 1980; 
Fukugita, Shimasaku \& Ichikawa 1995).  (We used N. Benitez's Bayesian 
Photometric Redshift package (Benitez 2000) to derive K-corrections 
for the Coleman, Wu \& Weedman early-type galaxy template in the SDSS 
passbands.)  If we were certain that evolution effects were not important, 
then these empirically determined K-corrections would allow us to work 
out the K-corrections we should apply to the high redshift population.  
However, if the stars in early-type galaxies formed at approximately 
the same time, and if the mass in the galaxies has remained constant, 
so the evolution is entirely due to the passive aging of the stellar 
population, then, the mass to light ratio of early-type galaxies is 
expected to vary approximately as $M/L\propto (t-t_{\rm form})^{-0.6}$ 
(e.g., Tinsley \& Gunn 1976), with the precise scaling being different in 
different bands.  Because our sample spans a relatively large range in 
redshift, we may be sensitive to the effects of this passive evolution.  
In addition, because the sample is large, we may be able to measure, and 
hence be sensitive to, even a relatively small amount of evolution.  
For this reason, it would be nice to have a prescription for making 
K-corrections which accounts for evolution.  Absent empirically determined 
templates for this evolution, we must use stellar population systhesis 
models to estimate this evolution, and so determine K-corrections for 
different bands.  

\begin{figure}[t]
\centering
\epsfxsize=\hsize\epsffile{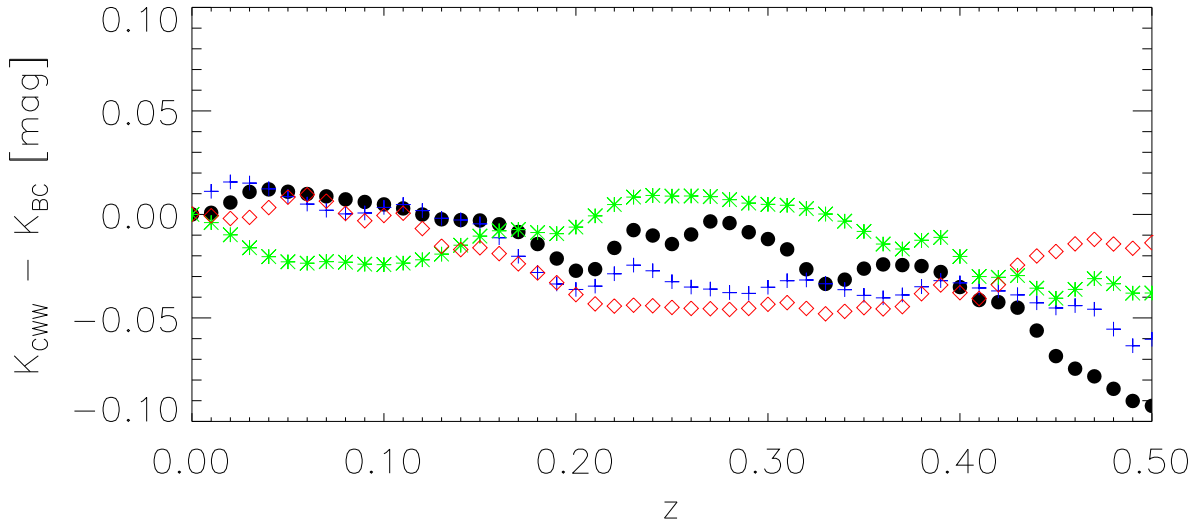}
\vspace{-8cm}
\caption[]{Difference between K-corrections based on two models 
(Coleman, Wu \& Weedman 1980 and Bruzual \& Charlot 2003) of the SDSS 
colors of early-type galaxies.  Filled circles, crosses, stars and 
diamonds are for the $g^*$, $r^*$, $i^*$ and $z^*$ bands.}
\label{kCWW-kBC}
\end{figure}

As a first example, we chose a Bruzual \& Charlot (2003) model for a 
$10^{11}M_\odot$ object which formed its stars with the universal 
IMF given by Kroupa (2001) in a single solar metallicity and abundance 
ratio burst 9~Gyr ago.  We then recorded how its colors, as observed 
through the SDSS filters, changed as it was moved through redshift without 
altering its age.  This provides what we will call the no-evolution 
K-correction.  Figure~\ref{kCWW-kBC} shows a comparison of this with 
the empirical Coleman, Wu \& Weedman (1980) nonevolving K-corrections.  
The two estimates are in good agreement in $g^*$ and $i^*$ out
to $z\sim 0.3$. They differ substantially at higher redshifts, 
but this is not a concern because none of the galaxies in our sample are 
so distant. In $r^*$ and $z^*$ the two estimates agree only at $z\le 0.15$.
Therefore, quantitative 
estimates of evolution in luminosity and/or color will depend on which 
K-correction we use.

\begin{figure}
\centering
\epsfxsize=\hsize\epsffile{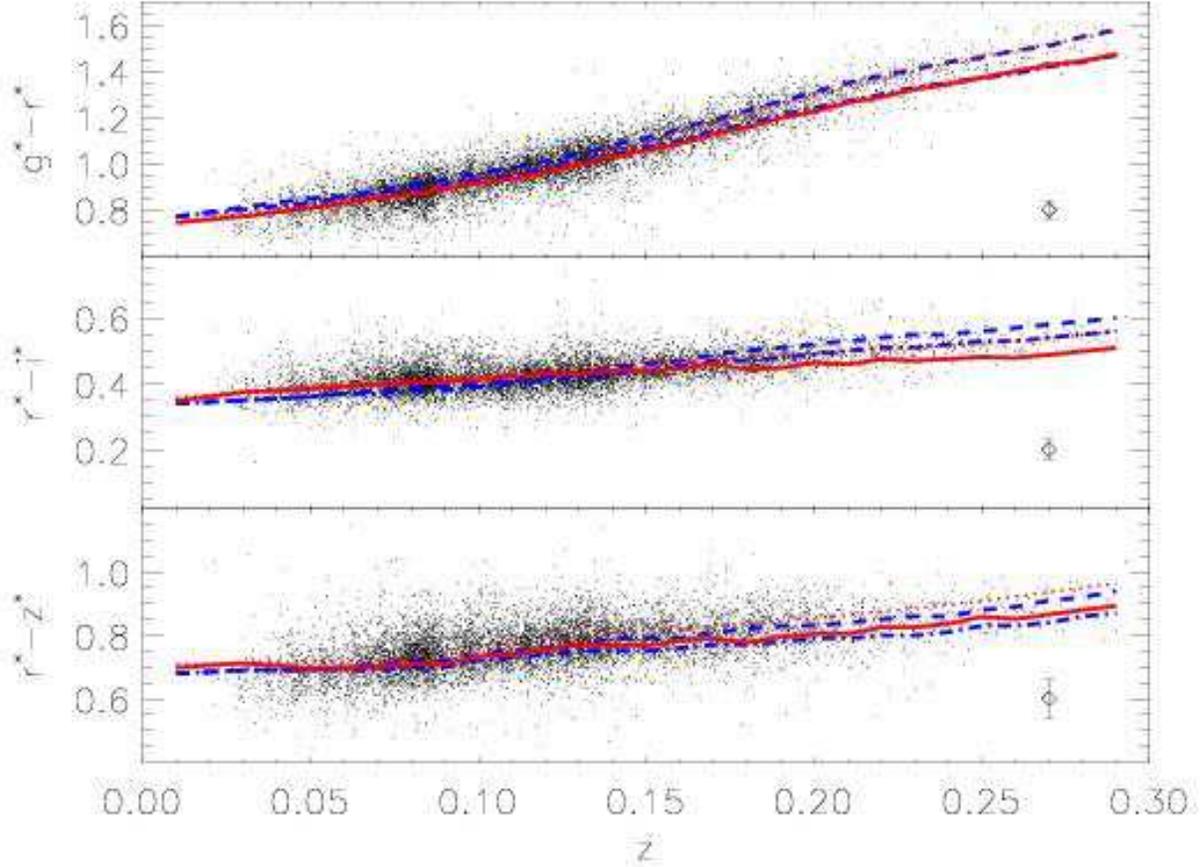}
\vspace{0cm}
\caption[]{Apparent colors of the galaxies in our sample.  In each panel, 
dotted and solid lines show the non-evolving and evolving Coleman et al. 
templates, whereas dashed and dot-dashed show Bruzual \& Charlot models.  
The upper set of curves in each panel show what one expects to see 
if the intrinsic colors of galaxies at higher redshifts are the same as 
they are nearby, whereas the lower sets of curves show the predictions 
if the higher redshift population is slightly younger, and so bluer.  
The magnitude limit of the sample makes it appear as though the no-evolution 
curves describe our data well.  In the main text, we use the lower 
solid curve to make K-corrections to the observed magnitudes. }
\label{fig:kcorr}
\end{figure}

Figure~\ref{fig:kcorr} compares both sets of nonevolving templates 
with the observed colors of the galaxies in our sample.  The upper 
set of curves in each panel show the colors associated with the 
nonevolving Coleman, Wu \& Weedman (1980) template (dotted) and the 
Bruzual \& Charlot (2003) no-evolution model (dashed).  
The figure shows that both predictions for $g^*-r^*$ are similar, but 
that they are different for $r^*-i^*$ and $r^*-z^*$, with the differences 
increasing with redshift.  
[In both cases, we have shifted the predicted $g^*-r^*$ blueward by 
0.08~mag at all $z$.  Such an offset appears to be required for the SDSS 
photometric calibrations in Stoughton et al. (2002) which we use here 
(also see Eisenstein et al. 2001 and Strauss et al. 2002), although the 
reason for it is not understood.]

\begin{figure}
\centering
\epsfxsize=\hsize\epsffile{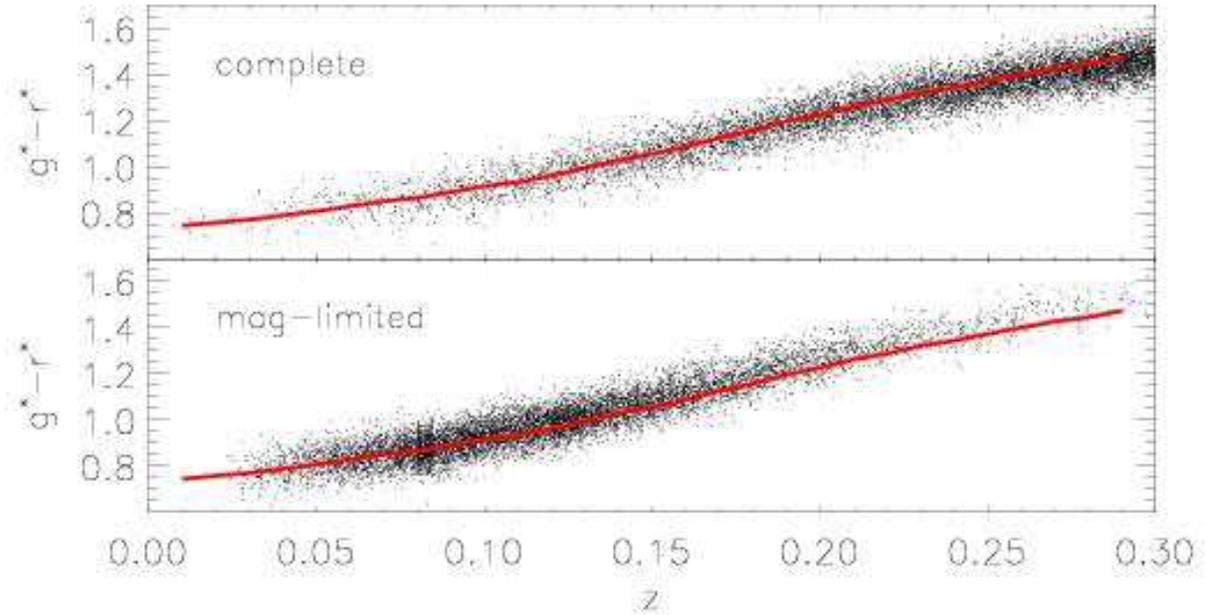}
\vspace{0cm}
\caption[]{Apparent $g^*-r^*$ colors of simulated galaxies in mock catalogs 
of a passively evolving population; the galaxies at higher redshift are 
younger and, in their rest-frame, bluer.  Top panel shows the expected 
distribution of observed colors if there were no magnitude limit; 
bottom panel shows the effect of imposing the same magnitude limit as in 
our SDSS sample.  Solid curves (same in both panels) show the trend of 
observed color with redshift of this evolving population.  
Although the smooth curve describes the complete catalog well, it is 
substantially bluer than the subset of objects which are included in the 
magnitude limited sample.  The difference between the curves and our 
magnitude limited mock catalogs is similar to that between the curves and 
the data (see previous Figure), suggesting that the colors of the 
galaxies in our data are evolving similarly to how we assumed in our 
simulations.}
\label{kcorrsim}
\end{figure}

We argue in Papers~II, III and~IV that the luminosities, colors and 
chemical abundance ratios of these objects show evidence for passive 
evolution: the higher redshift population appears to be slightly 
younger.  Therefore, our next step is to include the effects of 
evolution.  Because the predicted observed colors at redshift zero are 
in good agreement with our data, we took the same 
Bruzual \& Charlot (2003) model, 
but this time we recorded how its rest-frame colors evolve with redshift, 
and then computed what these evolved (i.e., $z$-dependent) colors look 
like when observed in the SDSS filters.  These evolving colors are shown 
as the dot-dashed lines in Figure~\ref{fig:kcorr}.  
In an attempt to include evolution in the CWW templates, we set 
$K_{\rm CWW}^{\rm evol}(z) = K_{\rm CWW}^{\rm noev}(z) + 
                             K_{BC}^{\rm evol}(z) - K_{BC}^{\rm noev}(z)$.  
The lower solid lines in each panel of Figure~\ref{fig:kcorr} show the 
observed $g^*-r^*$, $r^*-i^*$ and $r^*-z^*$ colors associated with these 
evolving models (and again, the predicted $g^*-r^*$ curves have been 
shifted blueward (downward) by 0.08~mag at all $z$).  
Comparing the evolving Bruzual--Charlot and the Coleman et al. colors 
with the upper set of no-evolution curves shows the evolution towards the 
blue at high redshift.  Although the data appear to be very well fit by 
the no-evolution curves, this agreement is slightly misleading.  
More luminous galaxies tend to be redder.  As a consequence, a magnitude 
limited catalog contains only the redder objects of the higher redshift 
population.  A curve which describes the colors of the population as a 
whole will therefore appear to be biased blue.  

Figure~\ref{kcorrsim} shows this explicitly.  The two panels were  
constructed by making mock catalogs of a passively evolving population 
(i.e., the higher redshift population is brighter and bluer) in which our 
estimates of the correlation between colors, luminosities, velocity 
dispersions and color and luminosity evolution were included
(see Appendix~A in Paper~II).  The top panel 
shows the distribution of observed colors if there were no magnitude 
limit, and the bottom panel shows the observed colors of a magnitude 
limited sample.  The solid curve, the same in both panels, is the 
predicted trend of color with redshift which we use to make our 
K-corrections; i.e., $K_{\rm CWW}^{\rm evol}(z)$.  
Notice that although it describes the complete simulations well, 
it is bluer than the higher redshift galaxies in the magnitude limited 
sample.  Comparison with the previous figure shows that the difference 
here is similar to that seen in the real data, suggesting that our 
K-corrections and evolution estimates of the mean of the population 
are self-consistent.  

Of course, if we do not observe the mean of the high redshift population, 
but only the redder fraction, then we must decide whether it is realistic 
to use a K-correction which has been constructed to fit the truely typical 
galaxy at each redshift.  For example, if color is an indicator of age 
and/or metallicity, then the results above suggest that our sample contains 
the oldest and/or most metal rich part of the high redshift population.  
If the objects which satisfy our apparent magnitude limit are in fact, 
older than the typical high redshift galaxy, then it may be that those 
objects are similar in age to the average object at lower redshifts in our 
sample.  If so, then we are better-off using a nonevolving K-correction 
even though the higher redshift sample as a whole is younger.  None of 
the results presented in the main text change drastically if we use 
non-evolving rather than evolving K-corrections.  


To decide which K-corrections to use, we computed the color-magnitude 
and color-$\sigma$ relations discussed in Paper~IV using both the 
Coleman et al. (1980) template and the Bruzual \& Charlot (2003) models.  
The slopes of the mean relations, and the scatter around the mean, 
remained approximately the same for both K-corrections, so we have chosen 
to not show them here.  This suggests that our ignorance of the true 
K-correction does not strongly compromise our conclusions about how color 
correlates with magnitude and velocity dispersion.  Conclusions about 
evolution, however, do depend on the K-correction.  

The color--$\sigma$ relations constructed using either 
K$^{\rm evol}_{\rm BC}$ or K$^{\rm evol}_{\rm CWW}$ show evidence 
for evolution.  However, K$^{\rm evol}_{\rm BC}$ yields evolution in 
$g^*-r^*$ of 0.04~mags, and in $r^*-i^*$ of 0.07~mags, whereas 
K$^{\rm evol}_{\rm CWW}$ has changes of 0.07 and 0.03, respectively.  
Thus, K$^{\rm evol}_{\rm BC}$ suggests that the evolution in $r^*-i^*$ 
is larger than in $g^*-r^*$.  
This is not the expected trend; the $g^*-r^*$ and $r^*-i^*$ wavelength 
baselines are about the same, so one expects more of the evolution to come 
in at the bluer color.  Using K$^{\rm evol}_{\rm CWW}$ instead suggests 
that most of the evolution is in $g^*-r^*$, which is more in line with 
expectations.  

We also tried K-corrections from Fukugita et al. (1995).  
At low redshifts, the predicted early-type colors are redder than those 
in our sample, the predicted S0 colors are bluer, and the differences 
depend on redshift.  A straight average of the two is an improvement, 
although the resulting low redshift $g^*-r^*$ is red by 0.05~mags.  
If we shift by this amount to improve the agreement at low redshifts, 
then the observed $g^*-r^*$ colors at $z=0.25$ are redder than the 
predicted no evolution curves by about 0.2 mags.  This is larger than 
the offset we expect for the selection effect introduced by the magnitude 
limit, so we decided against presenting further results from these 
K-corrections.  

\section{Velocity dispersion: methods and measurements}\label{vmethods}
This Appendix describes how we estimated the line-of-sight velocity 
dispersions $\sigma$ for the sample of galaxies selected for this paper.  

\begin{figure}[t]
\centering
\epsfxsize=\hsize\epsffile{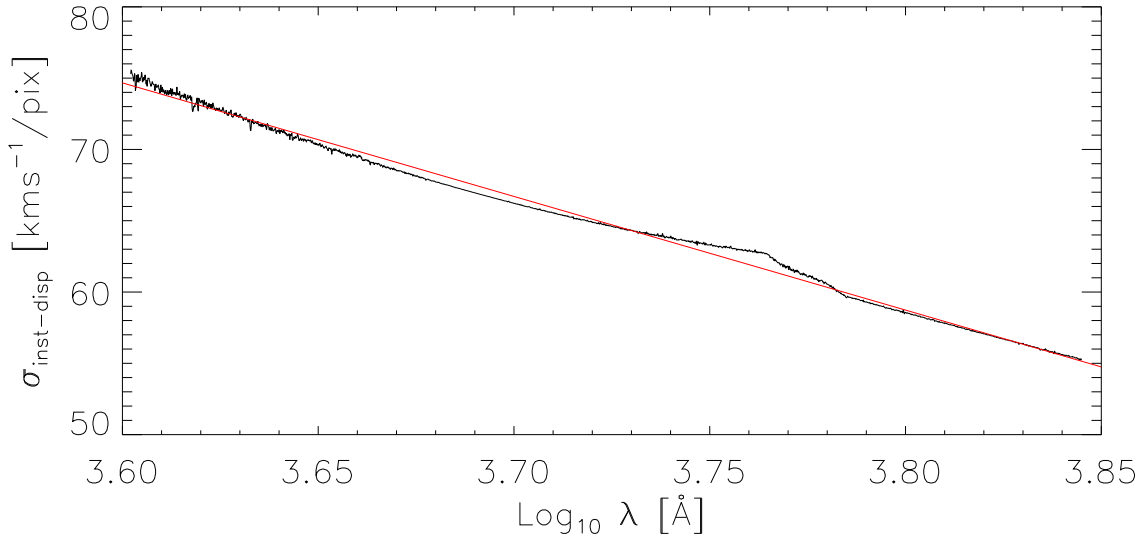}
\vspace{-8cm}
\caption[]{Variation of instrumental dispersion over the range in 
wavelengths used to measure velocity dispersions later in this paper.  
Solid line shows a linear fit. }
\label{fig:resol}
\end{figure}

Estimates of $\sigma$ are limited by the instrumental dispersion 
and resolution.  Recall that the instrumental dispersion of the SDSS 
spectrograph is 69 km~s$^{-1}$ per pixel, and the resolution is 
$\sim 90$ km~s$^{-1}$.  In addition, the instrumental dispersion may 
vary from pixel to pixel, and this can affect measurements of $\sigma$.
These variations are estimated for each fiber by using arc lamp 
spectra (upto 16 lines in the range 3800-6170~\AA\ and 39 lines 
between 5780-9230~\AA).  An example of the variation in instrumental 
dispersion for a single fiber is shown in Figure~\ref{fig:resol}.  The 
figure shows that a simple linear fit provides a good description of 
this variation.  This is true for almost all fibers, and allows us to 
remove the bias such variations may introduce when estimating galaxy 
velocity dispersions.  

A number of methods for making accurate and objective velocity dispersion 
measurements as have been developed (Sargent et al. 1977; 
Tonry \& Davis 1979; Franx, Illingworth \& Heckman 1989; Bender 1990; 
Rix \& White 1992).  These methods are all based on a comparison between 
the spectrum of the galaxy whose velocity dispersion is to be determined, 
and a fiducial spectral template. This can either be the spectrum of an 
appropriate star, with spectral lines unresolved at the spectra resolution 
being used, or a combination of different stellar types, or a high $S/N$ 
spectrum of a galaxy with known velocity dispersion. In this work, we use 
SDSS spectra of 32 K and G giant stars in M67 as stellar templates.  

Since different methods can give significantly different results, 
thereby introducing systematic biases especially for low $S/N$ spectra, 
we decided to use three different techniques for measuring the velocity 
dispersion.  These are 
1) the {\it cross-correlation} method (Tonry \& Davis 1979); 
2) the {\it Fourier-fitting} method (Tonry \& Davis 1979; 
Franx, Illingworth \& Heckman 1989; van der Marel \& Franx 1993); and 
3) a modified version of the {\it direct-fitting} method 
(Burbidge, Burbidge \& Fish 1961; Rix \& White 1992). Because a galaxy's 
spectrum is that of a mix of stars convolved with the
distribution of velocities within the galaxy, Fourier space is the natural
choice to estimate the velocity dispersions---the first two methods make 
use of this.  However, there are several advantages to treating the problem 
entirely in pixel space.  In particular, the effects of noise are much 
more easily incorporated in the pixel-space based {\it direct-fitting} 
method.  Because the $S/N$ of the SDSS spectra are relatively low, 
we assume that the observed absorption line profiles in early-type galaxies 
are Gaussian (see Rix \& White 1992 and Bender, Saglia \& Gerhard 1994 
for a discussion of how to analyze the line profiles of high $S/N$ spectra 
in the case of asymmetric profiles).  

It is well known that all three methods have their own particular biases, 
so that numerical simulations must be used to calibrate these biases.  
In our simulations, we chose a template stellar spectrum measured at high 
$S/N$, broadened it using a Gaussian with rms $\sigma_{input}$, added 
Gaussian noise, and compared the input velocity dispersion with the 
measured output value.  
The first broadening allows us to test how well the methods work as a 
function of velocity dispersion, and the addition of noise allows us to 
test how well the methods work as a function of $S/N$.

\begin{figure}
\centering
\epsfxsize=\hsize\epsffile{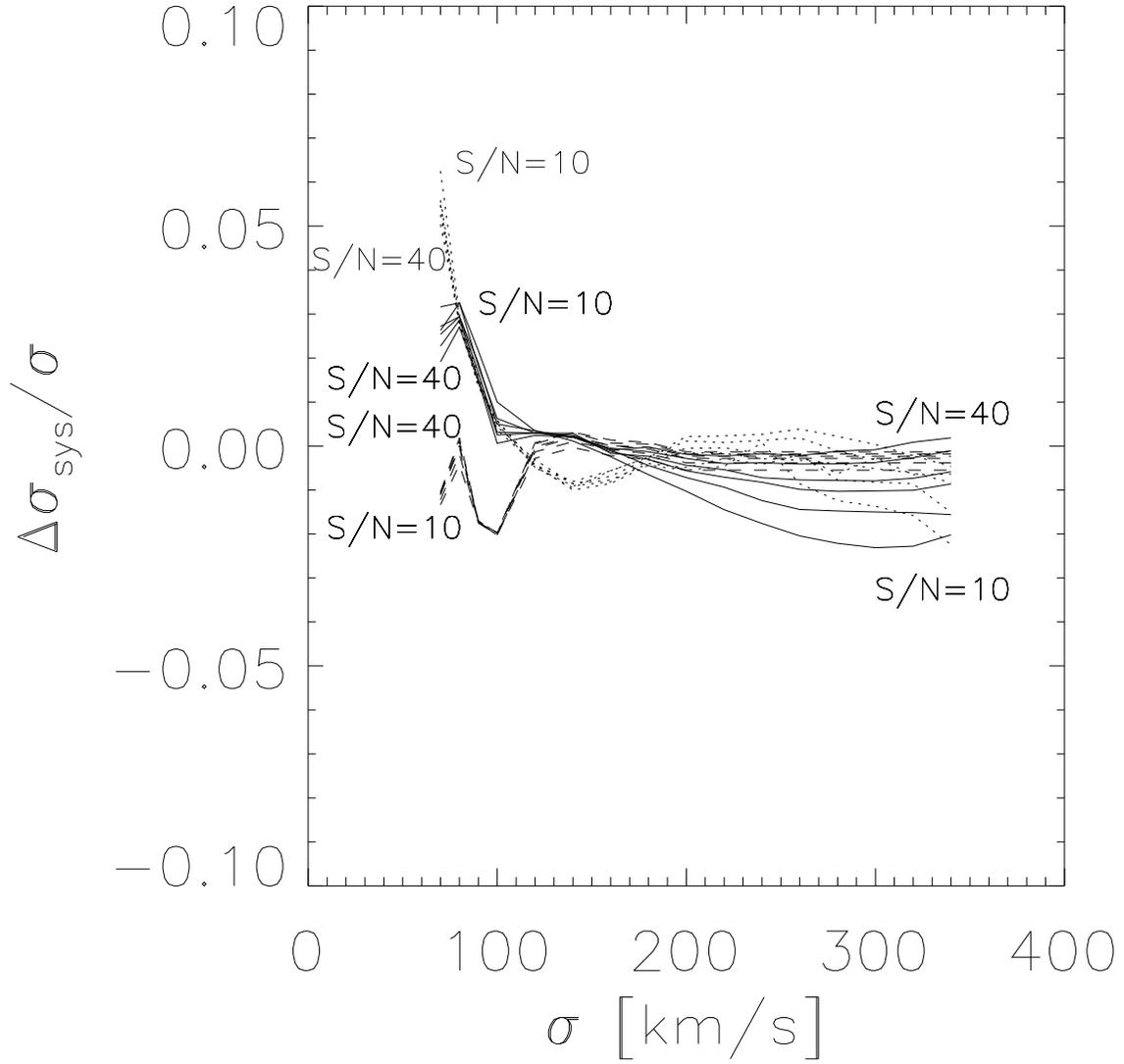}
\caption[]{Systematic biases in the three methods used to 
estimate the velocity dispersion.  Solid, dashed and dotted lines 
show the biases in the {\it Fourier-fitting}, {\it direct-fitting} and {\it
cross-correlation} 
methods, as a function of velocity dispersion and signal-to-noise.}
\label{fig:vsim}
\end{figure}

The best-case scenario is one in which there is no `template mismatch': 
the spectrum of the template star is exactly like that of the galaxy 
whose velocity dispersion one wishes to measure.  
Figure~\ref{fig:vsim} shows the fraction of systematic bias associated 
with each of the different methods in this best-case scenario.  
Slightly more realistic simulations, using a combination of stellar 
spectra as templates, were also done.  
The results are similar to those shown in Figure~\ref{fig:vsim}. 
With the exception of the {\it cross-correlation} method at low 
($\sigma < 100$ km~s$^{-1}$) velocity dispersion, the systematic 
errors on the velocity dispersion measurements appear to be smaller 
than $\sim 3\%$. 

\begin{figure}
\centering
\epsfxsize=\hsize\epsffile{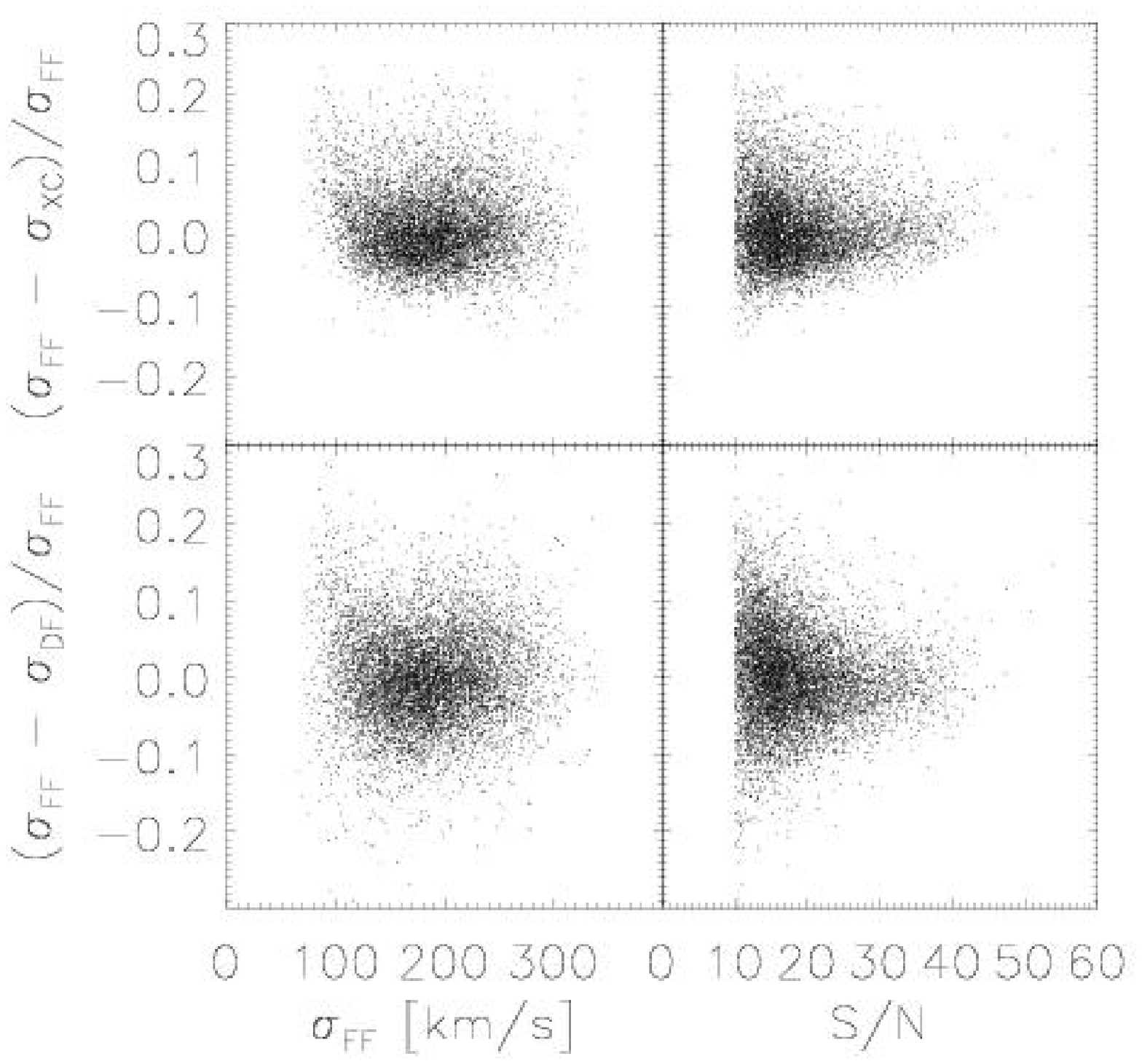}
\caption[]{Comparison of the various methods used to estimate 
the velocity dispersions; the agreement is quite good, with a 
scatter of about five percent.  Most of our spectra have 
$S/N \sim 15$, with approximately exponential tails on either 
side of this mean value.  }
\label{fig:vmeth}
\end{figure}

\begin{figure}
\centering
\epsfxsize=\hsize\epsffile{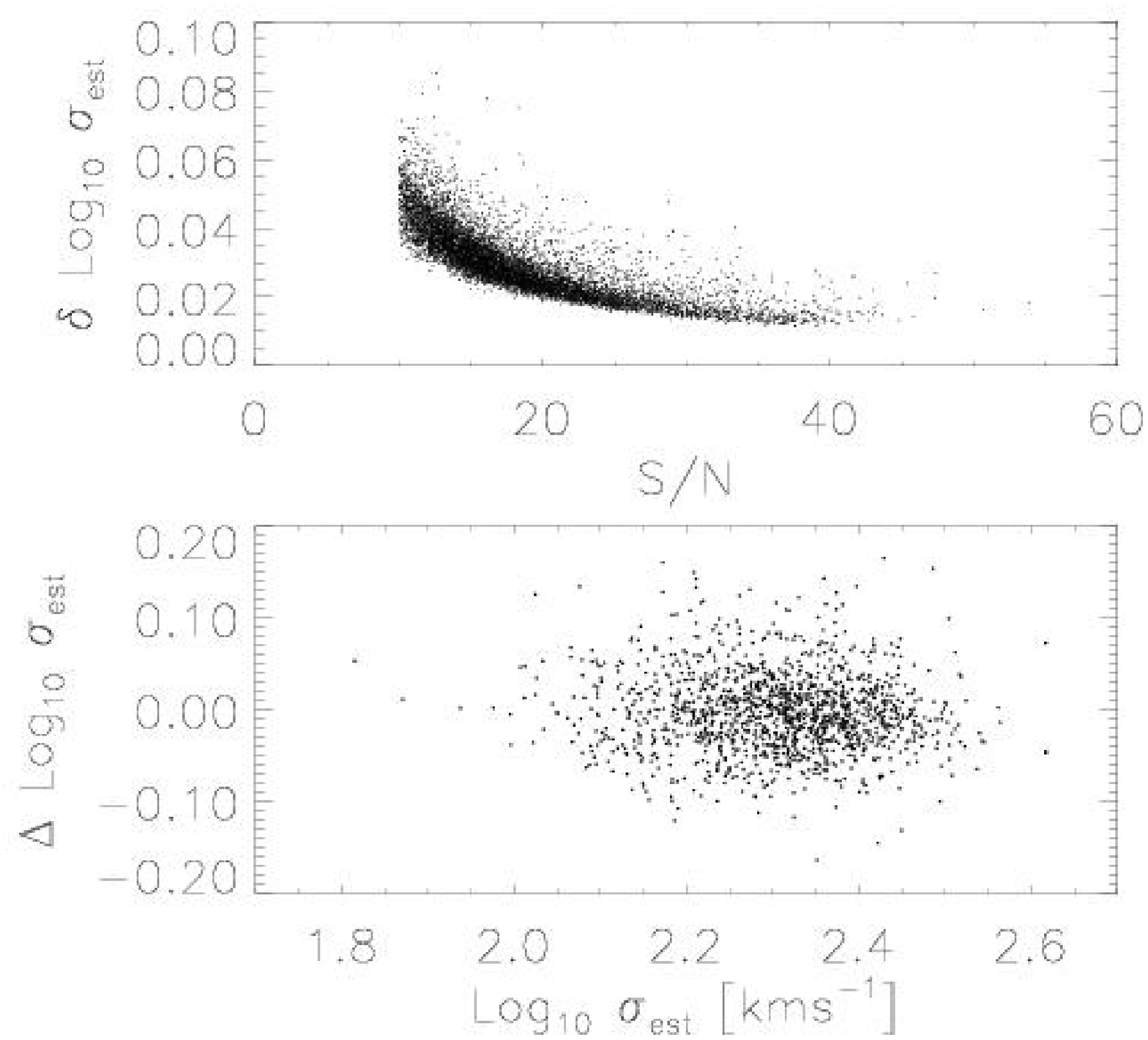}
\caption[]{Distribution of errors as a function of S/N (top) 
and comparison of estimates from repeated observations (bottom).  
Both panels suggest that, when the $S/N \ge 15$ then the typical 
error on an estimated velocity dispersion is 
$\delta \log_{10}\sigma < 0.04$.}
\label{fig:vrptd}
\end{figure}

Although the systematics are small, note that the measured velocity
dispersion is more biased at low velocity dispersions 
($\sigma < 100$ km~s$^{-1}$).  For any given $S/N$ and resolution, 
there is a lower limit on the velocity dispersion measurable without 
introducing significant bias.  Since the $S/N$ of the SDSS spectra is 
not very high (see, e.g., Figure~\ref{fig:vmeth}),
and because the instrumental resolution is $\sim 90$ km~s$^{-1}$, 
we chose 70 km~s$^{-1}$ as a lower limit.  
Figure~\ref{fig:vmeth} shows a comparison of the 
velocity dispersion estimates obtained from the three different methods 
for the galaxies in our sample.  The median offsets are not 
statistically significant and the rms scatter is $\sim 0.05$.  
On the other hand, the top panels suggest that the cross-correlation 
method sometimes underestimates the velocity dispersion, particularly 
at low $S/N$.  

We evaluate the dependence of the velocity dispersion on the wavelength 
range by fitting the spectra in different intervals: $4000 - 5800$~\AA\ 
which is the usual wavelength range used in the literature; 
$3900 - 5800$~\AA\ to test the effect of including the Ca H and K 
absorption lines (e.g., Kormendy 1982); 
$4000 - 6000$~\AA\ to test the effect of including the NaD line 
(e.g., Dressler 1984); and 
$4000 - 7000$~\AA\  and $4000 - 9000$~\AA\ to test the effect of 
including longer wavelengths. 
The velocity dispersion obtained with the Ca H and K is $\sim 2 \%$ 
larger than that obtained using the standard wavelength region 
$4000 - 5800$~\AA, and the rms difference between the three different 
methods increases to $\sim 7\%$. 
Including the NaD line increases the velocity dispersion by $\sim 3\%$ 
but does not increase the scatter between the different methods. 
Using the wavelength range $4000 - 7000$~\AA\  only provides velocity
dispersions which are $\sim 3\%$ larger than the values obtained if 
only $4000 - 5800$~\AA\  range is used.  On the other hand, in this 
wavelength region, the different methods (and measurements from repeated
observations) are in better agreement; the scatter is $\sim 8\%$ 
smaller than in the $4000 - 5800$~\AA\  region.  
In the range $4000 - 9000$~\AA, the velocity dispersion estimates 
increase by $\sim 7\%$. This last effect is probably due to the 
presence of molecular bands in the spectra of early-type galaxies at 
long wavelengths  (i.e., to the presence of cool stars).  
Furthermore, the scatter in this wavelength region increases
dramatically ($\sim 15\%$).  Presumably  this is due to the presence 
of higher sky-line residuals and lower $S/N$.  

The estimated velocity dispersion we use in the main text are obtained 
by fitting the wavelength range $4000- 7000$~\AA\ and then using the 
average of the estimates provided by the {\it Fourier-fitting} and 
{\it direct-fitting} methods to define what we call $\sigma_{\rm est}$.  
We do not use the cross-correlation estimate because of its behavior 
at low $S/N$ as discussed earlier.  

The top panel of Figure~\ref{fig:vrptd} shows the distribution of 
the errors on the velocity dispersion as a function of the $S/N$ of the 
spectra.  The errors for each method were computed by adding in quadrature 
the statistical error due to the noise properties of the spectrum, and 
the systematic error associated with the template and galaxy mismatches.  
The final error on $\sigma_{\rm est}$ is got by adding in quadrature 
the errors on the two estimates (i.e., the Fourier-fitting and 
direct-fitting) which we average.  
The resulting errors range from $0.02\le\delta\log_{10}\sigma\le 0.06$~dex, 
depending on the $S/N$ of the spectra, with a median value of 
0.03~dex.  

A few galaxies in our sample have been observed more than once.  
The bottom panel shows a comparison of the velocity 
dispersion estimates from multiple observations. The scatter between 
different measurements is $\sim 0.04$~dex, consistent with the amplitude 
of the errors on the measurements.  

\section{Velocity dispersion:  profiles and aperture corrections}\label{vr}

\begin{figure}
\centering
\epsfxsize=\hsize\epsffile{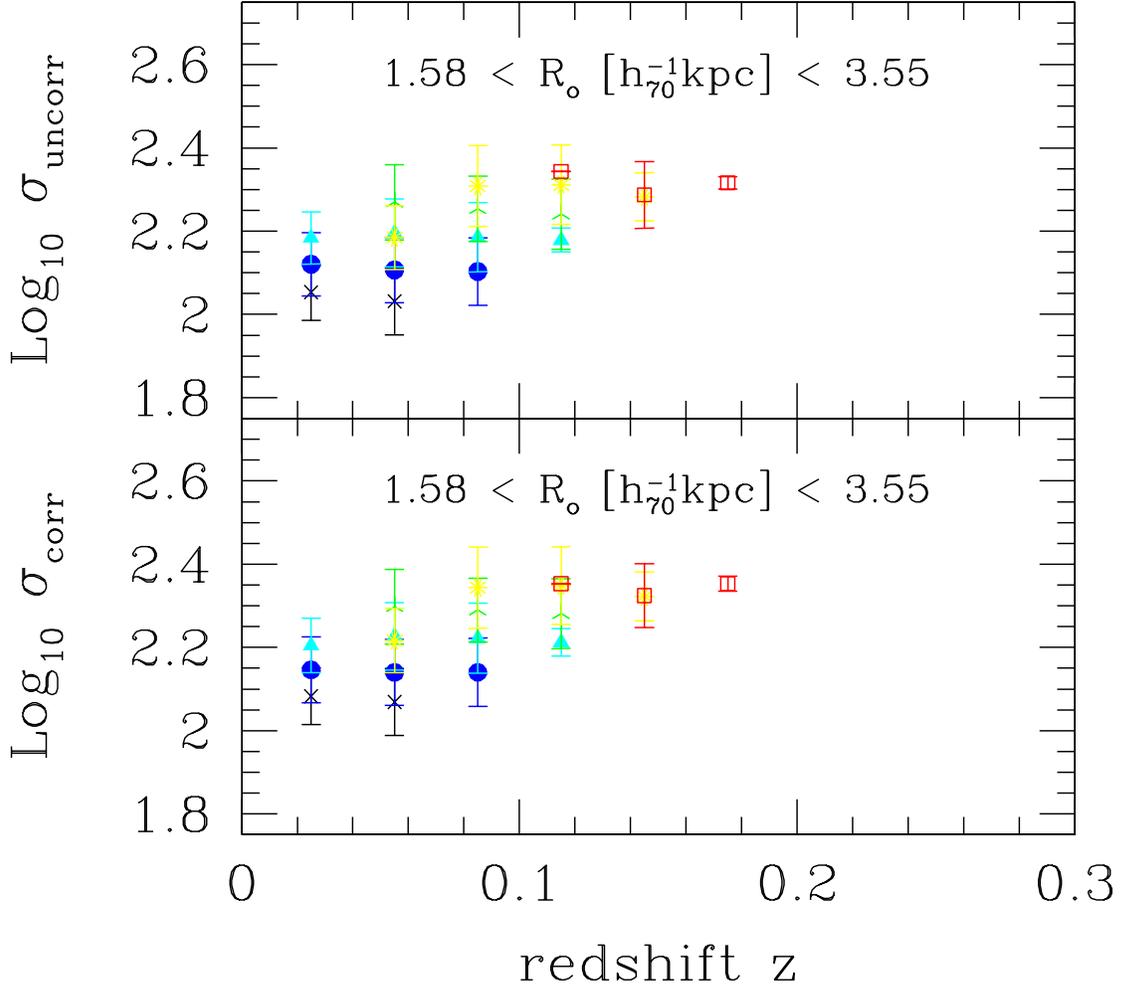}
\caption{Velocity dispersions of galaxies as a function of redshift.  
Top panel shows the estimated velocity dispersion, and bottom panel 
shows the values after correcting the estimate as described in the 
main text (equation~\ref{appcorr}).  
Different symbols show the result of averaging over 
volume-limited subsamples (same as in Figure~\ref{fig:XzR}) of 
galaxies having approximately the same luminosities and effective 
radii at each redshift.  (Error bars show the rms scatter around 
this mean value.)  The mean trends with redshift can be used to 
infer how, on average, the velocity dispersion changes with distance 
from the centre of the galaxy, and how this change depends on 
luminosity and effective radius.  }
\label{apcorr1}
\end{figure}

\begin{figure}
\centering
\epsfxsize=\hsize\epsffile{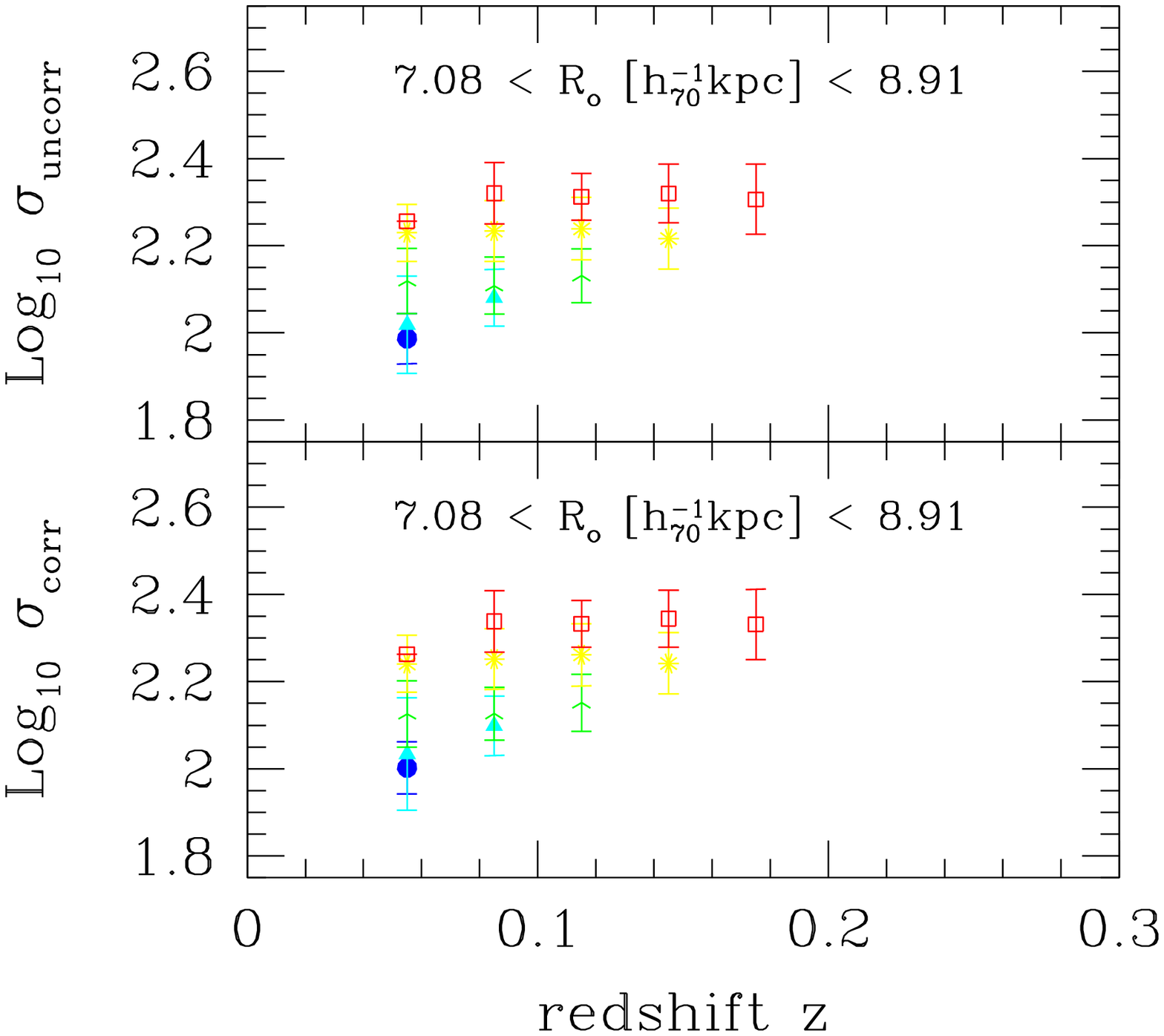}
\caption{As for previous figure, but for galaxies with larger radii.  }
\label{apcorr3}
\end{figure}

The SDSS spectra measure the light within a fixed aperture of radius 
1.5 arcsec.  Therefore, the estimated velocity dispersions of more 
distant galaxies are affected by the motions of stars at larger physical 
radii than for similar galaxies which are nearby.  If the velocity 
dispersions of early-type galaxies decrease with radius, then the 
estimated velocity dispersions (using a fixed aperture) of more distant 
galaxies will be systematically smaller than those of similar galaxies 
nearby.  

We have not measured the velocity dispersion profiles $\sigma(r)$ of 
any of the galaxies in our sample, so we cannot correctly account for 
this effect.  If we assume that the galaxies in our sample are similar 
to those for which velocity dispersion profiles have been measured, then 
we can use the published $\sigma(r)$ curves to correct for this effect.  
This is what equation~(\ref{appcorr}) in Section~\ref{vdisp} does.  

An alternative procedure can be followed if evolution effects are not 
important for the velocity dispersions in our sample.  To illustrate the 
procedure, the galaxies in each of the volume-limited subsamples 
shown in Figure~\ref{fig:XzR} were further classified into small bins 
in effective physical radius (i.e., $R_o$ in kpc/$h$, not $r_o$ in arcsec).  
Figures~\ref{apcorr1}--\ref{apcorr3} show the result of plotting the 
velocity dispersions of these galaxies versus their redshifts.  
Since the galaxies at low and high redshift are supposed to be similar, 
any trend with redshift can be used to infer and average velocity 
dispersion profile, and how the shape of this profile depends on 
luminosity and effective radius.  In this way, the SDSS data themselves 
can, in principle, be used to correct for the effects of the fixed 
aperture of the SDSS spectrograph.  

In practice, because there is substantial scatter in the velocity 
dispersions at fixed luminosity and size 
(Paper~II shows that this scatter is about 14\%), 
the trends in the present data set are relatively noisy.  
When the dataset is larger, it will be worth returning to this issue.  
For now, because the corrections are small anyway, we have chosen to 
use equation~(\ref{appcorr}) to correct the velocity dispersions.  
Nevertheless, curves like those presented above provide 
a novel way to study the velocity dispersion profiles of early-type 
galaxies.  

\section{Error estimates}\label{errors}
Let ${\cal E}$ denote the covariance matrix of the errors in 
our estimates of the the absolute magnitude $M$, the half light 
radius $R \equiv \log_{10}R_o$, and the velocity dispersion 
$V\equiv \log_{10}\sigma$:  
\begin{eqnarray}
{\cal E} &=& \left( \begin{array}{ccc} 
                      \epsilon^2_{MM} & \epsilon^2_{RM} & \epsilon^2_{VM} \\ 
                      \epsilon^2_{RM} & \epsilon^2_{RR} & \epsilon^2_{RV} \\
                      \epsilon^2_{VM} & \epsilon^2_{RV} & \epsilon^2_{VV} \\
                    \end{array}\right).
\end{eqnarray}
The elements of the error matrix ${\cal E}$ are obtained as follows.  

The photometric pipeline estimates the size $r_{\rm dev}$ and the 
apparent magnitude $m_{\rm dev}$ from the same fitting procedure.  
As a result, errors in these two quantities are correlated.  
Let $e_r$ denote the error in $\log_{10}r_{\rm dev}$, and $e_m$ the 
error in $m_{\rm dev}$.  The correlation means that we need three 
numbers to describe the errors associated with the fitting procedure, 
$\langle e_r e_r\rangle$, $\langle e_m e_m\rangle$, and 
$\langle e_m e_r\rangle$, but the pipeline only provides two.  
The error output by the pipeline in $r_{\rm dev}$, is correctly 
marginalized over the uncertainty in $m_{\rm dev}$, so it is 
essentially $\langle e_r e_r\rangle$.  
On the other hand, the quoted error in $m_{\rm dev}$, say 
$\langle e_{photo}^2\rangle$  is really 
$\langle e_m e_m\rangle -
\langle e_m e_r\rangle^2/\langle e_r e_r\rangle$.  
To estimate the values of $\langle e_m e_r\rangle$ and 
$\langle e_m e_m\rangle$ which we need, we must make an assumption 
about the correlation between the errors.  

Fortunately, this can be derived from the fact that, for a wide 
variety of galaxy profile shapes, the quantity 
$\xi\equiv e_r - \alpha e_\mu$, with $\alpha\approx 0.3$, 
has a very small scatter (e.g. Saglia et al. 1997).  Here 
$\mu \equiv m_{\rm dev} + 5\log_{10}r_{\rm dev} + 2.5\log_{10}(2\pi)$ 
is the surface brightness, and $e_\mu$ is the error in the surface 
brightness.  As a result, 
\begin{eqnarray}
\langle e_\mu e_\mu\rangle &=& {\langle e_r^2\rangle\over\alpha^2} + 
{\langle\xi^2\rangle\over\alpha^2}{(\alpha^2-1)\over (1+\alpha^2)} \nonumber\\
\langle e_\mu e_r\rangle &=& {\langle e_r^2\rangle\over\alpha} - 
{\langle\xi^2\rangle\over \alpha (1+\alpha^2)}
\end{eqnarray}
(Saglia et al. 1997).  This means that 
$\langle e_r\xi\rangle = \langle \xi^2\rangle/(1+\alpha^2)$, so that 
\begin{eqnarray}
\langle e_m e_r\rangle &=& 
      \langle e_r e_r\rangle \left(1-5\alpha\over\alpha\right) - 
      {\langle \xi^2\rangle\over \alpha(1+\alpha^2)} ,\nonumber\\
\langle e_m e_m\rangle &=& 
      \langle e_r e_r\rangle \left(1-5\alpha\over\alpha\right)^2 
      + {\langle \xi^2\rangle\over \alpha^2}\,
        \left[1 - 2\left(1-5\alpha\over 1+\alpha^2\right)\right],
\end{eqnarray}
and 
\begin{equation}
\langle e_m e_m\rangle - 
{\langle e_m e_r\rangle^2\over\langle e_r e_r\rangle} = 
 {\langle \xi^2\rangle\over \alpha^2}  
- {\langle \xi^2\rangle\over\alpha^2} 
  {\langle \xi^2\rangle/\langle e_r e_r\rangle \over(1+\alpha^2)^2} = 
   \langle e_{photo}^2\rangle.  
\end{equation}
The final equality shows that the error output from the pipeline 
provides an estimate of $\langle \xi^2\rangle$ which we can insert into 
our expressions for $\langle e_m e_m\rangle$, $\langle e_m e_r\rangle$, 
and $\langle e_\mu e_\mu\rangle$.  
(Notice that if $\langle \xi^2\rangle\ll \langle e_r e_r\rangle$, then 
it would be a good approximation to set 
 $\langle e_{photo}^2\rangle\approx \langle \xi^2\rangle/\alpha^2$.  
Since this is not always the case for our dataset, we must solve 
the quadratic.)  Once this has been done, we set 
\begin{eqnarray}
\epsilon_{MM}^2 &=& \langle e_m e_m\rangle  ,\nonumber\\
\epsilon_{RR}^2 &=& \langle e_r e_r\rangle + 
                    {\langle e_{ab}\,e_{ab}\rangle\over 4} ,\nonumber\\
\epsilon_{RM}^2 &=& \langle e_m e_r\rangle ,\nonumber\\
\epsilon_{VM}^2 &=& 0 ,\nonumber\\
\epsilon_{VV}^2 &=& \langle e_v e_v\rangle + (0.04\,\epsilon_{RR})^2,
 \nonumber\\
\epsilon_{RV}^2 &=& -0.04\,\epsilon_{RR}^2 ,
\end{eqnarray}
That is, we compute the error in the absolute magnitude by assuming 
that there are no errors in the determination of the redshift (and 
$K$ correction!) which would otherwise propagate through.  

Subsequent papers will focus almost exclusively on the circularly 
averaged radius $R_o$ defined in Section~\ref{LandR}.  The 
errors on it are given by adding the errors in the size 
$r_{\rm dev}$ to those which come from the error on the shape $b/a$.  
We assume that the errors in $b/a$ are neither correlated with those 
in $\log_{10}r_{\rm dev}$ nor with those in the absolute magnitude.  
Finally, we assume that errors in magnitudes are not correlated 
with those in velocity dispersion, so $\langle\epsilon^2_{VM}\rangle$ 
is set to zero, and that errors in size and velocity dispersion are 
only weakly correlated because of the aperture correction we apply.  
Here $\langle e_v e_v\rangle$ is the error in what was called 
$\log_{10} \sigma_{\rm est}$ in the main text.  

These error estimates are presented in Tables~\ref{tab:obs} 
and~\ref{tab:phys}, where we have set 
$\delta m_{photo} = \sqrt\langle e_m^2\rangle$, 
$\delta_r = \sqrt\langle e_r^2\rangle$, and 
$\delta M = \epsilon_{MM}$.

\end{document}